\documentclass[11pt,onecolumn]{article}
\usepackage{setspace,dcolumn}
\usepackage{subfig}
\usepackage{hyperref}
\usepackage{cite}
\usepackage{graphicx,psfrag,epsfig,graphbox}
\usepackage[font=small,format=plain,labelfont=bf,justification=raggedright,singlelinecheck=false]{caption}
\usepackage{amsmath,amsfonts,amssymb,amsthm,bm,upgreek}
\allowdisplaybreaks[2]
\usepackage{geometry}
\usepackage[mathscr]{eucal}
\usepackage{esvect}
\usepackage{booktabs,threeparttable}
\usepackage{indentfirst}
\usepackage[normalem]{ulem}
\usepackage{lscape}
\usepackage{physics}
\usepackage{subfiles}
\usepackage{multirow}
\usepackage{tikz}
\usepackage{tikz-cd}
\usepackage{quiver}

\usepackage{extarrows}

\usepackage{color}
\definecolor{darkgreen}{RGB}{40,150,60}
\definecolor{violet}{RGB}{140,50,230}
\definecolor{orange}{RGB}{230,150,0}

\newcommand{\glie}{\mathfrak{g}}

\newcommand{\solie}{\mathfrak{so}}
\newcommand{\isolie}{\mathfrak{iso}}

\newcommand{\llie}{\mathfrak{l}}
\newcommand{\extension}{\operatorname{Ext}}
\newcommand{\C}{\mathbb{C}}
\newcommand{\cO}{\mathcal{O}}
\newcommand{\wave}[1]{\widetilde{#1}}

\hypersetup{colorlinks=true,linkcolor=blue,anchorcolor=blue,citecolor=blue,urlcolor=blue}
\numberwithin{equation}{section}

\title{Constructing Carrollian Field Theories from Null Reduction}
\author{Bin Chen$^{1,2,3}$, Reiko Liu$^4$, Haowei Sun$^1$, Yu-fan Zheng$^1$}

\begin{document}
\maketitle
\begin{center}
	{\it
		$^{1}$School of Physics, Peking University, No.5 Yiheyuan Rd, Beijing 100871, P.~R.~China\\
		\vspace{2mm}
		$^{2}$Collaborative Innovation Center of Quantum Matter, No.5 Yiheyuan Rd, Beijing 100871, P.~R.~China\\
		$^{3}$Center for High Energy Physics, Peking University, No.5 Yiheyuan Rd, Beijing 100871, P.~R.~China\\
		$^{4}$Yau Mathematical Sciences Center, Tsinghua University, Beijing 100084, P.~R.~China\\
	}
	\vspace{10mm}
\end{center}

\begin{abstract}
    \vspace{5mm}
    \begin{spacing}{1.5}
        In this paper, we propose a novel way to construct off-shell actions of $d$-dimensional Carrollian field theories by considering the null-reduction of the Bargmann invariant actions in $d+1$ dimensions.  This is based on the fact that $d$-dimensional Carrollian symmetry is the restriction of the $(d+1)$-dimensional Bargmann symmetry to a null hyper-surface. We focus on free scalar field theory and electromagnetic field theory, and show that the electric and magnetic sectors of these theories originate from different Bargmann invariant actions in one higher dimension. In the cases of the massless free scalar field and $d=4$ electromagnetic field, we verify Carrollian conformal invariance of the resulting  theories,  and find that there appear naturally chain representations and staggered modules of Carrollian conformal algebra.
    \end{spacing}
\end{abstract}
\newpage

\setcounter{tocdepth}{2}
\tableofcontents

\section{Introduction}\label{sec:Introduction}
    
    Carrollian symmetry group was first found by L\'evy-Leblond in 1965\cite{Levy-Leblond:1965} (and independently by Sen Gupta \cite{Gupta:1966}) by studying the  ultra-relativistic ($c \to 0$) contraction of the Poincar\'e group.  Later on, Carroll group was discovered\cite{Bacry:1968zf} to be one of possible kinematical groups, which means that it could be the spacetime symmetry of a nonrelativistic manifold, the Carrollian manifold. The Carroll group is generated by the Carrollian boost
    \begin{equation}
        \vec{x}~'=\vec{x}, \hspace{3ex}t'=t-\vec{b}\cdot \vec {x},
    \end{equation}
    the translations, and the rotations among spatial directions. In the Carrollian limit, the lightcones collapse, and there appears the notion of absolute space. Consequently the motion of a free Carrollian particle is trivial: it runs without moving\cite{Levy-Leblond:1965,Duval:2014uoa}. For a massless Carrollian particle, it has infinite dimensional Carrollian conformal symmetry\cite{Bergshoeff:2014jla}. However, for interacting Carrollian particles, there could be  nontrivial dynamics. 
    
    Since its discovery, the Carrollian symmetry has been found in various physical systems. The Carrollian boost was discovered in the isometry of plane-gravitational wave\cite{Souriau:1973,Duval:2017els}. The Carrollian limit was found to control the dynamics of the gravitational field near a spacelike singularity\cite{Henneaux:1979vn,Belinski:2017fas}. The Carrollian symmetry appears in the near horizon of black hole as well\cite{Penna:2018gfx,Donnay:2019jiz,Freidel:2022vjq,Redondo-Yuste:2022czg}. Recently, it was applied to the study of physical problems in  cosmology\cite{deBoer:2021jej} and condensed matter physics\cite{Casalbuoni:2021fel,Pena-Benitez:2021ipo}. More importantly, the Carrollian conformal symmetry\cite{Duval:2014uva}, which is isomorphic to the BMS group\cite{Bondi:1962px,Sachs:1962wk,Sachs:1962zza,Barnich:2009se}, is essential in the study of $3d$ flat space holography\cite{Bagchi:2012xr,Hartong:2015usd, Ciambelli:2019lap} and celestial holography\cite{Donnay:2022aba,Bagchi:2022emh,Donnay:2022wvx}. \par
    
    The Carrollian invariant field theories can be constructed  by taking the ultra-relativistic  contraction of a Lorentz invariant field theories\cite{Duval:2014uoa,Henneaux:2021yzg,Bagchi:2016bcd,Bagchi:2019clu,Bagchi:2019xfx,Bergshoeff:2022qkx,Bergshoeff:2022eog}. One way is to taking the limit on the equations of motion directly. There are actually two inequivalent Carroll contractions, leading to two different Carrollian field theories, usually named by electric sector and magnetic sector respectively\cite{Duval:2014uoa}. From the construction, the theories are manifestly on-shell Carrollian invariant, but their off-shell actions need to be constructed separately. For the electric sector the construction is usually easy, but for the magnectic sector the construction turns out to be difficult\cite{Bagchi:2019clu}. Another way to construct the Carrollian invariant field theories by contraction is based on the Hamiltonian action principle\cite{Henneaux:2021yzg}. Even though this approach can yield Corrollian invariant action automatically, after gauging away the extra field and accordingly modifying the transformation rules,  it lacks manifest spacetime covariance\footnote{Some examples of Carrollian diffeomorphism invariant theory on general Carrollian manifolds are discussed in \cite{Rivera-Betancour:2022lkc,Hansen:2021fxi}.}.\par
    
    In this paper, we propose a novel method for constructing $d$-dimensional Carrollian field theories by performing null reduction on $(d+1)$-dimensional theories defined on the flat Bargmann manifold. The logic behind our construction is similar to the one of the null reduction technique employed in the Galilean case \cite{Julia_1995}. The key point is that the Carrollian symmetry is a subgroup of the Bargmann group. Specifically, by disregarding the translation along a null direction, the Bargmann group reduces to the Carroll group. Thus, if we begin with Bargmann-invariant theories and carry out reductions along the null direction, we will obtain the theories which is guaranteed to be Carrollian invariant. Our focus in this work is primarily on the free massless scalar and electromagnetic field theories. Previous discussions in the literature, such as \cite{Duval:2014uoa,Henneaux:2021yzg}, have addressed the existence of two distinct rescalings of the fields when taking the limit $c\to 0$, leading to the electric and magnetic sectors of the theories. In our approach, these two sectors arise from two different Bargmann field theories in one higher dimension. The resulting electric sector is exactly the same as the one found in \cite{Henneaux:2021yzg}, while the resulting magnetic sector differs slightly from the one in \cite{Duval:2014uoa,Henneaux:2021yzg}. We demonstrate that our action is off-shell Carrollian invariant\footnote{The on-shell reduction from Bargmann theories has been discussed in \cite{Duval:2014uoa}. For recent studies on BMS invariant theories, refer to \cite{Bagchi:2022eav,Bagchi:2022nvj,Baiguera:2022lsw,Saha:2022gjw,Liu:2022mne,Dutta:2022vkg}.}, and subsequently calculate the correlators using the path-integral formalism. 

    Another motivation for present study is to find Carrollian conformal invariant theories and study their properties. The Corrollian conformal field theory (CCFT) plays an important role in flat space holography\cite{Bagchi:2012xr,Hartong:2015usd, Ciambelli:2019lap} and celestial holography\cite{Donnay:2022aba,Bagchi:2022emh,Donnay:2022wvx}. In particular,  higher dimensional ($d\geq 3$) CCFT presents some novel features\cite{Chen:2021xkw}. First of all, the representations of the higher dimensional Carrollian conformal algebra(CCA) are much more involved. There appear the multiplet structure and staggered modules in the highest-weight representations. Secondly the constraints from the Ward identity on the two-point correlators are less restrictive. It is important to construct concrete CCFTs and study their properties.  As will be discussed in section \ref{subsec:CarrollianConformalSymmetry}, the $d$-dimensional Carrollian conformal group is not a subgroup of the $(d+1)$-dimensional Bargmann conformal group. This implies that the null reduction of a Bargmann conformal invariant theory does not automatically yield a CCFT, and we need to check the Carrollian conformal invariance of null-reduced theories case by case. In this work, we demonstrate  that the free Carrollian scalar theory and the Carrollian electromagnetic theory in $d=4$ are really Carrollian conformal invariant, by checking the invariance of the actions and the Ward identities of 2-point correlators of the primary operators. We also discuss the representations of the fields, and find that the staggered modules appear naturally in these theories. \par
    
    The remaining parts of this paper are organized as follows. In Section \ref{sec:CarrollianSymmety}, we give a brief review of the Carrollian symmetry, its conformal extension, and the representations of  Carrollian conformal algebra. Then in Section \ref{sec:CarrollianScalar} we construct the electric and magnetic sectors of free Carrollian scalar theory from null reduction of the Bargmann scalar theory. We read the 2-point correlators of the fundamental fields from the path-integral and check the conformal symmetries in both sectors. In section \ref{sec:CarrollianPForm}, we investigate more subtle and nontrivial models, including the electromagnetic theory and free $p$-form field theories. For the electromagnetic theory, we discuss in detail the related boost multiplet structures and compute the 2-point correlators of the fundamental fields in both sectors by using the path-integrals in suitable gauges. In section \ref{quotientreduction}, we discuss the further reduction of the 4d electromagnetic theories from quotient representations.  We conclude with some discussions in Section \ref{sec:Discussions}.  
    Some technical details are presented in the appendices. We briefly review the construction of staggered module, and discuss the possible staggered modules involving the scalars in CCFT in Appendix \ref{app:stagger}. We collect the path-integral computations of the 2-point correlators of Carrollian field theories  in Appendix \ref{app:PathInt}. After briefly reviewing the Ward identities of Carrollian conformal symmetry, we discuss carefully their restrictions on the 2-point correlators of the primary operators in various representations in Appendix \ref{app:2pt-ward}. Different from the discussions in \cite{Chen:2021xkw}, we pay more attention to the correlators with $\delta$-function distribution, which appear in the field theories studied in this paper and in the celestial holography\cite{Donnay:2022aba,Bagchi:2022emh,Donnay:2022wvx}. 

    \paragraph{Convention:} In the present work we use the convention that the Greek alphabets $\alpha, \beta, \cdots$ denote all the spacetime indices while the Latin alphabets  $i,j, \cdots$ only denote spacial indices. Moreover, we use $\alpha,\beta, \gamma$ to denote the Bargmann spacetime, $\mu,\nu,\delta$ to denote Carrollian spacetime, and $u$ and $v$ to denote null indices.

\section{Carrollian symmetry}\label{sec:CarrollianSymmety}
    In this section, we  briefly review the Carrollian symmetry and its relation with Bargmann symmetry. Additionally, we introduce the fundamentals of Carrollian conformal symmetry and its representations, which have been thoroughly studied in \cite{Chen:2021xkw}.

 \subsection{Carrollian symmetry from Bargmann symmetry}\label{subsec:CarrollianSymmetryformBargmannSymmetry}
   In \cite{Duval:1984cj}, it was shown that Newton-Cartan geometry is associated with the so-called Bargmann group. Then in \cite{Duval:2014uoa}, it was pointed out that  the Galilean and Carroll groups and their related geometries could be unified in relativistic Bargmann space.  A Bargmann manifold has three ingredients  $(\mathscr{B}, G, \xi)$: $\mathscr{B}$ is a $(d+1)$-dimensional manifold,  $G$ is a metric  of Lorentz signature, and $\xi$ is a nowhere vanishing null vector. In the flat case, it can be described in terms of the coordinates $x^\alpha = (u,\vec x, v),~(\alpha=0,1\cdots d)$ as
    \begin{equation}\label{eq:BargmannStructure}
        \mathscr{B}=\mathbb{R}\times\mathbb{R}^{d-1}\times\mathbb{R}, \qquad G = 2du dv + \delta_{ij} dx^i dx^j, \qquad \xi = \partial_u,
    \end{equation}
    where both $u,v$ are the lightcone coordinates and $\vec{x}$ is a $(d-1)$-vector. The Bargmann group is the isometry group of the flat Bargmann structure \eqref{eq:BargmannStructure} that keeps the metric $G$ and the null vector $\xi$ invariant. It is a subgroup of Poincar\'e group 
    \begin{equation}
        \mbox{Barg}(d,1) = ISO(d,1)\setminus \{J^0_{~d}, 1/\sqrt{2}\left(J^i_{~0}-J^i_{~d}\right)\}.
    \end{equation}
     The Bargmann generators are $P_\alpha, J^i_{~j}$, and $B^\mathscr{B}_i$, where $B^\mathscr{B}_i$ are Bargmann boosts. Their realizations as vector fields on the spacetime are shown in Table \ref{tb:BargmannAction}. The  commutation relations of the generators are
     \begin{equation}
         \begin{aligned}
            &[B^\mathscr{B}_i,P_v]=-P_i, \quad [B^\mathscr{B}_i,P_j]=\delta_{ij} P_u, \quad [B^\mathscr{B}_i,P_u]=0,\\
            &[J^{i}_{~j},J^{k}_{~l}]=\delta^{ik}J_{jl}-\delta^{i}_{l}J_j^{~k}+\delta_{jl}J^{ik}-\delta_j^{k}J^{i}_{~l}, \\
            &[J^{i}_{~j},P_k]=\delta^{i}_{~k}P_j-\delta_{jk}P^i, \quad [J^{i}_{~j},B^\mathscr{B}_k]=\delta^{i}_{~k}B^\mathscr{B}_j-\delta_{jk}B^{\mathscr{B}i}, \quad\text{others}=0.\\
         \end{aligned}
     \end{equation}
     
    \begin{table}[ht]
        \def\arraystretch{1.6}
        \centering
        \caption{\centering Generators of Bargmann symmetry as vector fields on the spacetime. }
        \label{tb:BargmannAction}
        \begin{tabular}{clc}
            \hline
            Generators & \quad Vector fields & Finite transformations\\
            \hline
            $P_\alpha$ & \quad$p_\alpha = \partial_\alpha$ & $x^\alpha+x^\alpha_0$ \\
            $J^{i}_{~j}$ & \quad$j^{i}_{~j} = x^i\partial_j-x_j\partial^i$ & $\left(u, \mathbf{J} \cdot \vec x, v\right)$ \\
            $B^\mathscr{B}_i$ & \quad$b^\mathscr{B}_i = v\partial_i - x_i\partial_u$ & \quad\quad$\left(u -\vec \nu \cdot \vec x -\frac{1}{2}\vec \nu^2 v, \vec x +\vec \nu v, v\right)$ \\
            \hline
        \end{tabular}
    \end{table}\par
    
    Geometrically, the Carroll group can be viewed as a subgroup of the Bargmann group that preserves the $v=0$ null hyper-surface. By restricting  to the null hyper-surface $v=0$, we see immediately that the flat Bargmann structure reduces to the flat Carrollian structure $(\mathscr{C},g,\xi)$ with the coordinates $x^\mu = (~t=u~, \vec x)$, the degenerated metric $g_{\mu\nu}=\left. G_{\mu\nu} \right|_{v=0}=\delta_{\mu}^i\delta_{\nu}^j\delta_{i j}$, and the timelike vector $\xi=\partial_t$. A Bargmann transformation naturally induces a transformations on the $v=0$ null hyper-surface if it leaves the $v=0$ null hyper-surface invariant.\par
    
    The commutation relations of the generators of the Carrollian algebra are
     \begin{equation}
         \begin{aligned}
            &[B_i,P_j]=\delta_{ij} P_0, \quad [B_i,P_0]=0\\
            &[J^{i}_{~j},J^{k}_{~l}]=\delta^{ik}J_{jl}-\delta^{i}_{l}J_j^{~k}+\delta_{jl}J^{ik}-\delta_j^{k}J^{i}_{~l}, \\
            &[J^{i}_{~j},P_k]=\delta^{i}_{~k}P_j-\delta_{jk}P^i, \quad [J^{i}_{~j},B_k]=\delta^{i}_{~k}B_j-\delta_{jk}B^i, \quad\text{others}=0.\\
         \end{aligned}
     \end{equation} \par

  \subsection{Carrollian conformal symmetry}  \label{subsec:CarrollianConformalSymmetry}
    
    Besides as the subgroup of the Bargmann group, the Carroll group can be obtained from the ultra-relativistic ($c\to0$) contraction of the Poincare group as well. Moreover, the Carrollian conformal symmetry comes naturally from the  $c\to0$ limit of relativistic conformal symmetry. The symmetry algebra of the Carrollian conformal group is generated by  $\{P_\mu, J^i_{~j}, B_i, D, K_\mu\}$, $\mu=0,1,\dots,d-1,~i,j=1,\dots,d-1$ with the commutation relations\footnote{The spacial indices are raised (lowered) by $\delta^{ij}$  ($\delta_{ij}$).}
    \begin{equation}\label{eq:CCACommutations}
        \begin{aligned}
            &[D,P_\mu]=P_\mu, ~~ [D,K_\mu]=-K_\mu, ~~ [D,B_i]=[D,J^{i}_{~j}]=0, \\
            &[J^{i}_{~j},G_k]=\delta^{i}_{~k}G_j-\delta_{jk}G^i, ~~ G\in\{P,K,B\}, ~~\\ &[J^{i}_{~j},P_0]=[J^{i}_{~j},K_0]=0,\\
            &[J^{i}_{~j},J^{k}_{~l}]=\delta^{ik}J_{jl}-\delta^{i}_{l}J_j^{~k}+\delta_{jl}J^{ik}-\delta_j^{k}J^{i}_{~l}, \\
            &[B_i,P_j]=\delta_{ij}P_0, ~~ [B_i,K_j]=\delta_{ij}K_0, ~~ [B_i,B_j]=[B_i,P_0]=[B_i,K_0]=0, \\
            &[K_0,P_0]=0, ~~ [K_0,P_i]=-2B_i, ~~ [K_i,P_0]=2B_i, ~~ [K_i,P_j]=2\delta^{i}_{j}D+2J^{i}_{~j}.
        \end{aligned}
    \end{equation}
    This algebra is isomorphic to the Poincar\'e algebra $\mathfrak{cca}_{d}\simeq \mathfrak{iso}(d,1)$, however their homogeneous vector space realizations are different. The actions of the generators of Carrollian conformal group on space-time point $(t,\vec x)$ are shown in Table \ref{tb:CCAAction}.
    
    \begin{table}[ht]
        \def\arraystretch{1.6}
        \centering
        \caption{\centering Generators of CCA  as the vector fields on the space-time.}
        \label{tb:CCAAction}
        \begin{tabular}{clc}
            \hline
            Generators & \qquad Vector fields & Finite transformations\\
            \hline
            $D$ & \qquad$d = t\partial_t+x^i\partial^i$ & $\lambda x^\mu$ \\
            $P_\mu$ & \qquad$p_\mu = \left(\partial_t ~, ~ \vec\partial\right)$ & $x^\mu+a^\mu$ \\
            $K_\mu$ & \qquad$k_\mu = \left(-\vec x^2\partial_t, 2\vec x x_\mu\partial^\mu-\vec x^2\vec\partial\right)$ & $\left(\frac{t-a^0\vec x^2}{1-2\vec a \cdot \vec x+\vec a^2\vec x^2},\frac{\vec x-\vec a\vec x^2}{1-2\vec a \cdot \vec x+\vec a^2\vec x^2}\right)$ \\
            $B_i$ & \qquad$b_i = x_i\partial_t$ & $\left(t+\vec v \cdot \vec x, \vec x\right)$ \\
            $J^{i}_{~j}$ & \qquad$j^{i}_{~j} = x^i\partial_j-x_j\partial^i$ & $\left(t, \mathbf{J} \cdot \vec x\right)$ \\
            \hline
        \end{tabular}
    \end{table}
    
    As shown in the last subsection, the Carroll group appears as the restriction of Bargmann group on a null hyper-surface of Bargmann manifold.  However, this is not true for the conformal case. Recall that the  (Lorentzian) conformal group is generated by the diffeomorphisms that transform the metric $g^L$ as
    \begin{equation}
        a^* g^L = \Omega^2 g^L.
    \end{equation}
    In the Bargmann case, there are two geometric notions, the metric $G$ and the invariant null-vector $\xi$. One can similarly define the conformal transformations of order $k$ as
    \begin{equation}\label{eq:BargmannConformalKillingEq}
        a^* G = \Omega^2 G, \qquad a^* \xi = \Omega^{-2/k}\xi.
    \end{equation}
    One may take $k=2$ to keep the scaling in the $\xi$ direction  the same as the ones in other directions. It turns out that the conformal version of the Bargmann group is a subgroup of Lorentzian conformal  group, keeping the null vector $\xi$ invariant. Indeed, the group is generated by $\{P_\alpha, J^i_{~j}, B_i, D\}$, where $D$ is the dilation generator. But now the Lorentzian special conformal transformations (SCT) do not satisfy the conformal condition \eqref{eq:BargmannConformalKillingEq}. In other words,  the $G$-preservering and $\xi$-preserving  subgroup of Lorentz conformal group consists of just the Bargmann transformations and a single dilation  without  special conformal transformations.
    
    However, things become different on the null hyper-surface $v=0$. Since the metric in the Carrollian manifold is degenerate, the Killing equation $a^* g = \Omega^2 g$ is less constrained. Actually the solutions to the Carrollian conformal Killing equations are
    \begin{equation}
        \begin{aligned}
            & p_i = \partial_i, \qquad m^i_{~j} = x^i\partial_j - x_j\partial^i,&\\
            & d = x^\mu\partial_\mu, \qquad k_i = 2 x_i x^\mu\partial_\mu - x^l x_l \partial_i,& m = g(x^i)\partial_0,
        \end{aligned}
    \end{equation}
    where $m$'s are the vector fields for infinite-dimensional extensions of the Carrollian conformal transformations and $g(x^i)$ is an arbitrary function of spacial coordinates. Especially, the global transformations in the $m$'s include the temporal translation, the boosts, and the temporally special conformal translation:
    \begin{equation}
       p_0 = \partial_0,\quad b_i = x_i\partial_0, \quad k_0 = - x^l x_l \partial_0. 
    \end{equation}
    Thus the (global) Carrollian conformal group\footnote{In this work, the Carrollian conformal symmetry always means the global one.} is generated by $\{P_\mu, J^i_{~j}, B_i, D, K_\mu\}$ with $K_0=(K^L_{d+1}+K^L_0)/\sqrt{2}|_{v=0}$ and $K_i=K^L_i|_{v=0}$. We see that Carrollian conformal group is not a subgroup of the Bargmann conformal  group.\par
    
\subsection{Representations of Carrollian conformal algebra} \label{subsec:CCAReps}
    
    The representations of the higher dimensional Carrollian conformal algebra (CCA) are much more involved than the ones of its Lorentzian cousin. The so-called scale-spin representation was used to study the representation of homogeneous Carrollian conformal group\cite{Bagchi:2019xfx}. However, this description is not precise enough to discuss the representations with a complicated boost multiplet structure. In \cite{Chen:2021xkw}, the authors discussed the highest-weight representations (HWR) of the Carrollian conformal group in detail. Here we outline some main results. \par

The stabilizer algebra of CCA is generated by 
dilation $D$, Carrollian rotations $M=\{J,B\}$ and special conformal transformations (SCTs) $K$. 
 The local operators $\mathcal{O}^a$ can be diagonalized simultaneously into the  eigenstates of the dilation and the representations of Carrollian rotations, 
\begin{equation}
    [D,\mathcal{O}]=\Delta_{\mathcal{O}} \mathcal{O},\hspace{3ex} [M,\mathcal{O}^a]=\Sigma^a_b \mathcal{O}^b,\end{equation}
    where $\Delta$ is the scaling dimension  of the operator. 
The highest-weight operator, which is often called primary operator, in a given representation is defined as the operator with the lowest eigenvalue of dilation generator $D$, satisfying the primary condition
\begin{equation}
[K, \mathcal{O}]=0.
\end{equation}
An operator in a highest-weight representation is therefore characterized by the scaling dimension and its representation with respect to the Carrollian rotations. 

Taken as an example, the scalar primary operator $\mathcal{O}_p$ at the origin, being the scalar under the Carrollian rotations,  has  scaling dimension $\Delta$ and commutes with all other generators, including the $K$ generators, 
    \begin{equation}
        \begin{aligned}
            &[D,\mathcal{O}_p]=\Delta\mathcal{O}_p, \quad [J_{ij}, \mathcal{O}_p]=[B_i,\mathcal{O}_p]=0, \quad [K_\mu,\mathcal{O}_p]=0. \\
        \end{aligned}
    \end{equation}
    The operators generated by acting one generator of $P_\mu$ on $\mathcal{O}_p$ are descendants and  have conformal dimension $(\Delta+1)$,
    \begin{equation}
        \begin{aligned}
            & P_\mu\mathcal{O}_p = [P_\mu, \mathcal{O}_p] = \partial_\mu\mathcal{O}_p, \quad
            [D, P_\mu\mathcal{O}_p] = (\Delta + 1) P_\mu\mathcal{O}_p. \\
        \end{aligned}
    \end{equation}
Acting the operator $K_\mu$ on $P_\mu\mathcal{O}_p$ leads back to $\mathcal{O}_p$:
\begin{equation}
   [K_i,P_j\mathcal{O}_p] = 2\Delta\delta_{ij}\mathcal{O}_p, \quad [K_0, P_i\mathcal{O}_p] =[K_i, P_0\mathcal{O}_p] =[K_0, P_0\mathcal{O}_p] = 0. 
\end{equation}
Similar to usual CFT, there are higher orders of descendant operator of $\mathcal{O}_p$ by acting more momentum operators. The primary operator $\mathcal{O}_p$ together with all its descendants are referred to as the conformal family of $\mathcal{O}_p$.
    
\begin{figure}[ht]
        \centering
        \includegraphics[width=7cm,align=c]{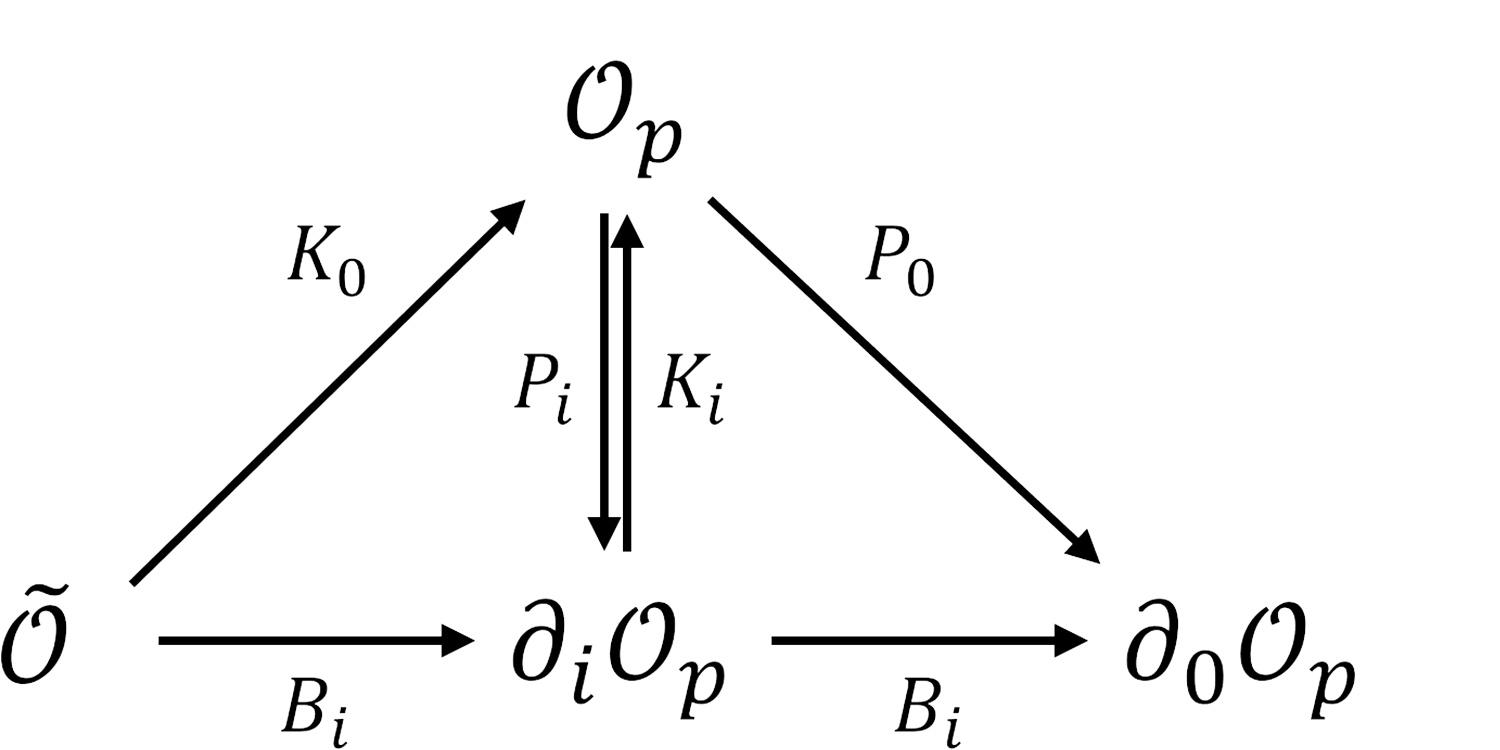}
        \caption{\centering The conformal family of $\mathcal{O}_p$ to the first order descendent level. This is a part of full staggered module \eqref{eq:LargeSaggeredModule}. }
        \label{fig:ConformalFamilyofScalar}
    \end{figure}

    One typical feature in the representations of CCA is the staggered structure. The staggered module has emerged in the studies of 2D LogCFT\cite{Rohsiepe:1996qj,Gaberdiel:2001tr,Kytola:2009ax,Creutzig:2013hma}, BMS free scalar\cite{Hao:2021urq} and BMS free fermions\cite{Yu:2022bcp,Hao:2022xhq}. Here we only give a simple example and leave the analysis of the staggered modules in higher dimensional CCFT to the appendix \ref{app:stagger}. For the above scalar case,  there could exist another scalar operator $\tilde{\mathcal{O}}$ with conformal dimension $\Delta + 1$ such that acting $K_0$ on  it gives $\mathcal{O}_p$. More precisely there are
    \begin{equation}
         [K_0,\tilde{\mathcal{O}}] =2\Delta \mathcal{O}_p, \quad [K_i, \tilde{\mathcal{O}}] = 0. 
    \end{equation}
    The relations among $\mathcal{O}_p$, $\partial_\mu\mathcal{O}_p$ and $\tilde{\mathcal{O}}$ are shown in Figure \ref{fig:ConformalFamilyofScalar}. The operators $\{\mathcal{O}_p, \partial_\mu\mathcal{O}_p, \tilde{\mathcal{O}}\}$ form a staggered structure in which $\{\mathcal{O}_p, \partial_\mu\mathcal{O}_p\}$ form a submodule and $\tilde{\mathcal{O}}$ is a quotient. The full staggered module containing the higher-order descendants is shown in \eqref{eq:LargeSaggeredModule}.  \par
    
    \par
    
    In order to study the other primary operators besides the scalar, we need to understand the representations of Carrollian rotations. Because the algebra of Carrollian rotations is not semi-simple, its finite dimensional representations are generically reducible but indecomposable, and are much more complicated than the ones of the usual Lorentzian rotations.
    
\begin{figure}[ht]
        \centering
        \includegraphics[width=5cm,align=c]{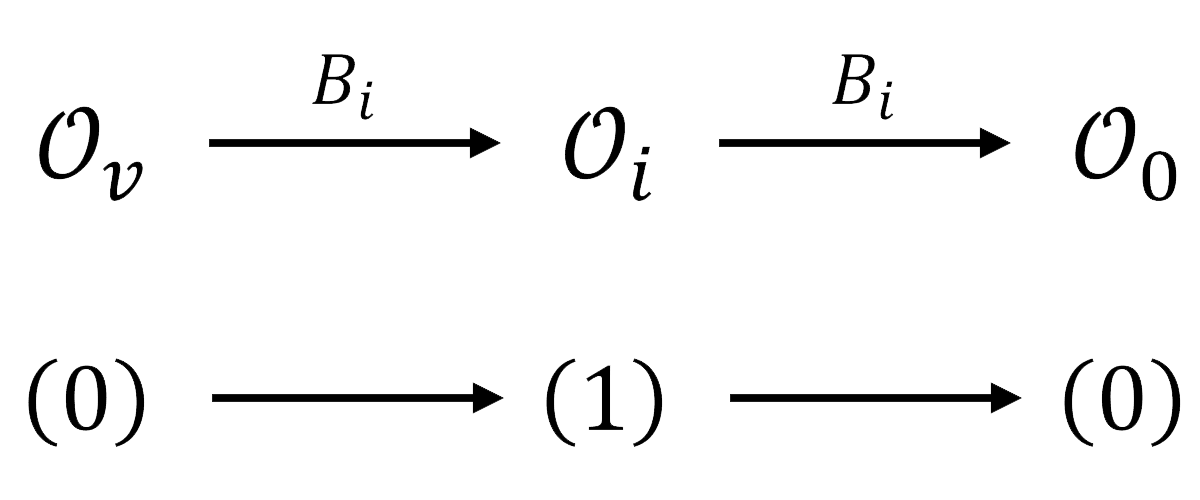}
        \caption{\centering The structure of Bargmann vector operator. }
        \label{fig:VectorAsChainRepresentation}
    \end{figure}

    One nontrivial representation for $d=4$ Carrollian rotation group, which will appear in the following study of this work,  is given by primary operators $\mathcal{O}_\alpha=\{\mathcal{O}_v, \mathcal{O}_i, \mathcal{O}_0\}$ with $\alpha = v,1,2,3,0$, and $i=1,2,3$. $\mathcal{O}_\alpha$ corresponds to a vector representation in $d=5$ Bargmann space. With respect to three-dimensional spacial rotations $J_{ij}$, the operators $\mathcal{O}_v$ and $\mathcal{O}_0$ are scalars, $\mathcal{O}_i$ is a vector, and they are related by the boost generators as follows, 
    \begin{equation}\label{eq:NontrivalExampleofCarrollianRotation}
        \begin{aligned}
            &\left[J_{kl}, \mathcal{O}_v \right] = 0, \qquad \left[J_{kl}, \mathcal{O}_i \right] = \delta_{ik} \mathcal{O}_l - \delta_{il} \mathcal{O}_k, \qquad \left[J_{kl}, \mathcal{O}_0 \right] = 0, \\
            &\left[B_k, \mathcal{O}_v \right] = -\mathcal{O}_k, \qquad \left[B_k, \mathcal{O}_i \right] = \delta_{ik} \mathcal{O}_0, \qquad \left[B_k, \mathcal{O}_0 \right] = 0, \\
        \end{aligned}
    \end{equation}
    as illustrated in Figure \ref{fig:VectorAsChainRepresentation}. This representation is denoted as $(0)\to(1)\to(0)$: $0$ and $1$ are the angular quantum number of $\mathfrak{so}(3)$ so that the first $(0)$ stands for $\mathcal{O}_v$, the last $(0)$ stands for $\mathcal{O}_0$, and $(1)$ stands for $\mathcal{O}_i$ operators; the arrows represent the action of boost generator $B_i$. This representation is reducible because that the $(1)\to(0)$ part is a sub-representation. The boost generators map $\mathcal{O}_v$ to $\mathcal{O}_i$, thus the representation is not decomposable. There are also descendent operators of $\mathcal{O}_\alpha$, which together with $\mathcal{O}_\alpha$ form the conformal family. The conformal family structure could be of staggered type, similar to the scalar case. \par
    
    \par
    
    The generic representation of Carrollian conformal group is more involved. The representation such as $(0)\to(1)\to(0)$ is called a multiplet in the sense that under the action of $B_i$ generators the representation contains multiple $SO(d-1)$ covariant primary operators. In contrast, the scalar primary operator discussed before is called a singlet. With the help of a theorem by Jakobsen\cite{Jakobsen:2011zz}, the finite dimensional representations of the Carrollian rotations are all multiplet representations with every sub-sector being the irreducible representation of $SO(d-1)$. The multiplet representations for $d\geq 3$  have complicated structures, including not only the usual chain representations like $(0)\to(1)\to(0)$, which appear in logCFT\cite{Hogervorst:2016itc} and $2d$ CCFT\cite{Chen:2020vvn,Hao:2021urq,Chen:2022cpx,Chen:2022jhx,Yu:2022bcp,Hao:2022xhq,Banerjee:2022ocj}, but also novel net representations. In a chain representation, the subsectors are connected in a linear fashion through the boost action, whereas in a net representation the subsectors exhibit a more intricate structure resembling a network.  In \cite{Chen:2021xkw}, the possible chain representations have been classified (see \eqref{Rank2} and \eqref{Rank3}). Here we would not go into the details, and the interested readers could find them in \cite{Chen:2021xkw}. \par 

    It should be stressed that the above discussions about the representations of Carrollian rotation group $M=\{J,B\}$ are independent of the conformal part of the symmetry. Thus it can be applied to the study of general (non-conformal) Carrollian field theories as well.\par

\section{Construction of free Carrollian  scalar theories}\label{sec:CarrollianScalar}

    In section \ref{subsec:CarrollianSymmetryformBargmannSymmetry}, we have seen that the Carrollian structure arises as the restriction of the Bargmann structure to the null hyper-surface $v=0$. This motivates us to construct Carrollian field theories by reducing the Bargmann field theories to the null hyper-surface. Our strategy is straightforward: firstly write an action of the fields using Bargmann geometric invariants; secondly do the null reduction to get the action of Carrollian field theory; moreover check the Carrollian conformal invariance.
    
    In this section, we introduce the procedure of null reduction and illustrate it with the example of a free scalar theory. We defer the applications to electromagnetic theory and general $p$-form theories to the next section.\par

\subsection{Construction of Carrollian theories}

    To construct the $d$-dimensional Carrollian field theories from $(d+1)$-dimensional Bargmann field theories, we implement the following null reduction procedure. In principle, we may insert a delta-function distribution $\delta(v)$ into a Bargmann action, in order to confine the theory to the $v=0$ null hyperplane. In practice, we use a taking-limit procedure to reach the delta function and show the Carrollian invariance of the confined theory. Starting from a Bargmann invariant action, we can modify the Bargmann theory by multiplying the Lagrangian $\mathcal{L}$  by an arbitrary function $h(v)$. Such a function  is invariant under all the Bargmann transformations listed in Table \ref{tb:BargmannAction}  except the translation along $v$-direction. As a consequence, even though the new Lagrangian $\mathcal{L}_h = h(v)\mathcal{L}$ is not fully Bargmann invariant, it is still invariant under spatial rotations, Bargmann boosts, and translations along other directions. Finally, if we choose $h(v)$ as a family of functions approaching the delta function, we can confine the integration to the hyper-surface $v=0$ by taking the limit. The resulting action is then naturally Carrollian invariant. \par
    
    To illustrate the reduction procedure,  let us consider a functional $S$ defined on $d+1$ dimensions
    \begin{equation}
        S[\Phi] = \int d^{d+1}x ~ \mathcal{L}(\Phi(x^\alpha), x^\alpha).
    \end{equation}
    We assume that the integrand $\mathcal{L}$ is well-behaved near $v=0$. For simplicity,  we choose $h(v)$ to be a uniformly distributed function over a small interval of $v$, 
    \begin{equation}\label{eq:hepsilon}
        h_\epsilon(v)=\left\{\begin{aligned}
            &\frac{1}{2\epsilon} && -\epsilon\le v\le\epsilon \\
            &0 && \text{otherwise.} \\
        \end{aligned}\right.
    \end{equation}
    Then we can define a smeared functional
    \begin{equation}\label{eq:ModifiedAction}
        S_\epsilon[\Phi] \equiv \int d^{d+1}x~ h_\epsilon(v) ~ \mathcal{L}(\Phi(x^\alpha), x^\alpha),
    \end{equation}
   and expand $\mathcal{L}$ to the powers of $v$ 
    \begin{equation}
        \begin{aligned}
            S_\epsilon[\Phi] & = \int du d^{d-1}x ~ \frac{1}{2\epsilon} \int_{-\epsilon}^\epsilon dv ~ \mathcal{L}_0 + \mathcal{L}_1 v + \mathcal{O}(v^2)\\
                & = \int du d^{d-1}x ~ \mathcal{L}_0 + 0 + \mathcal{O}(\epsilon^2),
        \end{aligned}
    \end{equation}
    where $\mathcal{L}_0 = \mathcal{L}(\Phi|_{v=0}, (u,\vec x, v=0))$ is the restriction of $\mathcal{L}$ to the $v=0$ surface. Thus taking the $\epsilon\to 0$ limit singles out the contribution of the fields on $v=0$ surface
    \begin{equation}
      S^\mathcal{C}[\Phi] \equiv  \lim_{\epsilon\to 0} S_\epsilon[\Phi] = \int du d^{d-1}x ~ \mathcal{L}_0.  
    \end{equation}
    If we start with a Bargmann invariant action $S^\mathscr{B}$, this manipulation will yield an Carrollian invariant action $S^\mathscr{C}$  on $v=0$.\par

    However, as have discussed in section \ref{subsec:CarrollianConformalSymmetry}, the Carrollian conformal symmetry is a different story. As shown in \cite{Chen:2021xkw},  we can construct $d$-dimensional Carrollian conformal theories from $(d+1)$-dimensional conformal theories on a null hyper-surface, since the $d$-dimensional (global) Carrollian conformal group is a subgroup of $(d+1)$-dimensional conformal group. Thus using the above procedure we get 
    \begin{equation}
      S^{\text{con}\mathcal{C}}[\Phi] \equiv  \lim_{\epsilon\to 0} S^{\text{con}}_\epsilon[\Phi] = \int du d^{d-1}x ~ \mathcal{L}^{\text{con}}_0,
    \end{equation}
    the leading expansion in $v$ of $(d+1)$-dimensional conformal Lagrangian produces a $d$-dimensional Carrollian conformal Lagrangian. However, there are  subtleties in null reductions, due to  geometric invariants. For Bargmann theories, as the geometric invariants are the metric $G$ and the time-like vector $\xi$,  we can construct  Carrollian invariant theories in both electric sector and magnetic sector.  By contrast,  the only geometric invariant  for a relativistic conformal theory is the metric $G$, and the same kind of null reduction only leads to  a Carrollian conformal theory in electric sector. In other words, there could exist Carrollian conformal theory in magnetic sector, which cannot be obtained by doing null reduction from a parent CFT.  We summarise these subtleties in Table \ref{nullreduction}.\par
    
    \begin{table}[ht]
        \centering
        \begin{tabular}{c|c|c}\hline
            Parent theory & kinds of theories & preserving conformal symmetry \\\hline
            Bargmann theory & electric and magnetic sector & not automatically conformal  \\
            Conformal theory & electric sector & automatically conformal\\\hline
        \end{tabular}
        \caption{\centering Null reduction from different parent theories.}\label{nullreduction}
    \end{table}
    
    In this work, we focus on the  construction of both electric and magnetic sector  theories from null reduction of  Bargmann theories. We will verify the Carrollian conformal symmetry of the resulting theories case by case. \par
    
\subsection{Carrollian free scalar theories}
    We aim to construct the Carrollian massless free scalar theories in $d\geq 3$. 
    The building blocks of Bargmann field theories are geometric invariants $G^{\alpha\beta}$ and $\xi^\alpha$. For a massless free scalar field $\Phi$, there are only two kinds of Bargmann invariant actions: 
    \begin{equation}
        S^\mathscr{B}_E=-\frac{1}{2}\int d^{d+1}x ~ \xi^\alpha\xi^\beta \partial_\alpha\Phi \partial_\beta\Phi, 
        \qquad S^\mathscr{B}_M=-\frac{1}{2}\int d^{d+1}x ~ G^{\alpha\beta} \partial_\alpha\Phi \partial_\beta\Phi. 
    \end{equation}
    The subscript $M$ and $E$ stand for magnetic sector and electric sector, which correspond to magnetic and electric Carrollian field theories\cite{Henneaux:2021yzg}, respectively. Let us start with the simpler one, i.e., the electric sector first. \par

\subsubsection*{Electric sector}
    In this sector, the Bargmann action is  \begin{equation}\label{eq:ActionOfBargmannScalarElectricSector}
        S^\mathscr{B}_E =-\frac{1}{2} \int d^{d+1}x ~ \xi^\alpha\xi^\beta \partial_\alpha\Phi \partial_\beta\Phi.
    \end{equation}
    Expanding $\Phi$ near $v=0$, we have
    \begin{equation}\label{eq:PhiExpansion}
        \Phi(u,\vec x,v) = \phi(u,\vec x) + v\pi(u,\vec x) + \mathcal{O}(v^2).
    \end{equation}
    Inserting this into the action, and choosing $\xi^\alpha=(1,\vec 0,0)$, we have
    \begin{equation}
        S^\mathscr{B}_E =-\frac{1}{2} \int d^{d+1}x ~ \partial_u\Phi\partial_u\Phi =-\frac{1}{2} \int d^{d+1}x ~ \partial_u\phi\partial_u\phi + 2v \partial_u\pi\partial_u\phi + \mathcal{O}(v^2).
    \end{equation}
    Thus we get the Carrollian invariant action
\begin{equation}\label{eq:ActionOfCarrollianScalarElectricSector}
        S^\mathscr{C}_E = \lim_{\epsilon\to 0}S^\mathscr{B}_{E,\epsilon} = -\frac{1}{2}\int d^d x ~ \partial_0\phi\partial_0\phi.
    \end{equation}
    This is actually the electric Carrollian conformal scalar theory with $\phi$ being the fundamental field. Under an infinitesimal symmetry transformation generated by $G$, the field changes as $\delta_G\phi(x) = -\xi_G^\mu(x)\partial_\mu\phi(x) + [G, \phi(x)]$, where $\xi_G$ is the vector field corresponding to $G$ in Table \ref{tb:CCAAction} and the term $[G, \phi(x)]$ is thus the representation of the symmetry on the field. Using the notation $\phi=\phi(0)$ introduced in the section \ref{subsec:CCAReps}, we have the actions of the CCA generators on $\phi$:
    \begin{equation}
        \begin{aligned}
            &\left[D,\phi\right]= \Delta_\phi \phi, \qquad \left[K_\mu,\phi\right]=0, \qquad \left[P_\mu,\phi\right]=\partial_\mu\phi, \\
            &\left[J^i_{~j},\phi\right]=0, \qquad \left[B_i,\phi\right]=0, \qquad \left[B_i,\partial_j\phi\right]=\delta_{ij}\partial_0\phi.\\
        \end{aligned}
    \end{equation}
    It can be checked that the action \eqref{eq:ActionOfCarrollianScalarElectricSector} is indeed invariant under the Carrollian conformal transformations.\par
    
    \begin{figure}[ht]
        \centering
        \includegraphics[width=4cm,align=c]{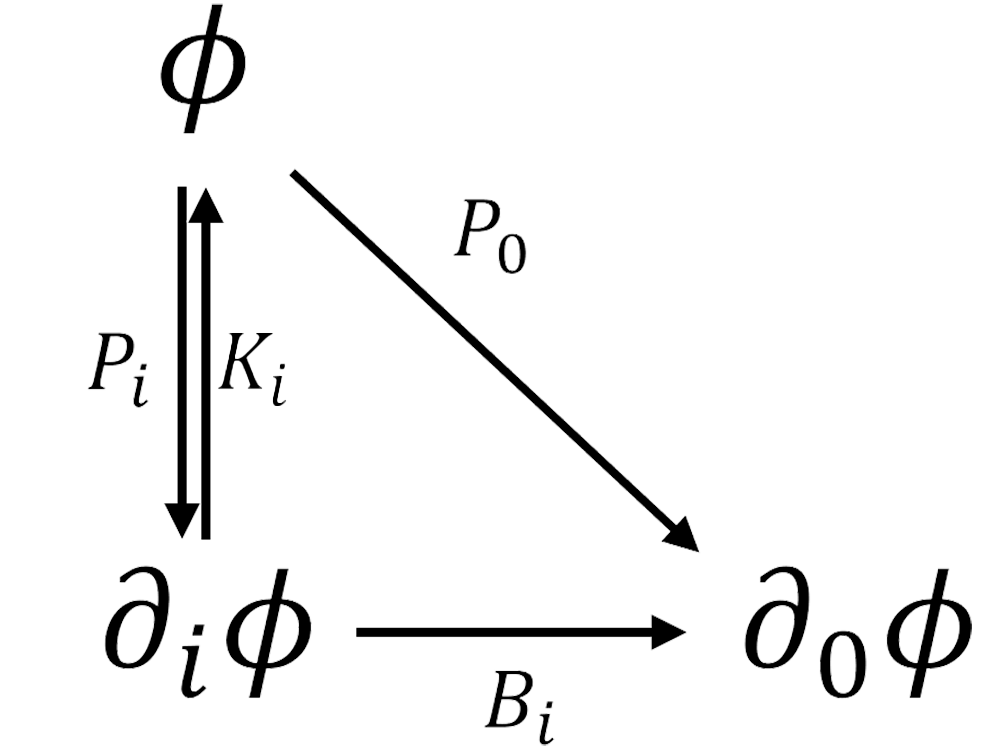}
        \caption{\centering The Carrollian conformal family of $\phi$ to the first order.}
        \label{fig:ScalarFirstdescendents}
    \end{figure}\par
    
    The field $\phi$ is a primary operator and the relations among its first-order descendants are shown in Figure \ref{fig:ScalarFirstdescendents}. Noticing that $\partial_0\phi$ is a descendent as well as a primary operator, because under an action of $K_\mu$, we have
    \begin{equation}
        \left[K_0,\left[P_0,\phi\right]\right]=\left[\left[K_0,P_0\right],\phi\right]=0, ~~~ \left[K_i,\left[P_0 ,\phi\right]\right]=\left[\left[K_i,P_0\right],\phi\right]=2\left[B_i,\phi\right]=0.
    \end{equation}
    The 2-point correlator of $\phi$ can be calculated via the path-integral
    \begin{equation}
        \left<\phi(x)\phi(0)\right>=\frac{i\abs{t}}{2} \delta^{(d-1)}(\vec x).
    \end{equation}
    It satisfies the Ward identity of CCA generators,  and this fact confirms again that $\phi$ is a primary operator. The computation details and the discussions on the 2-pt correlators in a CCFT can be found in Appendix \ref{app:2pt-ward}. \par
    
\subsubsection*{Magnetic sector}
    We now consider the more non-trivial magnetic sector. The Bargmann action is now given by
    \begin{equation}\label{eq:ActionOfBargmannScalarMagneticSector}
        S^\mathscr{B}_M=-\frac{1}{2}\int d^{d+1}x ~ G^{\alpha\beta} \partial_\alpha\Phi \partial_\beta\Phi.
    \end{equation}
    Using the expansion of $\Phi$ \eqref{eq:PhiExpansion}, we get
    \begin{equation}
        S^\mathscr{B}_M = -\frac{1}{2}\int d^{d+1}x ~ 2\partial_u\Phi\partial_v\Phi + \partial_i\Phi\partial_i\Phi = -\frac{1}{2}\int d^{d+1}x ~ 2\pi\partial_u\phi + \partial_i\phi\partial_i\phi + \mathcal{O}(v).
    \end{equation}
    Thus we find the action of magnetic Carrollian scalar theory, \begin{equation}\label{eq:ActionOfCarrollianScalarMagneticSector}
        S^\mathscr{C}_M = -\frac{1}{2}\int d^d x ~ 2\pi\partial_0\phi + \partial_i\phi\partial_i\phi.
    \end{equation}
    The fundamental fields in this theory are $\phi$ and $\pi$, which are in the expansion of the field $\Phi$.  Classically they are totally independent fields. In this theory, the canonical momentum of scalar field $\phi$ is $\Pi_\phi = 2\pi$, which matches perfectly with \cite{Henneaux:2021yzg}. The Carrollian invariance of this action is quite non-trivial. Under the Bargmann boost $B^\mathscr{B}_i$, the Bargmann scalar transforms as
    \begin{equation}
        \delta_{B^\mathscr{B}_i}\Phi  = -v\partial_i \Phi + x_i\partial_u \Phi,
    \end{equation}
    which can be expanded as 
    \begin{equation}\label{eq:MagneticScalarExpansionofBoostAction}
        \begin{aligned}
            \delta_{B^\mathscr{B}_i}\phi + v\delta_{B^\mathscr{B}_i}\pi + \mathcal{O}(v^2)  = x_i\partial_u\phi + v x_i\partial_u \pi - v\partial_i\phi + \mathcal{O}(v^2).\\
        \end{aligned}
    \end{equation}
    In the leading order of $v$, we get the infinitesimal transformation of the fields $\phi, \pi$ under the Carrollian boost $B_i$
    \begin{equation}
        \begin{aligned}
            \delta_{B_i}\phi & = x_i\partial_0\phi,\qquad
            \delta_{B_i}\pi & = x_i\partial_0 \pi - \partial_i\phi,\\
        \end{aligned}
    \end{equation}
    and furthermore we have
    \begin{equation}
        \delta_{B_i}\partial_j \phi = x_i\partial_0 \partial_j \phi + \delta_{ij}\partial_0\phi.
    \end{equation}
    This means $\phi$ transforms as a scalar under Carrollian boost and $(\pi,\partial_i\phi,\partial_0\phi)$ as a $(scalar)\to(vector)^{(d-1)}\to(scalar)$ representation, which is in $(0)\to(1)\to(0)$ chain representation. This is somehow expected since $\partial_\alpha$ could be seen as a contravariant Bargmann vector, and thus $(\partial_v)\to(\partial_i)\to(\partial_u)$ form a representation of the Carroll group.\par
    
    \begin{figure}[ht]
        \centering
        \includegraphics[width=6cm,align=c]{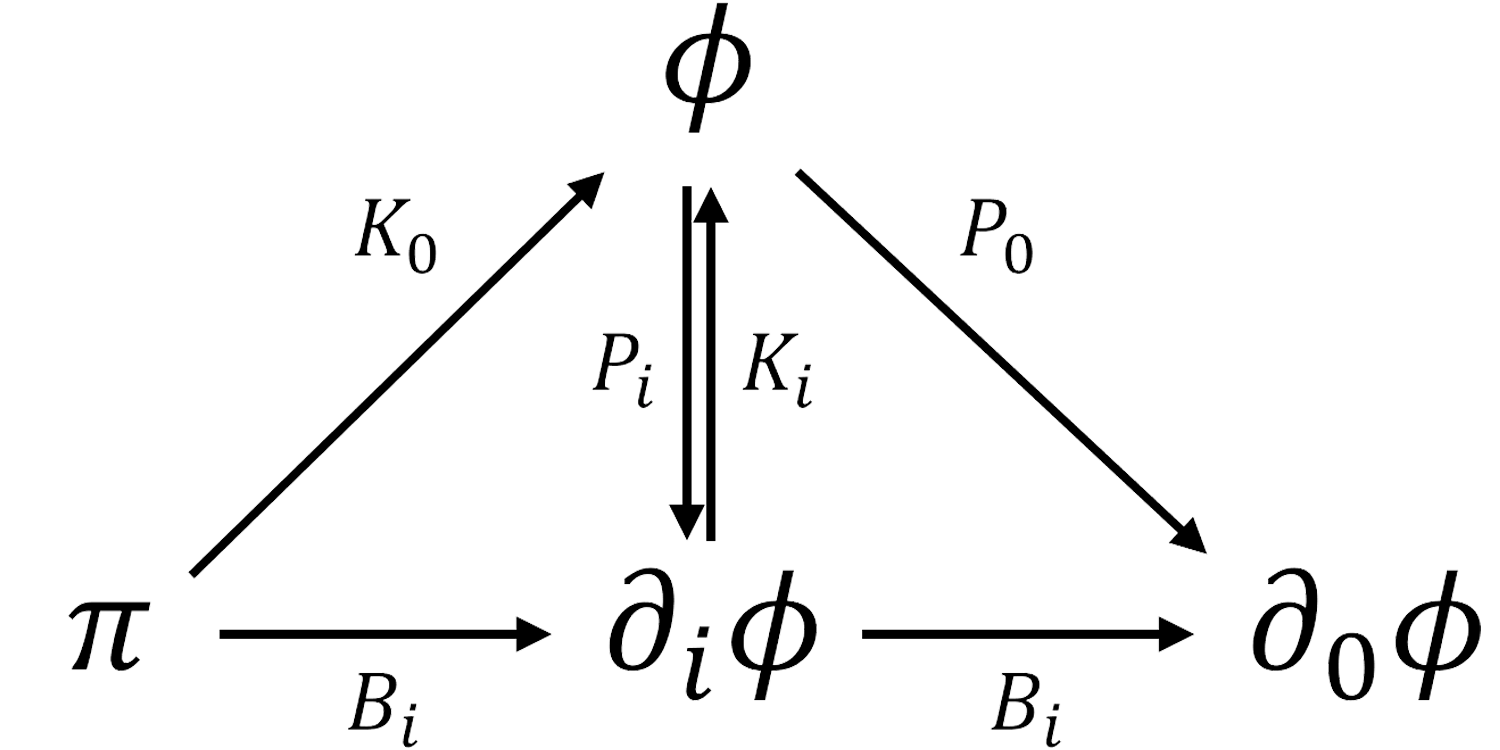}
        \caption{\centering The staggered structure of fields $\phi$, $\partial_\mu\phi$ and $\pi$.}
        \label{fig:ScalarStaggeredModule}
    \end{figure}

    Next we would like to check the conformal invariance of this action. Obviously, equipping $\phi$ with conformal dimension $\Delta_\phi = d/2-1$ and $\pi$ with $\Delta_\pi = d/2$, the action is invariant under the dilation $D$. The scalar $\phi$ is still a primary operator as in relativistic CFT, and the field $\pi$ appears as a part of staggered module of $\phi$'s conformal family as shown in Figure \ref{fig:ScalarStaggeredModule}. More explicitly, the generators act on the fields as
    \begin{equation}
        \begin{aligned}
            &\left[D,\phi\right]= \Delta_\phi \phi, \qquad \left[K_\mu,\phi\right]=0, \qquad \left[P_\mu,\phi\right]=\partial_\mu\phi, \\
            &\left[J^i_{~j},\phi\right]=0, \qquad \left[B_i,\phi\right]=0, \qquad \left[B_i,\partial_j\phi\right]=\delta_{ij}\partial_0\phi,\\
            &\left[D,\pi\right]= (\Delta_\phi+1) \pi, \qquad \left[K_0,\pi\right]=2\Delta_\phi \phi, \qquad \left[K_i,\pi\right]=0, \\
            &\left[J^i_{~j},\pi\right]=0, \qquad \left[B_i,\pi\right]=-\partial_i\phi.
        \end{aligned}
    \end{equation}
    The field $\pi$ is neither primary nor descendent, as it cannot be generated by acting the generators $P$ on $\phi$, while $\partial_0\phi$ is both a primary operator and a descendent of $\phi$. Thus we should treat $\pi$ as an independent field, and the fields $\phi, \partial_i \phi, \partial_0 \phi$ and $\pi$ constitute a staggered module.  The emergence of staggered module is a common feature of magnetic sector of Carrollian theories where the next-to-leading-order field in the expansion of Bargmann field shows up as an independent field in the action. \par
    
    For the special conformal transformation $K_\mu$, there is  
    \begin{equation}
        \delta_{K_\nu}S^\mathscr{B}_M = -\frac{1}{2} \int d^d x ~\partial_\mu(k^\mu_\nu (2\pi\partial_0\phi + \partial_i\phi\partial_i\phi)) + \partial_0 (2\Delta_\phi \phi^2),
    \end{equation}
    where $k^\mu_\nu$ is the $\mu$ component of the generator  $K_\nu$. This variation of the action differs from the usual structure for a generator $G$,
    \begin{equation}
        \delta_{G} S = \int d^d x ~\partial_\mu(g^\mu \mathcal{L}),
    \end{equation}
    only by a total derivative $\partial_0 (2\Delta_\phi \phi^2)$. 
    Thus the magnetic theory is Carrollian conformal invariant as well.   \par

    We can check the conformal family structure by calculating the correlation functions in the path-integral formalism. The details can be found in Appendix \ref{appsubsec:ScalarMagneticSector}.  We read the 2-pt correlators of the fundamental fields \begin{equation}\label{eq:CorrelatorsOfCarrollianScalarMagneticSector}
    \begin{aligned}
        \left<\phi(\vec x_1,t_1)\phi(\vec x_2,t_2)\right> &=0\\
        \left<\phi(\vec x_1,t_1)\pi(\vec x_2,t_2)\right> &= -\left<\pi(\vec x_1,t_1)\phi(\vec x_1,t_2)\right>= -\frac{i}{2}\mbox{Sign}(t)\delta^{(d-1)}(\vec x)\\ \left<\pi(\vec x_1,t_1)\pi(\vec x_2,t_2)\right>  &= \frac{i \abs{t}}{2} \vec\partial^2\delta^{(d-1)}(\vec x)
    \end{aligned}
    \end{equation}
    where $t=t_1-t_2$ and $\vec x =\vec x_1 -\vec x_2$. The correlator $\left<\phi\phi\right>=0$ obviously satisfies the Ward identities. However $\pi$ is not a primary operator, so the correlators of $\pi$ do not satisfy the constraints on the primary operators discussed in \ref{app:2pt-ward}. Nevertheless, these correlators indeed satisfy the Ward identities \eqref{app:ward-id} with non-vanishing $\left<[K_0,\mathcal{O}_1]\mathcal{O}_2\right>$ or $\left<\mathcal{O}_1[K_0,\mathcal{O}_2]\right>$ terms.\par
    
\subsection{Relations between Bargmann correlators and Carrollian correlators}

    Since the fundamental fields of Carrollian theories $\phi, \pi,$ etc. are the components of Bargmann field $\Phi$, it is sensible to expect their correlators to be the components in the expansion of the correlator of $\Phi$. In this section, we show that this is true for the free scalar theory. Our strategy is to express the derivative operator as the  integral kernel, from which we can easily find its ``inverse"  as the correlator. The relation between the Bargmann correlators and the Carrollian correlators \eqref{eq:RelationBetweenModifiedCorrelatorAndBargmannCorrelator} is shown by interchanging the order of taking $\epsilon\to 0$ limit and taking functional derivatives of the source in the path-integral. At the end of this subsection, we apply these discussions to the electric sector and magnetic sector of free scalar theories to reproduce the Carrollian correlators in the last subsection, which were derived via the path integral.\par
    
    Firstly, let us show how to express the derivative operator as the integral kernel by considering some $(d+1)$-dimensional discrete toy models. The $(d+1)$-dimensional action \eqref{eq:ModifiedAction} considered in this paper is inserted with the function $h_\epsilon(v)$, thus the corresponding kernel is different from the normal one, and we denote the kernel as $D_\epsilon$ where the subscript $\epsilon$ indicates the presence of $h_\epsilon(v)$. Starting from the first-order derivative, we consider the following model, whose continuous limit is $\int d^{(d+1)} x ~h_\epsilon(v) \Phi\partial_\mu\Phi$,  
    \begin{equation}\label{eq:DiscretePmuModel}
        \begin{aligned}
            & \sum_{\{n\}} ~ h^\epsilon_{n_v}\Phi_{\{n\}} (\Phi_{\{n\}} - \Phi_{\{n_\mu-1\}}) \\
                &= \sum_{\{n\}\{m\}} ~ \Phi_{\{n\}} ~ \delta_{n_0,m_0}\cdots(\delta_{n_\mu,m_\mu}-\delta_{n_\mu,m_\mu-1})\cdots\delta_{n_{d-1},m_{d-1}}h^\epsilon_{n_v}h^\epsilon_{m_v} ~ \Phi_{\{m\}} \\
                &= \sum_{\{n\}\{m\}} ~ \Phi_{\{n\}} ~ \delta_{n_0,m_0}\cdots(-\delta_{n_\mu,m_\mu}+\delta_{n_\mu-1,m_\mu})\cdots\delta_{n_{d-1},m_{d-1}}h^\epsilon_{n_v}h^\epsilon_{m_v} ~ \Phi_{\{m\}}. \\
        \end{aligned}
    \end{equation}
    Here $\{n\}$ stands for the point at $(n_0,...,n_\mu, ...,n_{d-1})$, $\{n_\mu-1\}$ is short for $(n_0,...,n_\mu-1,...,n_{d-1},n_{v})$,  $\Phi_{\{n\}}$ is the field defined at $\{n\}$, whose continuous limit is $\Phi(x^\alpha)$, and $h^\epsilon_{n_v}$ is defined such that
    \begin{equation}
        \begin{aligned}
            &\qquad h^\epsilon_{n_v}=\sum_{\abs{l_v}\le \epsilon}\frac{1}{2\epsilon}\delta_{n_v,l_v}, 
            &h^\epsilon_{n_v} &\xrightarrow[]{\text{continuous limit}} h_\epsilon(v).\\
        \end{aligned}
    \end{equation}
    The function $h^\epsilon_{n_v}$ restricts the action by only counting the interaction in $\abs{n_v}<\epsilon$. In the continuous limit, the relation  \eqref{eq:DiscretePmuModel} becomes
    \begin{equation}
        \begin{aligned}
            &\int d^{d+1} x ~ h_\epsilon(v) \Phi(x)\partial_\mu\Phi(x) \\
            & = \int d^{d+1} x_1 d^{d+1} x_2 ~ \Phi(x_1) \left(\frac{1}{2}(-\partial_{x^\mu_1}+\partial_{x^\mu_2} )\left(\delta(x^\mu_1-x^\mu_2)h_\epsilon(v_1)h_\epsilon(v_2)\right)\right) \Phi(x_2).
        \end{aligned}
    \end{equation}
    In other words, the modified Bargmann action can be rewritten in terms of the integral kernel $D_{1,\epsilon}(x^\mu_1-x^\mu_2,v_1,v_2)$:
    \begin{equation}
            \int d^{d+1} x ~ h_\epsilon(v) \Phi(x)\partial_\mu\Phi(x) = \int d^{d+1} x_1 d^{d+1} x_2 ~ \Phi(x_1) D_{1,\epsilon}(x^\mu_1-x^\mu_2,v_1,v_2) \Phi(x_2),\end{equation}
            with 
            \begin{equation}
              D_{1,\epsilon}(x^\mu_1-x^\mu_2,v_1,v_2) = -\frac{1}{2}(\partial_{x^\mu_1}-\partial_{x^\mu_2} ) (\delta(x^\mu_1-x^\mu_2)h_\epsilon(v_1)h_\epsilon(v_2)).
         \end{equation}
    Similarly, we may consider the following model, whose continuous limit is $\int d^{(d+1)} x ~h_\epsilon(v) \Phi\partial_v\Phi$,
    \begin{equation}
        \begin{aligned}
            & \sum_{\{n\}} ~ h^\epsilon_{n_v}\Phi_{\{n\}} (\Phi_{\{n\}} - \Phi_{\{n_v-1\}}) \\
            &= \sum_{\{n\}\{m\}} ~ \Phi_{\{n\}} ~ \delta_{n_0,m_0}\cdots\delta_{n_{d-1},m_{d-1}}h^\epsilon_{n_v}(h^\epsilon_{m_v}-h^\epsilon_{m_v-1}) ~ \Phi_{\{m\}} \\
            &= \sum_{\{n\}\{m\}} ~ \Phi_{\{n\}} ~ \delta_{n_0,m_0}\cdots\delta_{n_{d-1},m_{d-1}}(-h^\epsilon_{n_v}+h^\epsilon_{n_v-1})h^\epsilon_{m_v} ~ \Phi_{\{m\}}. \\
        \end{aligned}
    \end{equation}
    In this case, the action in the continuous limit could be rewritten as 
    \begin{equation}
        \begin{aligned}
            \int d^{d+1} x ~ h_\epsilon(v) \Phi(x)\partial_v\Phi(x) 
                &= \int d^{d+1} x_1 d^{d+1} x_2 
                    ~ \Phi(x_1) D_{2,\epsilon}(x^\mu_1-x^\mu_2,v_1,v_2) \Phi(x_2),\end{aligned}
    \end{equation}
    with 
     \begin{equation}
            D_{2,\epsilon}(x^\mu_1-x^\mu_2,v_1,v_2) = -\frac{1}{2}(\partial_{v_1}-\partial_{v_2})(\delta(x^\mu_1-x^\mu_2)h_\epsilon(v_1)h_\epsilon(v_2)).
    \end{equation} \par

    The above argument can be generalized to the higher-derivative operators. The reason behind is that higher-derivative operators correspond to longer range interactions in the lattice model, and the corresponding kernel can be derived in the same way. Now let us consider a quadratic action with a generic derivative operator $\int d^{d+1} x ~ h_\epsilon(v) \Phi(x)\hat{D}(\partial_\alpha)\Phi(x)$, where $\hat{D}(\partial_\alpha)$ is a function of derivative operator $\partial_\alpha$ that can be expanded to the powers of $\partial_\alpha$. The action could be written in terms of the corresponding integral kernel $D_\epsilon(x^\mu_1-x^\mu_2,v_1,v_2)$
    \begin{equation}
        \begin{aligned}
            \int d^{d+1} x ~ h_\epsilon(v) \Phi(x)\hat{D}(\partial_\alpha)&\Phi(x) = \int d^{d+1} x_1 d^{d+1} x_2 ~ \Phi(x_1) D_\epsilon(x^\mu_1-x^\mu_2,v_1,v_2) \Phi(x_2),\\
        \end{aligned}
    \end{equation}
    where 
    \begin{equation}\label{eq:DerivativeOperatorKernal}
        \begin{aligned}
        &D_\epsilon(x^\mu_1-x^\mu_2,v_1,v_2) = \hat{D}\left(-\frac{1}{2}(\partial_{x^\alpha_1}-\partial_{x^\alpha_2})\right)(\delta(x^\mu_1-x^\mu_2)h_\epsilon(v_1)h_\epsilon(v_2)).
   \end{aligned}
    \end{equation}
   It is illuminating to write the kernel in the momentum space
    \begin{eqnarray}
        \lefteqn{D_\epsilon(x^\mu_1-x^\mu_2,v_1,v_2)}\nonumber\\
        & = &\int \frac{d^{d} p_\mu d p_{1v} d p_{2v}}{(2\pi)^d (2\pi)^2} 
            ~ \hat{D}\left(-i p_\mu, -\frac{i}{2}(p_{1v}-p_{2v})\right)
            ~ \frac{\sin{\epsilon p_{1v}}\sin{\epsilon p_{2v}}}{\epsilon^2 p_{1v} p_{2v}}
            ~e^{ip_\mu(x_1-x_2)^\mu}e^{ip_{1v}v_1}e^{ip_{2v}v_2}\nonumber\\
        & =& \int \frac{d^{d+1} p_\alpha }{(2\pi)^{d+1}}\frac{d p_{v+} }{(2\pi)} 
            ~ \hat{D}(-ip_\alpha) 
            ~ \frac{\cos{2\epsilon p_v}-\cos{2\epsilon p_{v+}}}{\epsilon^2~(p_{v+}^2-p_v^2)}
            ~e^{ip_\alpha(x_1-x_2)^\alpha} e^{ip_{v+}(v_1+v_2)},
    \end{eqnarray}
    were $p_{v}\equiv (p_{1v}-p_{2v})/2$ and $p_{v+}\equiv (p_{1v}+p_{2v})/2$. 
    
    As the action is not invariant under the translation in $v$ direction, there is no mathematically rigorous definition for the inverse of $D_\epsilon$, namely there does not exist $D^{-1}_\epsilon$ such that 
    \begin{equation}
        \int d^{(d+1)} x_2 ~ D_\epsilon(x_1,x_2) D^{-1}_\epsilon(x_2,x_3) = \delta(x^\mu_1-x^\mu_3)\delta(v_1-v_3).
    \end{equation}
    The kernel $D_\epsilon(x_1,x_2)$ only counts the interaction in $\abs{v}<\epsilon$, thus the  translation symmetry along $v$ is broken, and the inverse can not be properly defined for $\abs{v}>\epsilon$. However, we may impose a looser inverse condition 
    \begin{equation}
        \begin{aligned}
        &\int d^{(d+1)} x_2 ~ D_\epsilon(x^\alpha_1,x^\alpha_2) D^{-1}_\epsilon(x^\alpha_2,x^\alpha_3) = \delta(x^\mu_1-x^\mu_3)\delta(v_1)\delta(v_3),\\
        \end{aligned}
    \end{equation}
    such that we can find the ``inverse" $D^{-1}_\epsilon(x^\mu_1-x^\mu_2,v_1,v_2)$. Noticing that this $D^{-1}_\epsilon(x^\mu_1-x^\mu_2,v_1,v_2)$ is not the rigorous inverse of $D_\epsilon(x_1,x_2)$, but rather an inverse on $v=0$ hypersurface. It takes the following form in the momentum space 
    \begin{equation}
        \begin{aligned}
        &D^{-1}_\epsilon(x^\mu_1-x^\mu_2,v_1,v_2) \\
        &= \int \frac{d^{d+1} p_\alpha }{(2\pi)^{d+1}}\frac{d p_{v+} }{(2\pi)} 
            ~ \frac{1}{\hat{D}(-ip_\alpha)} 
            ~ \frac{\epsilon^2~(p_{v+}^2-p_v^2)}{\cos{2\epsilon p_v}-\cos{2\epsilon p_{v+}}}
            ~ e^{ip_\alpha(x_1-x_2)^\alpha} e^{ip_{v+}(v_1+v_2)}. \\
        \end{aligned}
    \end{equation}
    It turns out that this form of inverse $D^{-1}_\epsilon$ is useful in our study. The reason we use the modified normalization condition is that the Carrollian action is simply defined on the null hyper-surface and we need not require the translation symmetry along $v$.\par
    
    With the integral kernel and its inverse, we can discuss the relation between the Bargmann correlators and Carrollian correlators. In general, the modified quadratic Bargmann action can be written as
    \begin{equation}
        S^\mathscr{B}_\epsilon=\int d^{(d+1)} x ~h_\epsilon(v) \Phi\hat{D}\Phi = \int d^{d+1} x_1 d^{d+1} x_2 ~ \Phi(x^\alpha_1) D_\epsilon(x^\mu_1-x^\mu_2,v_1,v_2) \Phi(x^\alpha_2),
    \end{equation}
    with $D_\epsilon(x^\mu_1-x^\mu_2,v_1,v_2)$ being defined in \eqref{eq:DerivativeOperatorKernal}. The corresponding modified generating functional, labeled with subscript $\epsilon$, is 
    \begin{equation}\label{eq:ModifiedBargmannFunctional}
        \begin{aligned}
            \mathcal{Z}^\mathscr{B}_\epsilon[J] &= \int \mathcal{D}\Phi \exp \left(i S^\mathscr{B}_\epsilon + i\int d^{d+1} x ~ \delta(v) J(x^\alpha) \Phi(x^\alpha)\right) \\
                &= \int \mathcal{D}\Phi \exp \left(i \int d^{d+1} x_1 d^{d+1} x_2 ~ \Phi(x^\alpha_1) D_\epsilon(x^\mu_1-x^\mu_2,v_1,v_2) \Phi(x^\alpha_2) \right.\\
                & \qquad\qquad\qquad\qquad\qquad +\left.i \int d^{d+1} x_1 d^{d+1} x_2 J(x^\alpha_1) \delta(v_1)\delta(x^\mu_1-x^\mu_2)\delta(v_2) \Phi(x^\alpha_2)\right) \\[10pt]
                & =N \exp (- \frac{i}{4} \int d^{d+1} x_1 d^{d+1} x_2 ~ J(x^\alpha_1) D^{-1}_\epsilon(x^\mu_1-x^\mu_2,v_1,v_2)J(x^\alpha_2)), \\
        \end{aligned}
    \end{equation}
    and the modified correlator is
    \begin{equation}\label{eq:ModifiedCorrelator}
        \begin{aligned}
            \left<\Phi(x^\mu_1,v_1)\Phi(x^\mu_2,v_2)\right>_\epsilon  =\left. \frac{1}{Z^\mathscr{B}_\epsilon[0]} \frac{(-i)^2 \delta^2}{\delta J(x_1^\alpha)\delta J(x_2^\alpha)}\mathcal{Z}^\mathscr{B}_\epsilon [J]\right|_{J=0} = -\frac{1}{2}D^{-1}_\epsilon(x^\mu_1-x^\mu_2,v_1,v_2).\\
        \end{aligned}
    \end{equation}
    In the limit $\epsilon\to 0$, this correlator reduces to the Bargmann correlator $\left<\Phi(x^\mu_1,v_1)\Phi(x^\mu_2,v_2)\right>$,
    \begin{equation}\label{eq:RelationBetweenModifiedCorrelatorAndBargmannCorrelator}
        \begin{aligned}
            &\lim_{\epsilon\to 0} \left<\Phi(x^\mu_1,v_1)\Phi(x^\mu_2,v_2)\right>_\epsilon = \lim_{\epsilon\to 0} \left. \frac{(-i)^2 \delta^2}{\delta J(x_1^\alpha)\delta J(x_2^\alpha)}\frac{\mathcal{Z}^\mathscr{B}_\epsilon [J]}{Z^\mathscr{B}_\epsilon[0]}\right|_{J=0}\\[10pt]
                &\qquad\qquad = \int \frac{d^{d+1} p_\alpha }{(2\pi)^{d+1}} -\frac{1}{2}\hat{D}^{-1}(-ip_\alpha) ~ e^{ip_\alpha(x_1-x_2)^\alpha} \int \frac{d p_{v+} }{2\pi}~e^{ip_{v+}(v_1+v_2)}\\[10pt]
                &\qquad\qquad = 2\delta(v_1+v_2)\left<\Phi(x^\alpha_1)\Phi(x^\alpha_2)\right>,
        \end{aligned}
    \end{equation} 
    where $\left<\Phi(x^\alpha_1)\Phi(x^\alpha_2)\right>$ is the normal Bargmann correlator without inserting $h_\epsilon(v)$ in the action. On the other way, we may interchange the order of taking the $\epsilon\to 0$ limit and taking the functional derivatives so that we can find the relation between modified correlator \eqref{eq:ModifiedCorrelator} and the Carrollian correlators. Taking the $\epsilon\to 0$ limit first, we see that the modified Bargmann generating functional \eqref{eq:ModifiedBargmannFunctional} becomes the Carrollian generating functional,
    \begin{equation}
        \begin{aligned}
            &Z^\mathscr{B}_\epsilon[J]\xrightarrow{\epsilon\to 0}Z^\mathscr{C}[J] = \int \mathcal{D}\phi\mathcal{D}\pi\cdots \exp \left(i S^\mathscr{C} + i \int d^{d+1} x J(x) \delta(v) (\phi(x^\mu)+\pi(x^\mu) v +\cdots)\right). 
        \end{aligned}
    \end{equation}
    Here we have expanded the Bargmann field $\Phi(x)= \phi(x^\mu)+\pi(x^\mu) v +\cdots$. Note that the source terms corresponding to the component fields are $J_\phi=\int dv ~ J(x) \delta(v)$, $J_\pi=\int dv ~vJ(x) \delta(v)$, etc. Next, we take functional derivatives to read the correlation functions,
    \begin{equation}\label{eq:RelationBetweenModifiedCorrelatorAndCarrollianCorrelator}
        \begin{aligned}
            &\frac{-i \delta}{\delta J(x^\alpha_1)}\frac{-i \delta}{\delta J(x^\alpha_2)}\lim_{\epsilon\to 0}Z^\mathscr{B}_\epsilon[J]|_{J=0} \\
            &= \int \mathcal{D}\phi\mathcal{D}\pi\cdots \delta(v_1) (\phi(x^\mu_1)+\pi(x^\mu_1) v_1 +\cdots) \delta(v_2)(\phi(x^\mu_2)+\pi(x^\mu_2) v_2 +\cdots) \exp i S^\mathscr{C} \\
            &= \delta(v_1)\delta(v_2) (\left<\phi(x^\mu_1)\phi(x^\mu_2)\right> +v_1 \left<\pi(x^\mu_1)\phi(x^\mu_2)\right> \\
                &\qquad\qquad\qquad\qquad\qquad\qquad\qquad\qquad +v_2 \left<\phi(x^\mu_1)\pi(x^\mu_2)\right> +v_1 v_2 \left<\pi(x^\mu_1)\pi(x^\mu_2)\right>+\cdots)
        \end{aligned}
    \end{equation}
    Matching the last line in \eqref{eq:RelationBetweenModifiedCorrelatorAndBargmannCorrelator} with the one of \eqref{eq:RelationBetweenModifiedCorrelatorAndCarrollianCorrelator}, we find
    \begin{equation}\label{eq:MatchingBargmannCorrelatorAndCarrollianCorrelator}
        \begin{aligned}            &2\delta(v_1+v_2)\left<\Phi(x^\mu_1,v_1)\Phi(x^\mu_2,v_2)\right> = \delta(v_1)\delta(v_2) (\left<\phi(x^\mu_1)\phi(x^\mu_2)\right> \\ &\qquad\qquad\qquad\qquad +v_1 \left<\pi(x^\mu_1)\phi(x^\mu_2)\right> +v_2 \left<\phi(x^\mu_1)\pi(x^\mu_2)\right> +v_1 v_2 \left<\pi(x^\mu_1)\pi(x^\mu_2)\right>+\cdots).\\
        \end{aligned}
    \end{equation}
    It shows that the Carrollian correlators are given by the expansion of the Bargmann correlator, as expected. This matching is feasible only if assuming that taking $\epsilon\to 0$ limit and taking the functional derivatives are commutative, and we believe this assumption is generally correct. \par
     
    Now we show the relation \eqref{eq:MatchingBargmannCorrelatorAndCarrollianCorrelator} explicitly in the electric and magnetic sector of scalar theories. For the electric scalar sector (labeled by superscript $E$) \eqref{eq:ActionOfBargmannScalarElectricSector}, the 2-point correlators of Bargmann field $\Phi$ and Carrollian field $\phi$ are respectively
    \begin{equation}
        \begin{aligned}
            &\left<\Phi(x^\mu_1,v_1)\Phi(x^\mu_2,v_2)\right>^E = i \int \frac{d^{(d+1)}p}{(2\pi)^{(d+1)}} e^{ip_\alpha x^\alpha} ~ \frac{-1}{p_0^2} = \frac{i\abs{t_1-t_2}}{2} \delta^{(d-1)}(\vec x_1-\vec x_2)\delta(v_1-v_2),\\[10pt]
            &\left<\phi(x^\mu_1)\phi(x^\mu_2)\right>^E = \frac{i\abs{t_1-t_2}}{2} \delta^{(d-1)}(\vec x_1-\vec x_2).\\
        \end{aligned}
    \end{equation}
    Noticing the fact that 
    \begin{equation}
        2\delta(v_1+v_2)\delta(v_1-v_2)= \delta(v_1)\delta(v_2),
    \end{equation}
    we find that the relation \eqref{eq:MatchingBargmannCorrelatorAndCarrollianCorrelator} indeed holds.\par
    
    For the magnetic sector (labeled by superscript $M$) of scalar \eqref{eq:ActionOfBargmannScalarMagneticSector}, the 2-point Bargmann correlator of $\Phi$ is 
    \begin{equation}
        \begin{aligned}
            \left<\Phi(x^\mu_1,v_1)\Phi(x^\mu_2,v_2)\right>^M = \int \frac{d^{(d+1)}p}{(2\pi)^{(d+1)}} e^{i p_\alpha (x_1 - x_2)^\alpha}~ \frac{-1}{{\vec p}^2+2p_0 p_v}.
        \end{aligned}
    \end{equation}
    Integrating $p_v$ and expanding in $v_i$, we find
    \begin{equation}
        \begin{aligned}
            &\left<\Phi(x^\mu_1,v_1)\Phi(x^\mu_2,v_2)\right>^M = \int \frac{d^{d}p}{(2\pi)^{d}} ~ \frac{-i}{2 p_0} e^{-\frac{i{\vec p}^2(v_1-v_2)}{2p_0}}\mbox{Sign}(v_1-v_2)\\[10pt]
     	    &=\frac{\mbox{Sign}(v_1-v_2)}{v_1-v_2} ~ (v_1-v_2)\int \frac{d^{d}p}{(2\pi)^{d}} ~\left(\frac{-i}{2 p_0} - \frac{{\vec p}^2}{4 p_0^2}(v_1-v_2)+\cdots \right)\\[10pt]
     	    &=\abs{v_1-v_2}^{-1} \left(0 + v_1 \int \frac{d^{d}p}{(2\pi)^{d}} ~\frac{-i}{2 p_0} + v_2 \int \frac{d^{d}p}{(2\pi)^{d}} ~\frac{i}{2 p_0} + v_1 v_2 \int \frac{d^{d}p}{(2\pi)^{d}} ~\frac{{\vec p}^2}{2 p_0^2}+\cdots\right)\\
        \end{aligned}
    \end{equation}
    Viewing $\abs{v_1-v_2}^{-1}$ as a generalized function, it is proportional to $\delta(v_1-v_2)$ by the canonical regularization \cite{gel1964properties}
    \begin{equation}
        \frac{1}{\Gamma(0) |x|} = \delta(x),
    \end{equation}
    thus the equation \eqref{eq:MatchingBargmannCorrelatorAndCarrollianCorrelator} matches the Bargmann correlator with the correlators of Carrollian fields \eqref{eq:mag-corr},
    \begin{equation}
        \begin{aligned}
            &2\delta(v_1+v_2)\left<\Phi(x^\mu_1,v_1)\Phi(x^\mu_2,v_2)\right>^M \\
            &\propto  2\delta(v_1+v_2)\delta(v_1-v_2)~ \left(0 + v_1 \int \frac{d^{d}p}{(2\pi)^{d}} ~\frac{-i}{2 p_0} + v_2 \int \frac{d^{d}p}{(2\pi)^{d}} ~\frac{i}{2 p_0} + v_1 v_2 \int \frac{d^{d}p}{(2\pi)^{d}} ~\frac{{\vec p}^2}{2 p_0^2}+\cdots\right) \\[10pt]
            &= \delta(v_1)\delta(v_2) ~ \left(\left<\phi(x_1)\phi(x_2)\right>^M + v_1 \left<\pi(x_1)\phi(x_2)\right>^M \right.\\
            &\qquad\qquad\qquad\qquad\qquad\qquad \left.+ v_2 \left<\phi(x_1)\pi(x_2)\right>^M + v_1v_2\left<\pi(x_1)\pi(x_2)\right>^M +\cdots\right).\\
        \end{aligned}
    \end{equation}
     Strictly speaking, the relation \eqref{eq:MatchingBargmannCorrelatorAndCarrollianCorrelator} for the magnetic sector of scalar theory is not exact since there is a divergent overall factor $\Gamma(0)$. Despite of the overall divergent factors, the Bargmann correlators could be related to  the Carrollian correlators with correct  patterns and relative coefficients.\par

It is not hard to find that the above discussions also apply to the massive case. Since turning on the mass amounts to adding a constant term in the derivative operator $\hat{D}(\partial_\alpha) \to \hat{D}(\partial_\alpha) + m^2$, it does not spoil the discussions.\par

\section{Carrollian \texorpdfstring{$p$}~-form Field Theories}\label{sec:CarrollianPForm}

    The above construction of Carrollian field theory from Bargmann field theories can be extended to other kinds of field theories. In this section, we study the construction of Carrollian $p$-form field theories, which has been discussed in \cite{Henneaux:2021yzg}. The $p$-form field theories appears in the low energy effective action of superstring theory, and could be important in studying the physics near singularity. We start from Carrollian $1$-form field, which gives rise to Carrollian electromagnetic theories. These theories were first studied in \cite{Duval:2014uoa}. 

\subsection{Electromagnetic theories}
    In this subsection, we construct Carrollian electromagnetic theories from electromagnetic theories in Bargmann space. There are two kinds of Bargmann invariant actions for a vector field in general\footnote{For $d=3, 4$ there are extra topological terms in one higher dimension constructed by contraction with the Levi-Civita tensor:
    \begin{equation*}\label{eq:barg-ext-action}
        \begin{aligned}
            S^\mathscr{B}_{top}&=-\frac{1}{4}\int d^{4}x ~ \epsilon^{\alpha\beta\gamma\delta} F_{\alpha\beta}F_{\gamma\delta}, \qquad (d=3)\\
            S^\mathscr{B}_{top}&=-\frac{1}{4}\int d^{5}x ~ \xi_\lambda\epsilon^{\lambda\alpha\beta\gamma\delta} F_{\alpha\beta}F_{\gamma\delta}, \qquad (d=4).\\
        \end{aligned}
    \end{equation*} 
    In this paper, we do not discuss these topological terms. For higher dimensions, there could be the action involved more than two field-strength tensors, e.g., $d=5$ with $\mathcal{L}=\epsilon^{\alpha\beta\gamma\delta\sigma\rho} F_{\alpha\beta}F_{\gamma\delta}F_{\sigma\rho}$. }
    \begin{equation}
        S^\mathscr{B}_E=-\frac{1}{4}\int d^{d+1}x ~ G^{\alpha\gamma}\xi^\beta\xi^\delta F_{\alpha\beta}F_{\gamma\delta}, 
        \qquad S^\mathscr{B}_M=-\frac{1}{4}\int d^{d+1}x ~  G^{\alpha\gamma} G^{\beta\delta} F_{\alpha\beta}F_{\gamma\delta},
    \end{equation}
    Here $F_{\alpha\beta}=\partial_\alpha a_\beta - \partial_\beta a_\alpha$ is the usual field-strength tensor, and $a_\alpha$ is the gauge potential, which is a vector in Bargmann space. As we have done in the scalar case, we first expand the fundamental field to the powers of $v$ near $v=0$,
    \begin{equation}
        a_\alpha (u,\vec x, v) = A_\alpha (u,\vec x) + v~ \pi_\alpha (u,\vec x) +\mathcal{O}(v^2),
    \end{equation}
    and insert this expansion into the action to read Carrollian theories. Consequently, we obtain the actions of electric sector and magnetic sector of Carrollian electromagnetic theory, from $S^\mathscr{B}_E$ and $S^\mathscr{B}_M$, respectively.\par
    
\subsubsection*{Electric sector}
    The action of electric Carrollian $U(1)$ gauge theory is  \begin{equation}\label{eq:ActionOfCarrollianU1GaugeElectricSector}
        S_E = -\frac{1}{2}\int d^d x ~ F_{0i}F_{0i},
    \end{equation}
    where $F_{0i} = \partial_0 A_i - \partial_i A_0$, and the fundamental fields are $A_\mu = (A_0, A_i)$. In this theory, the field $A_v$ and the second-order field $\pi_\alpha$ get decoupled, and only the leading-order gauge symmetry survives. From the perspective of the representations of Carrollian rotations, $A_\mu$ is in the sub-representation $(1)\to(0)$ of the full $A_\alpha$ representation $(0)\to(1)\to(0)$.  \par
    
    It is well known that $4d$ Lorentzian electromagnetic theory is conformal invariant with the field-strength tensor  $F_{\mu\nu}$ (other than the vector potential fields $A_\mu$) being the primary operators. However, the story is different in the Carrollian case. As will be shown later, both the electric and magnetic sectors  are Carrollian conformal invariant. Moreover,  the gauge potential $A_\mu$ itself is now the primary operators with conformal dimension $\Delta_A=1$ in $d=4$. The actions of the symmetry generators on $A_\mu$ are
    \begin{equation}
        \begin{aligned}
            &\left[D, A_\mu \right] = \Delta_A A_\mu, \qquad \left[K_\mu, A_\nu \right] =0, \qquad \Delta_A = \frac{d-2}{2}, \\
            &\left[J^i_{~j}, A_k \right] = \delta^i_{k} A_j - \delta_{jk} A^i, \qquad \left[J^i_{~j}, A_0 \right] =0, \\
            &\left[B_k, A_i \right] = \delta_{ik} A_0, \qquad \left[B_k, A_0 \right] = 0. \\
        \end{aligned}
    \end{equation}
    Remarkably, the field-strength tensor $F_{0i}$ is a primary operator as well since it satisfies $[K_\mu,F_{0i}]=0$. Thus in the electric sector of Carrollian electromagnetic theory given by  \eqref{eq:ActionOfCarrollianU1GaugeElectricSector}, both $A_\mu$ and $F_{0i}$ are primary operators. This is supported by the fact that both the correlators of $A_\mu$ and the correlators of $F_{0i}$  satisfy the Ward identities of $d=4$ Carrollian conformal symmetry. \par
    
    The action \eqref{eq:ActionOfCarrollianU1GaugeElectricSector}  has a gauge symmetry which transforms the potential $A_\mu$ as
    \begin{equation}\label{eq:u1-gauge-sym-e}
        \begin{aligned}
            & A_\mu(x) \to A_\mu(x) + \partial_\mu \omega_0(x).
        \end{aligned}
    \end{equation}
    We may select a gauge $\partial_0 A_0 = 0$ which is Carrollian conformal invariant  to compute the path-integral. When we take this gauge and select the Landau gauge $\xi=0$, the correlators are simply proportional to the Dirac $\delta$-functions (see Appendix \ref{appsubsec:U1ElectricSector} for details)
    \begin{equation}
        \left<A_i(x)A_j(0)\right> = \frac{i}{2}\delta_{ij}\abs{t}\delta^{(3)}(\vec x), \qquad \text{others}=0,
    \end{equation}
    which obey the Ward identities for the Carrollian conformal symmetries. \par
    
    In fact, $\xi^\mu \xi^\nu \partial_\mu a_\nu = \partial_u a_u$ is an appropriate Bargmann gauge-fixing term whose leading  order in $v$ is exactly $\partial_0 A_0=0$. Using this gauge-fixing term and taking the gauge $\xi=0$, we find that the path-integral gives the Bargmann correlator 
    \begin{equation}
        \left<a_i(x)a_j(0)\right> = \frac{i}{2}\delta_{ij}\abs{u}\delta^{(3)}(\vec x)\delta(v),
    \end{equation}
   whose leading  order in $v$ gives the Carrollian correlator
    \begin{equation}
        \left<a_i(x)a_j(0)\right> = \left<A_i(x)A_j(0)\right> + \mathcal{O}(v).
    \end{equation} \par

    Finally, let us consider the Carrollian Maxwell equations. Denote the magnetic and electric field as
    \begin{equation}\label{eq:F-def-e}
        \bm{B}_k=\frac{1}{2}\epsilon^{ijk}F_{ij}, \qquad \bm{E}_k = F_{0k}.
    \end{equation} 
    Although not appearing in the action,  $F_{ij}=\partial_i A_j-\partial_j A_i$ can be verified to be a primary operator. It is clear that the on-shell equations in the electric sector give the first line of the electric Carroll contraction of the Maxwell equations given in \cite{Duval:2014uoa},
    \begin{equation}\label{Maxwell-e}
        \begin{aligned}
            \nabla \cdot \mathbf{E} &= 0, & \frac{\partial\mathbf E}{\partial t}&= 0, \\
            \nabla \cdot \mathbf{B} &= 0, & \nabla \times \mathbf{E} +\frac{\partial\mathbf B}{\partial t}&= 0.
        \end{aligned}
    \end{equation}
    The other two equations in the second line are indeed automatically satisfied, and in this sense we say this action is a realization of the electric Carrollian electromagnetism and call it the electric sector.\par 

\subsubsection*{Magnetic sector}
    The action of the magnetic Carrollian $U(1)$ gauge theory is
    \begin{equation}\label{eq:ActionOfCarrollianU1GaugeMagneticSector}
        \begin{aligned}
            S_M & = -\frac{1}{4}\int d^d x ~ \delta^{ik}\delta^{jl}F_{ij}F_{kl} + 4\delta^{ij}F_{0i}F_{vj} - 2F_{0v}F_{0v}\\
                & = -\frac{1}{4}\int d^d x ~ \delta^{ik}\delta^{jl}(\partial_i A_j - \partial_j A_i)(\partial_k A_l - \partial_l A_k) \\
                    &\qquad\qquad\qquad\qquad +4 \delta^{ij} (\partial_0 A_i - \partial_i A_0)(\pi_j - \partial_j A_v) -2 (\pi_0 - \partial_0 A_v)^2.\\
        \end{aligned}
    \end{equation}
    The fundamental fields appearing in the action are $A_\alpha=(A_0,A_i,A_v)$ and $\pi_\mu=(\pi_0,\pi_i)$, with $\pi_v$ being decoupled.\par

    \begin{figure}[ht]
        \centering
        \includegraphics[width=11cm,align=c]{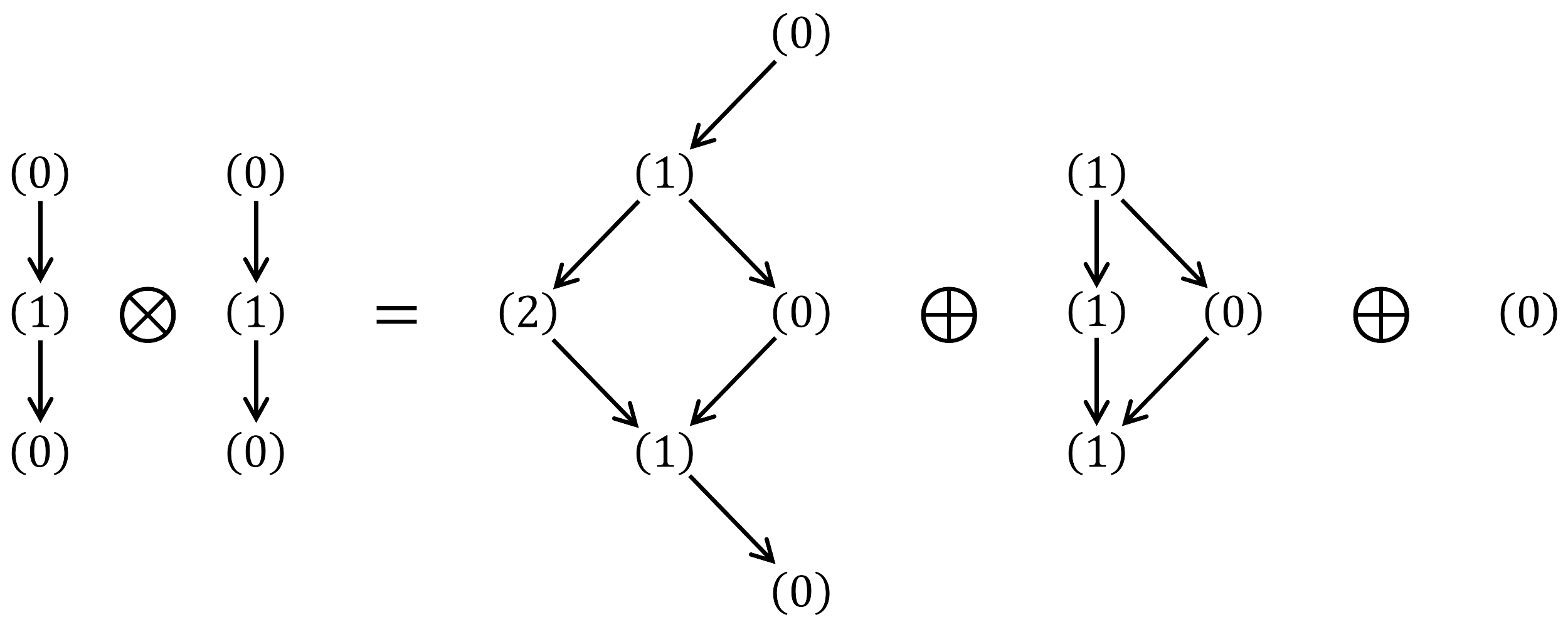}
        \includegraphics[width=15cm,align=c]{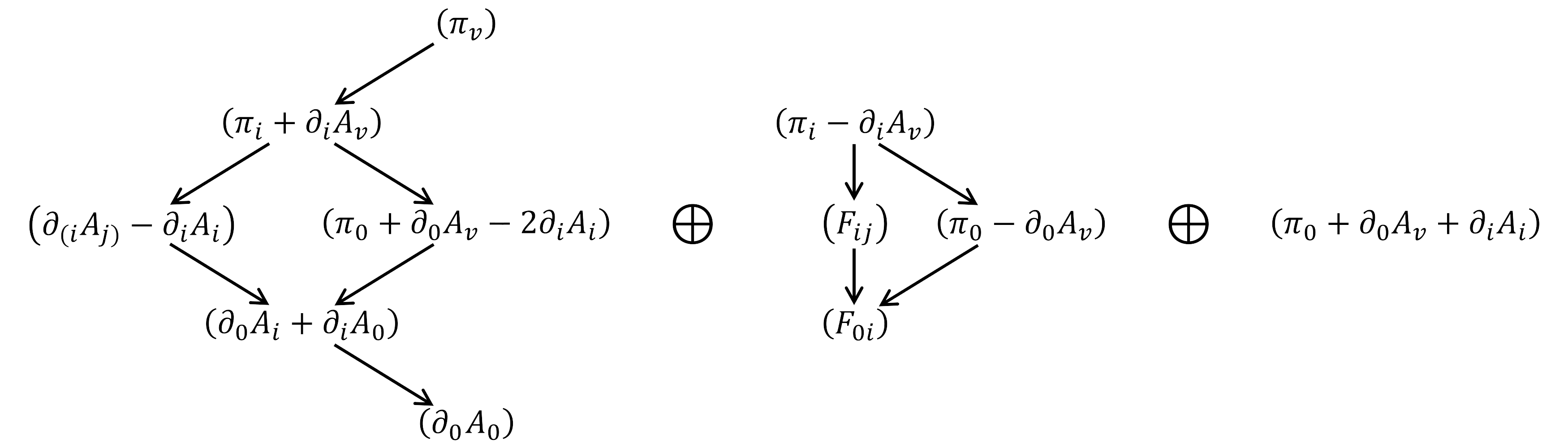}
        \caption{\centering The representation structure of field strength tensors under Carrollian boost $B_i$ (arrows) for $d=4$.}
        \label{fig:RepsOfU1StrengthTensors}
    \end{figure}
    
    Under the Carrollian rotations, the potential fields $A_\alpha$ form a $(0)\to(1)\to(0)$ representation, but the representation of $\pi$ fields is more complex. Actually,  under Carrollian rotations, and especially under the boost generators, the $\pi$ fields together with the first-order derivatives of $A_\alpha$, i.e., $\partial_\mu A_\alpha$, form the tensor product of two $(0)\to(1)\to(0)$ representations, as shown in  Figure \ref{fig:RepsOfU1StrengthTensors}. The upper part of Figure \ref{fig:RepsOfU1StrengthTensors} shows the decomposition of the tensor product, while the lower part  shows the field realizations.  The action of the Carrollian rotations on the fields are: \begin{equation}\label{eq:MagneticEMCarrollianRotationActions}
        \begin{aligned}
            &\left[J_{kl}, A_v \right] = 0, \qquad \left[J_{kl}, A_i \right] = \delta_{ik} A_l - \delta_{il} A_k, \qquad \left[J_{kl}, A_0 \right] = 0, \\
            &\left[J_{kl}, \pi_i \right] = \delta_{ik} \pi_l - \delta_{il} \pi_k, \qquad \left[J_{kl}, \pi_0 \right] = 0, \qquad \left[J_{kl}, \pi_v \right] = 0, \\
            &\left[B_k, A_v \right] = -A_k, \qquad \left[B_k, A_i \right] = \delta_{ik} A_0, \qquad \left[B_k, A_0 \right] = 0, \\
            &\left[B_k, \pi_i \right] = \delta_{ik}\pi_0 - \partial_k A_i, \qquad \left[B_k, \pi_0 \right] = - \partial_k A_0, \quad\left[B_k, \pi_v \right] = -\pi_k - \partial_k A_v, \\
        \end{aligned}
    \end{equation}
    which can be read off from the expansion of Bargmann field $a_\alpha$,  similar to equation \eqref{eq:MagneticScalarExpansionofBoostAction}.  It is easy to see that  only the middle part in Figure \ref{fig:RepsOfU1StrengthTensors}, the anti-symmetric part, appears in the  action of the magnetic sector. It is straightforward to verify that the action \eqref{eq:ActionOfCarrollianU1GaugeMagneticSector} is invariant under these actions and thus represents a Carrollian field theory. \par

    Now we consider the Carrollian conformal symmetries. Unlike the free scalar  theories that are Carrollian conformal in generic dimension, the magnetic sector of Carrollian electromagnetic theory is conformal only in $d=4$. The potential $A_\mu$ is a primary operator with conformal dimension $\Delta_A=1$, while the combinations of $\pi_\mu$ fields and $\partial_\mu A_\nu$ are descendent operators with dimension $\Delta_\pi = \Delta_A+1= 2$. The explicit actions of the CCA generators on the fields are given by  \begin{equation}\label{eq:MagneticEMCarrollianConformalActions}
        \begin{aligned}
            &\left[D,A_\alpha \right]=\Delta_A A_\alpha=A_\alpha,  \quad \left[D,\pi_\alpha\right]=\Delta_\pi \pi_\alpha=2 \pi_\alpha, \quad \left[K_\mu,A_\alpha \right]=0, \\
            & \left[K_i,\pi_j\right]=2\delta_{ij}A_v, \quad \left[K_i,\pi_0\right]=-2A_i, \quad \left[K_i,\pi_v\right]=0,\\
            & \left[K_0,\pi_i\right]=2\Delta_A A_i, \quad \left[K_0,\pi_0\right]=2(\Delta_A-1) A_0=0, \quad \left[K_0,\pi_v\right]=2(\Delta_A+1) A_v=4 A_v,
        \end{aligned}
    \end{equation}
    where $\alpha = 0, i, v,$ and $\mu=0, i$. It can be further checked that the field-strength tensors $F_{vi}, F_{v0}, F_{ij},F_{0i}$ are all primary operators as well. 

    \begin{figure}[ht]
        \centering
        \includegraphics[width=15cm,align=c]{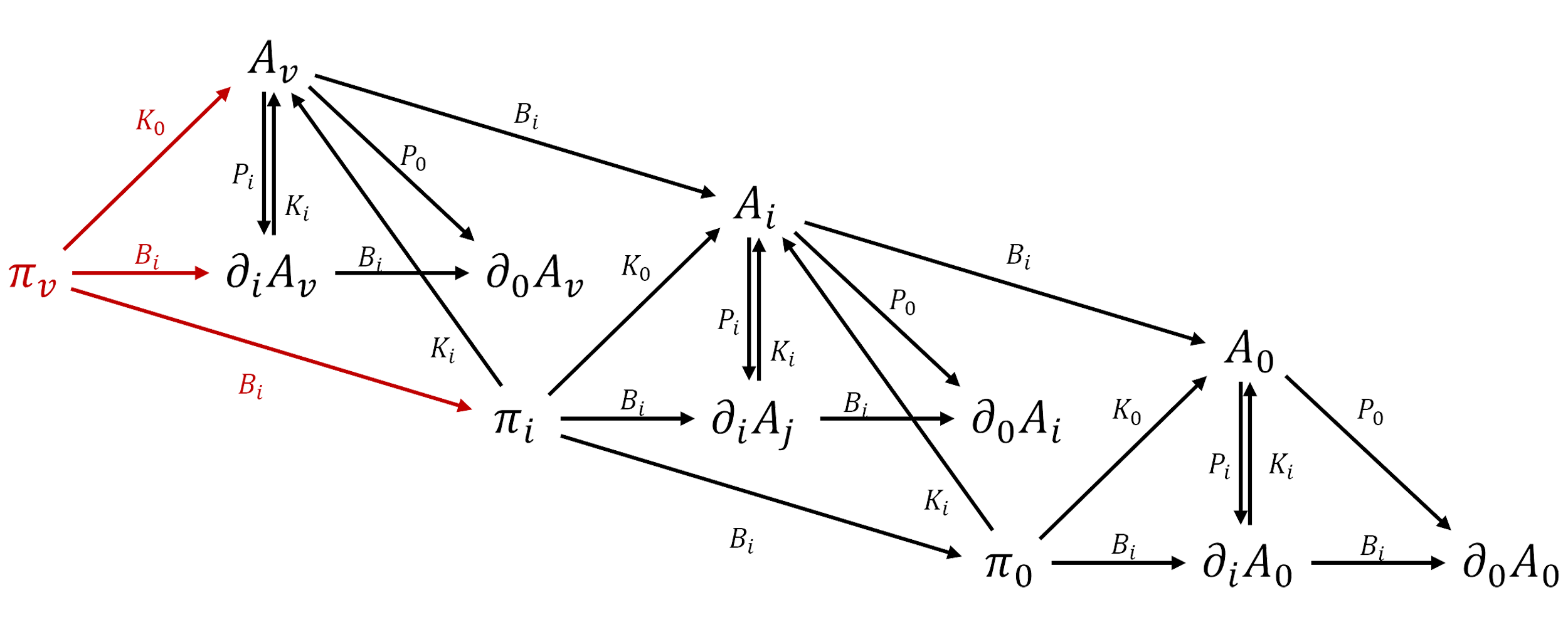}
        \caption{\centering The full structure of  the fields $\{A_\alpha, \pi_\alpha, \partial_\mu A_\alpha\}$. The arrows labels the action of some Carrollian conformal generators, whose explicit actions are shown in \eqref{eq:MagneticEMCarrollianRotationActions} and \eqref{eq:MagneticEMCarrollianConformalActions}. The black part shows the relations among the fields in the magnetic sector of electromagnetic theory, while the red part involving $\pi_v$ does not appear in the theory. }
        \label{fig:U1StaggeredModule}
    \end{figure}

    Different from the electric sector, the fields in the magnetic sector do not form a staggered structure. The relations between the fields are illustrated in Figure \ref{fig:U1StaggeredModule}. The representations of the operators $\{A_\alpha, \pi_\mu, \partial_\mu A_\alpha\}$ do not have a staggered structure. The fields $\pi_i$ which potentially lead to a staggered structure are combinations of the descendent operators $\partial_i A_v$ and the primary operator $F_{vi}$, in the form of  $\pi_i= F_{vi} + \partial_i A_v$. Similarly, $\pi_0= F_{v0} + \partial_0 A_v$ does not cause a staggered structure. As shown in Fig. \ref{fig:U1StaggeredModule}, the field $\pi_v$ may lead to a staggered module, but it does not appear in the action. The actions of the CCA generators on $\pi_v$ are listed in \eqref{eq:MagneticEMCarrollianRotationActions} and \eqref{eq:MagneticEMCarrollianConformalActions}, as well as in Figure \ref{fig:U1StaggeredModule}. Thus we can safely say that the magnetic sector of electromagnetic theory does not contain staggered module. \par
    
    In this case, the gauge fixing should be  treated carefully. The gauge transformation in Bargmann $U(1)$ gauge theory is $a_\mu(u,\vec x,v) \to a_\mu(u,\vec x,v) + \partial_\mu \Omega(u,\vec x,v)$, where $\Omega(x^\alpha)$ is the gauge parameter. Considering the expansion to the powers of $v$ and keeping the leading-order term in $v$,  we find that there are two set of gauge transformations for the magnetic Carrollian $U(1)$ theory
    \begin{equation}\label{eq:u1-gauge-sym}
        \begin{aligned}
            &A_i(x) \to A_i(x) + \partial_i \omega_0(x), \quad A_0(x) \to A_0(x) + \partial_0 \omega_0(x),\\
            &A_v(x) \to A_v(x) + \omega_1(x), \quad \pi_i(x) \to \pi_i(x) + \partial_i \omega_1(x), \quad \pi_0(x) \to \pi_0(x) + \partial_0 \omega_1(x).\\
        \end{aligned}
    \end{equation}
    where $\omega_0(x),\omega_1(x)$ appearing in the expansion of the Bargmann gauge parameters $\Omega(x^\alpha) = \omega_0(x^\mu) +\omega_1(x^\mu) v +\cdots$ are referred to as the first-order and the second-order  gauge parameters, respectively. We find that the following gauge-fixing term is manifestly Carrollian conformal invariant 
    \begin{equation}\label{eq:covariant-gauge-fixing}
        \mathcal{L}_{gf} = -\frac{1}{2 \xi_1}\left(\partial_0 A_0\right)^2-\frac{1}{2 \xi_2}\left(3 \partial_i A_0 \partial_i A_0 + 2 \partial_0 A_0 \left(2\pi_0 - \partial_0 A_v - \partial_i A_i\right)\right).
    \end{equation}
    In this expression we have summed over repeated indices. With the help of this gauge-fixing term, the correlators are now computable in the path-integral formalism, and by selecting the Landau-type gauge $\xi_2 =0$, the correlators can be organized in a relatively compact form 
    \begin{equation}\label{eq:u1-mag-corr}
        \begin{aligned}
            &\left<A_v(x)A_v(0)\right> =-2i \abs{t} \delta^{(3)}(\vec x), \quad \left<A_v(x)\pi_i (0)\right> = -\left<\pi_i (x)A_v(0)\right>=\frac{3i}{2} \abs{t} \partial_i \delta^{(3)}(\vec x),\\
            &\left<A_v(x)\pi_0 (0)\right> = -\left<\pi_0 (x)A_v(0)\right>=i \mbox{Sign}(t) \delta^{(3)}(\vec x),\\
            &\left<A_i(x)\pi_j(0)\right> = -\left<\pi_j(x)A_i(0)\right> = -\frac{i}{2} \delta_{ij} \mbox{Sign}(t) \delta^{(3)}(\vec x),\\
            &\left<\pi_i(x)\pi_j(0)\right> = \left<\pi_j(x)\pi_i(0)\right> = \frac{i}{2}\abs{t}  \left( \partial_i\partial_j\delta^{(3)}(\vec x)+ \delta_{ij} \vec{\partial}^2 \delta^{(3)}(\vec x)\right),\\
            &\left<\pi_i(x)\pi_0(0)\right> = \left<\pi_0(x)\pi_i(0)\right> = \frac{i}{2}\mbox{Sign}(t)  \partial_i \delta^{(3)}(\vec x), \qquad \left<\pi_0(x)\pi_0 (0)\right> = i \delta(t) \delta^{(3)}(\vec x),
        \end{aligned}
    \end{equation}
    with all other correlators being vanishing. It should be noted that even if we use the gauge-fixing Lagangian without the $\xi_1$-term, the path-integral is still well-defined and the final result is the same. The details of calculations can be found in Appendix \ref{appsubsec:U1MagneticSector}.\par
    
    Different from the scalar case, the above correlators cannot be reduced directly from those in the Bargmann theory. This is because the complicated gauge choice lacks a corresponding consistent choice on the Bargmann manifold. \par

\begin{figure}[ht]
        \centering
        \includegraphics[width=6cm,align=c]{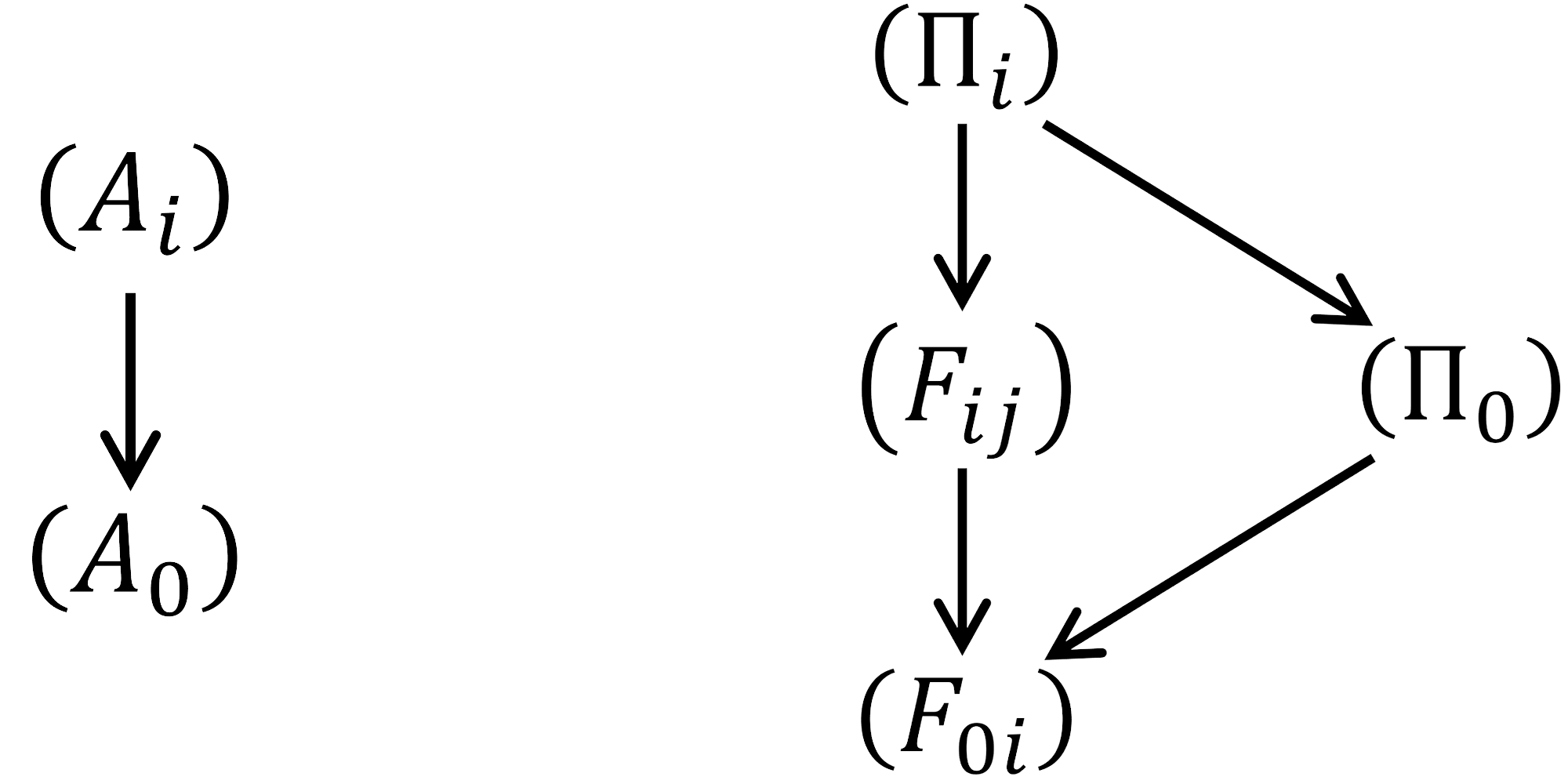}
        \caption{\centering The representations of redefined fields in the magnetic sector of electromagnetic theory. The $A_\mu$ fields are in $(1)\to (0)$ representation, while the $\Pi$ fields as well as the field strength tensors are in a net representation. }
        \label{fig:RepsofRedefinedMagneticEM}
    \end{figure}

    With all field components reduced from the Bargmann fields unmodified, the approach adopted above preserves the Carrollian symmetry and gauge symmetry as much as possible. However, it is also possible to perform the following field redefinition first 
    \begin{equation}\label{eq:redef-u1m}
        \Pi_i = \pi_i -\partial_i A_v, \qquad \Pi_0 = \pi_0 -\partial_0 A_v,
    \end{equation} 
    to simplify the action
    \begin{equation}\label{eq:ActionOfCarrollianU1GaugeMagneticSector-mod}
        \begin{aligned}
            S_M^\prime[A_i,A_0,\Pi_i,\Pi_0] = -\frac{1}{4}\int d^d x ~ (\partial_i A_j - \partial_j A_i)^2+4 \Pi_i (\partial_0 A_i - \partial_i A_0) -2 \Pi_0^2.\\
        \end{aligned}
    \end{equation}
    After this field transformation, the $A_v$ field is implicit as it is partially absorbed into the $\Pi$-fields. The $\Pi$-fields and thus the action \eqref{eq:ActionOfCarrollianU1GaugeMagneticSector-mod} are invariant under gauge transformation generated by $\omega_1$ in \eqref{eq:u1-gauge-sym}. Actually, the $\Pi$-fields lose the interpretation as the sub-leading-order term of the Bargmann fields $a_\mu$ and match the leading-order term of Bargmann field strength tensors $F_{v\mu}$, even though they are not fundamental fields in Bargmann $U(1)$ gauge theory. Owing to this fact, the second-order gauge symmetry is hidden because it leaves all the remaining fields invariant, while the first-order gauge symmetry and the Carrollian symmetry continue to be manifest. The action \eqref{eq:ActionOfCarrollianU1GaugeMagneticSector-mod} suggests that the $\Pi_0$ field is decoupled from other fields, and appears only as a mass term without dynamics. Nevertheless, it is helpful to keep it in the action to show the structure of representation and to explicitly exhibit the  Carrollian (conformal) symmetry without using equations of motion. The transformation rules for the modified fields under the Carrollian boosts and SCTs are
    \begin{equation}\label{eq:u1m-nv-trans}
        \begin{aligned}
            &\left[B_k, A_i \right] = \delta_{ik} A_0, \qquad \left[B_k, A_0 \right] = 0, \\
            &\left[B_k, \Pi_i \right] = \delta_{ik}\Pi_0 + (\partial_i A_k-\partial_k A_i) = \delta_{ik}\Pi_0 + F_{ik}, \\
            &\left[B_k, \Pi_0 \right] = \partial_0 A_k-\partial_k A_0 = F_{0k}, \\
            &\left[K_\mu,A_\nu \right]=\left[K_\mu,\Pi_\nu \right]=0, \qquad \mu,\nu=0,i.
        \end{aligned}
    \end{equation}
    These relations can be illustrated in Figure \ref{fig:RepsofRedefinedMagneticEM}.
    
    The modified fields are all Carrollian primaries, and  $\Pi_i$ are the conjugate momenta of $A_i$. The correlation functions  can be derived both from the combination of the previous correlators in \eqref{eq:u1-mag-corr}, or directly from the path-integral by adding the gauge fixing term $-\frac{1}{2\xi}(\partial_0 A_0)^2$ to \eqref{eq:ActionOfCarrollianU1GaugeMagneticSector-mod} and taking $\xi=0$ in the end. The two approaches turn out to be consistent, and lead to the correlators of the following forms 
    \begin{equation}\label{eq:u1-mag-corr-mod}
        \begin{aligned}
            &\left<\Pi_0(x)\Pi_0 (0)\right> =i \delta(t) \delta^{(3)}(\vec x),\\
            &\left<A_i(x)\Pi_j(0)\right> =-\left<\Pi_j(x) A_i (0)\right>= -\frac{i}{2} \delta_{ij} \mbox{Sign}(t) \delta^{(3)}(\vec x),\\ 
            &\left<\Pi_i(x)\Pi_j(0)\right> = \left<\Pi_i(x)\Pi_j(0)\right> =\frac{i}{2}\abs{t}\left(\delta_{ij} {\vec \partial}^2-\partial_i \partial_j\right)\delta^{(3)}(\vec x).\\
        \end{aligned}
    \end{equation}
    As expected, the $\Pi_0$ field behaves as a non-kinematic field. Nevertheless, we should still keep it in the theory if we want to write down the Ward identities associated to Carrollian boost symmetries, since $\left[B_k, \Pi_i \right] = \delta_{ik}\Pi_0 + F_{ik}$.\par

    We can  recover the magnetic Carrollian electromagnetism found in \cite{Duval:2014uoa}, where the Maxwell equations are
    \begin{equation}\label{Maxwell-m}
        \begin{aligned}
            \nabla \cdot \mathbf{E} &= 0, & \nabla \times \mathbf{B} - \frac{\partial\mathbf E}{\partial t}&= 0, \\
            \nabla \cdot \mathbf{B} &= 0, & \frac{\partial\mathbf B}{\partial t}&=0.
        \end{aligned}
    \end{equation}
    To reveal it, we need to define
    \begin{equation}\label{F-def-m}
        \bm{B}_k=\frac{1}{2}\epsilon^{ijk}F_{ij}, \qquad \bm{E}_k = F_{vk} = \Pi_k
    \end{equation}
    The equations of motions for \eqref{eq:ActionOfCarrollianU1GaugeMagneticSector-mod} are given by:
    \begin{equation}\label{eq:eom-u1-mag}
        \begin{aligned}
            \delta A_i:& \qquad \partial_j F_{ij}=\partial_0 \Pi_i\\
            \delta A_0:& \qquad \partial_i \Pi_i = 0\\
            \delta \Pi_i:& \qquad F_{0i} = 0\\
            \delta \Pi_0:& \qquad \Pi_0=0\\
        \end{aligned}
    \end{equation}
    The on-shell equations with regard to $\pi_i, \pi_0/\Pi_i, \Pi_0$ provide us with $F_{0i}=\Pi_0=0$, and the equations with regard to $A_0,A_i$ give the first line in \eqref{Maxwell-m}, after using $(\nabla \times \mathbf{B})_i =  \partial_j F_{ij}. $ \\
    A simple calculation shows that $\nabla \cdot \mathbf{B} =\epsilon^{ijk}\partial_k F_{ij} =0$. Finally, from $F_{0i}=0$ we have $$\partial_0 F_{ij} =\partial_i F_{0j} - \partial_j F_{0i}=0,$$ which implies the last Maxwell equation $\partial_0 \mathbf{B}=0$. Under $F_{0i}=\Pi_0=0$, the on-shell boost transformation rule becomes just the correct one
    \begin{equation}
        [B_i,\bm{E}_j] = \epsilon^{ijk}\bm{B}_k = - (\bm{b_i} \cross \bm{B})_j.
    \end{equation}\par
    
    Besides, it is well worth mentioning that in the case $d=3$, the corresponding $(d+1)=4$ Bargmann theory possesses the electromagnetic duality $F \leftrightarrow *F$. After restriction to the null hyperplane, the magnetic $U(1)$ Lagrangian \eqref{eq:ActionOfCarrollianU1GaugeMagneticSector} is invariant under the transformations
    \begin{equation}\label{electromagnetic-duality}
        \begin{aligned}
            \pi_0-\partial_0A_v \leftrightarrow F_{12},\quad\pi_1-\partial_1A_v \leftrightarrow F_{02},\quad\pi_2-\partial_2A_v \leftrightarrow F_{01}.
        \end{aligned}
    \end{equation}
   Note that this kind of duality is intrinsic in the magnetic sector.

\subsection{\texorpdfstring{$p$}~-form theories}
    It is straightforward to extend the above construction to the $p$-form free theory. Here we only briefly introduce the construction of the Carrollian action, without discussing the Carrollian conformal symmetry. For a general $p$-form field $a$, the field strength $F$ is a $(p+1)$-form. Similar to the electromagnetic case, there are only two kinds of Bargmann invariant actions:
    \begin{equation}
        \begin{aligned}
            S^\mathscr{B}_E & =-\frac{1}{2(p+1)!}\int d^{d+1}x ~ \xi^{\alpha_1}\xi^{\beta_1} G^{\alpha_2\beta_2}\cdots G^{\alpha_{p+1}\beta_{p+1}}F_{\alpha_1\cdots\alpha_{p+1}}F_{\beta_1\cdots\beta_{p+1}}, \\
            S^\mathscr{B}_M & =-\frac{1}{2(p+1)!}\int d^{d+1}x ~  G^{\alpha_1\beta_1}\cdots G^{\alpha_{p+1}\beta_{p+1}}F_{\alpha_1\cdots\alpha_{p+1}}F_{\beta_1\cdots\beta_{p+1}}.
        \end{aligned}
    \end{equation}
     Other combinations of $G, \xi$ and $F$ would be vanishing since the field strength is anti-symmetric. In the above actions, $F_{\alpha_1\cdots\alpha_{p+1}}=(p+1) \partial_{[\alpha_1} a_{\alpha_2\cdots\alpha_{p+1}]}$ is the field strength tensor, and $a_{\alpha_1\cdots\alpha_{p}}$ is a $p$-form gauge potential in Bargmann space. The expansion of the field $a$ to the powers of $v$ near $v=0$ is 
    \begin{equation}
        a_{\alpha_1\cdots\alpha_{p}} (u,\vec x, v) = A_{\alpha_1\cdots\alpha_{p}} (u,\vec x) + v~ \pi_{\alpha_1\cdots\alpha_{p}} (u,\vec x) +\mathcal{O}(v^2).
    \end{equation} \par
    
   For the $p$-form gauge theory, the action of electric sector is
    \begin{equation}\label{eq:ActionOfCarrollianPFormElectricSector}
        S_E = -\frac{1}{2(p+1)!}\int d^d x ~ F_{0i_2\cdots i_{p+1}}F_{0i_2\cdots i_{p+1}},
    \end{equation}
    and the action of magnetic sector is
    \begin{equation}\label{eq:ActionOfCarrollianPFormMagneticSector}
        \begin{aligned}
            S_M & = -\frac{1}{2(p+1)!}\int d^d x ~ F_{i_1\cdots i_{p+1}}F_{i_1\cdots i_{p+1}} \\ 
                &\qquad\qquad\qquad\qquad + 2(p+1) F_{0 i_2\cdots i_{p+1}}F_{v i_2\cdots i_{p+1}} - p(p+1) F_{0v i_3\cdots i_{p+1}}F_{0v  i_3\cdots i_{p+1}}\\
                & = -\frac{1}{2(p+1)!}\int d^d x ~  (p+1)^2 (\partial_{[i_1} A_{i_2\cdots i_{p+1}]})^2 \\ 
                    &\qquad\qquad + 2(p+1)^2 \partial_{[0} A_{i_2\cdots i_{p+1}]} (\pi_{i_2\cdots i_{p+1}} + \partial_{[i_2} A_{i_3\cdots i_{p+1}]v})\\
                    &\qquad\qquad - p(p+1) (\pi_{0 i_3\cdots i_{p+1}} + \partial_{[0} A_{i_3\cdots i_{p+1}] v})^2.
                    \end{aligned}
    \end{equation}
   After integrating out  $\pi_{u i_2\cdots i_{p}}$and $A_{0 i_2\cdots i_{p}}$, we find the following action  
   \begin{equation}
       S_M=\int d^d x ~ \frac{(p+1)}{p!} \Pi_{i_2\cdots i_{p+1}}\partial_{[0} A_{i_2\cdots i_{p+1}]}  -\frac{1}{2~(p-1)!} (\partial_{[i_1} A_{i_2\cdots i_{p+1}]})^2, 
   \end{equation}
    which is the same as the one in  \cite{Henneaux:2021yzg}. The fundamental fields in this action are $A_{\alpha_1\cdots \alpha_{p}}$ and $\pi_{\mu_1\cdots \mu_{p}}$, where $\Pi_{i_1\cdots i_{p}}$ are canonical momentum of fields $A_{i_1\cdots i_{p}}$. The representation of $A$ is the totally anti-symmetric part of $[(0)\to(1)\to(0)]^{\otimes p}$, and the representation of $(\pi,\partial A)$ is the totally anti-symmetric part of $[(0)\to(1)\to(0)]\otimes[A]$ in $d=4$.\par

    \section{Carrollian field theories from further reduction}\label{quotientreduction}
    
    In the last few sections, we have constructed Carrollian invariant field theories from null reduction of Bargmann field theories. Although we managed to obtain a bunch of Carrollian field theories including scalar theories, $U(1)$ theories and $p$-form theories, we can take a further step. In this section, we will show that some of these theories can be modified by removing some of its field components, and the resulting theories are still  Carrollian invariant. The essential point is that  both the removed fields and the remaining fields still form  {\it bona fide} sub-representations of  Carrollian rotations. Such modifications result in intrinsically different theories. In this way, we are able to discuss Carrollian field theories which can not be directly reduced from Bargmann field theories.\par

    As reviewed in section \ref{subsec:CCAReps}, a multiplet  representation of Carrollian rotation group is reducible but indecomposible. It could be organized as either chain representation or net representation.  An interesting property is that we can always find sub-representations in a multiplet representation. The presence of such sub-representations allows to remove them and construct another shorter but well-defined representation of Carrollian rotations using the remaining fields. We have seen in \cite{Chen:2021xkw} that all possible chain representations with at least rank $2$ are in one of the following patterns:\par
    \paragraph{Rank 2:}
    \begin{equation}\label{Rank2}
        \begin{aligned}
            (j) \to& (j),~~ j\neq0\\
            (j) \to& (j+1),\\
            (j) \to& (j-1).\\
        \end{aligned}
    \end{equation}\par
    \paragraph{$\geq$ Rank 3:}
    \begin{equation}\label{Rank3}
        \begin{aligned}
            (0) \to (1)& \to (0),\\
            \cdots \to (j) \to (j+1)& \to (j+2) \to \cdots,\\
            \cdots \to (j) \to (j-1)& \to (j-2) \to \cdots.\\
        \end{aligned}\\
    \end{equation}
    For $(0) \to (1) \to (0)$ representation, both $(1) \to (0)$ and $(0)$ are its sub-representations. After removing one of these sub-representations from the chain, the resulting quotient representation $(0)$ or $(0) \to (1)$ still belongs to  the allowed patterns. Similar results hold for the increasing and decreasing chains. Let us consider the  chain $(j)\to(j\pm1)\to(j\pm2)\to\cdots\to(j\pm n)$ of length $n$, for any integer $m < n$ the sub-chain starting at $(j\pm(m+1))$ and ending at $(j\pm n)$ will give a sub-representation. Upon removing this sub-representation from the full representation we get a shorter chain $(j)\to(j\pm1)\to(j\pm2)\to\cdots\to(j\pm m)$ as a quotient representation. Moreover, we can generalize to net representations and notice that removing a sub-net from the whole net will give another legal net representation. Although this seems not to be the unique way to get {\it bona fide} representation of  Carrollian rotations by removing fields, because for example we can also remove the first few terms and the last few terms of a  chain representation to get another representation, it is necessary to remove just sub-representations to get quotient representations in order that the reduced action is Carrollian invariant. 

    Consequently,  we are able to reduce field components from already established Carrollian field theories so as to obtain new Carrollian field theories of the remaining field components. The only requirement is that the removed fields  should consist of only fundamental fields with no derivatives and they form a sub-representation. Here we use $\bm{\Phi}=(\Phi_i)_{i\in\mathcal{I}}$ to stand for the original Carrollian fields and $S[\bm{\Phi}]$ for the original action, $\bm{\Psi}$ for the fields making up the sub-representation, $\bar{\bm{\Phi}}=(\bar{\Phi}_j)_{j\in\mathcal{I'}\subset \mathcal{I}}$ for the remaining fields after removing $\bm{\Psi}$ from $\bm{\Phi}$, and $\bar{S}[\bar{\bm{\Phi}}]$ for the modified action. For any generator $g$ of Carrollian (conformal) group, the invariance of modified action under the transformation can be seen from the  variation
    \begin{equation}\label{var-of-reduced-action}
        \bar{\delta}_g \bar{S}[\bar{\bm{\Phi}}] = \sum_{j\in\mathcal{I'}} \left.  \frac{\delta S}{\delta \bar{\Phi}_j}\right|_{\bm{\Psi}=0}\bar{\delta}_g \bar{\Phi}_j = \left. \left( \sum_{i\in\mathcal{I}}  \frac{\delta S}{\delta \Phi_i}\delta_g \Phi_i \right) \right|_{\bm{\Psi}=0} = \left. \delta_g S[\bm{\Phi}] \right|_{\bm{\Psi}=0} =0,
    \end{equation}
    where $\bar{\delta}_g \bar{\Phi}_j = \left. \delta_g \Phi_j \right|_{\bm{\Psi}=0}$. The removed fields must fit into a sub-representation in this argument, because $ \delta_g \bm{\Psi}$ should only depend on $\bm{\Psi}$ and becomes vanishing after reduction. This does not only make sense for Carrollian symmetry, but also for Carrollian conformal symmetry.\par

    Now we take the $U(1)$ magnetic sector as a non-trivial example. In this case, as  analyzed in previous sections, the potential fields $A_\alpha=(A_v,A_i,A_0)$ form a $(0) \to (1) \to (0)$ representation. As discussed above, $(A_0)$ and $(A_i) \to (A_0)$ are two sub-representations, which can be removed from the theory. Let us consider the resulting Lagrangians case by case.\par
    
{\noindent$\bm{A_0 = 0:}$}\par
    The simplest choice is $A_0=0$, and the reduced action is given by
    \begin{equation}\label{eq:u1m-rd-action}
        \bar{S}_M[A_v,A_i,\pi_i,\pi_0] = -\frac{1}{4}\int d^d x ~ (\partial_i A_j - \partial_j A_i)^2 +4(\partial_0 A_i)(\pi_i - \partial_i A_v) -2 (\pi_0 - \partial_0 A_v)^2.
    \end{equation}
    The actions of the boost generators and SCTs in $d=4$ are modified to be
    \begin{equation}\label{eq:u1m-rd-trans}
        \begin{aligned}
            &\left[B_k, A_v \right] = -A_k, \quad \left[B_k, A_i \right] = 0, \quad \left[B_k, \pi_i \right] = \delta_{ik}\pi_0 - \partial_k A_i, \quad \left[B_k, \pi_0 \right] = 0,\\
            &\left[K_0,\pi_i\right]=2A_i,\quad\left[K_i,\pi_0\right]=-2A_i, \quad \left[K_i,\pi_j\right]=2\delta_{ij}A_v.\\
        \end{aligned}
    \end{equation}
    The action is still invariant under all the Carrollian conformal transformations. However, the gauge transformation on $A_0$ is no longer a symmetry, and the action is now only invariant under the second order gauge transformation
    \begin{equation}\label{eq:u1-gauge-quotient}
        \begin{aligned}
            &A_v(x) \to A_v(x) + \omega_1(x), \quad \pi_i(x) \to \pi_i(x) + \partial_i \omega_1(x), \quad \pi_0(x) \to \pi_0(x) + \partial_0 \omega_1(x).\\
        \end{aligned}
    \end{equation}
    It is not difficult to work out 2-point correlators via the path-integral, if we carefully choose the gauge-fixing term to be $-\frac{1}{2\xi}\left(2\pi_0 - \partial_0 A_v - \partial_i A_i\right)^2$,   
    \begin{equation}\label{eq:u1-mag-corr-noA0}
        \begin{aligned}
            &\left<A_v(x)A_v(0)\right> =-i (2-\frac{\xi}{2}) \abs{t} \delta^{(3)}(\vec x), \quad \left<A_v(x)\pi_i (0)\right> = -\left<\pi_i (x)A_v(0)\right>= \frac{i}{2} (3 - \xi) \abs{t} \partial_i \delta^{(3)}(\vec x),\\
            &\left<A_v(x)\pi_0 (0)\right> = -\left<\pi_0 (x)A_v(0)\right>=i (1-\frac{\xi}{2}) \mbox{Sign}(t) \delta^{(3)}(\vec x),\\
            &\left<A_i(x)\pi_j(0)\right> = -\left<\pi_j(x)A_i(0)\right> = -\frac{i}{2} \delta_{ij} \mbox{Sign}(t) \delta^{(3)}(\vec x),\\
            &\left<\pi_i(x)\pi_j(0)\right> = \left<\pi_j(x)\pi_i(0)\right> = \frac{i}{2}  \abs{t}  \left((1- \xi) \partial_i\partial_j\delta^{(3)}(\vec x)+ \delta_{ij} \vec{\partial}^2 \delta^{(3)}(\vec x)\right),\\
            &\left<\pi_i(x)\pi_0(0)\right> = \left<\pi_0(x)\pi_i(0)\right> = \frac{i}{2}(1-\xi)\mbox{Sign}(t)  \partial_i \delta^{(3)}(\vec x), \quad \left<\pi_0(x)\pi_0 (0)\right> = i (1-\xi) \delta(t) \delta^{(3)}(\vec x).
        \end{aligned}
    \end{equation}\par

    Similarly, after redefining $\Pi$-fields as in \eqref{eq:redef-u1m} to absorb $A_v$,  we get the action
    \begin{equation}\label{eq:u1m-rd-action-nv}
        \bar{S}_M[A_i,\Pi_i,\Pi_0] = -\frac{1}{4}\int d^d x ~ (\partial_i A_j - \partial_j A_i)^2 +4\Pi_i \partial_0 A_i -2 \Pi_0^2.
    \end{equation} 
    The corresponding action under the symmetry generators are 
    \begin{equation}\label{eq:u1m-rd-trans-nv}
        \begin{aligned}
            &\left[B_k, A_i \right] = 0, \\
            &\left[B_k, \Pi_i \right] = \delta_{ik}\Pi_0 + (\partial_i A_k-\partial_k A_i), \qquad \left[B_k, \Pi_0 \right] = \partial_0 A_k, \\
            &\left[K_\mu,A_\nu \right]=\left[K_\mu,\Pi_\nu \right]=0, \qquad \mu,\nu=0,i.
        \end{aligned}
    \end{equation}
In this formulation, we are free of gauge redundancy, and do not need to impose gauge fixing. The correlators are now
    \begin{equation}\label{eq:u1m-rd-corr-nv}
        \begin{aligned}
            &\left<\Pi_0(x)\Pi_0 (0)\right> =i \delta(t) \delta^{(3)}(\vec x),\\
            &\left<A_i(x)\Pi_j(0)\right> =-\left<\Pi_j(x) A_i (0)\right>= -\frac{i}{2} \delta_{ij} \mbox{Sign}(t) \delta^{(3)}(\vec x),\\ 
            &\left<\Pi_i(x)\Pi_j(0)\right> = \left<\Pi_i(x)\Pi_j(0)\right> =\frac{i}{2}\abs{t}\left(\delta_{ij} {\vec \partial}^2-\partial_i \partial_j\right)\delta^{(3)}(\vec x),\\
        \end{aligned}
    \end{equation}
which  are the same as the $\xi=0$ correlators in \eqref{eq:u1-mag-corr-mod} if neglecting the ones involving $A_0$. In fact, the resulting theory is intrinsically different from the original theory. This can be seen from the equations of motion 
    \begin{equation}\label{eq:quotient-eom-u1-mag}
        \begin{aligned}
            \delta A_i:& \qquad \partial_i F_{ij}=\partial_0 \Pi_i,\\
            \delta \Pi_i:& \qquad \partial_0 A_i = 0,\\
            \delta \Pi_0:& \qquad \Pi_0=0.\\
        \end{aligned}
    \end{equation}
    Here is no equation with respect to $A_0$, which leads to $\partial_i\Pi_i=0$.
    
As it stands, there are four apparently different Lagrangians starting from Bargmann $U(1)$ theory, including \eqref{eq:ActionOfCarrollianU1GaugeMagneticSector}, \eqref{eq:ActionOfCarrollianU1GaugeMagneticSector-mod}, \eqref{eq:u1m-rd-action}, and \eqref{eq:u1m-rd-action-nv}, each having different numbers of fundamental fields and gauge symmetries.
    
{\noindent$\bm{A_i = A_0 = 0:}$}\par
    If the field components $A_i,A_0$ in  the $(0)\to (1)\to (0)$ representation are removed, the action will reduce to a simpler one,
    \begin{equation}
        \bar{S}_M[A_v,\pi_0] = \frac{1}{2}\int d^d x ~ (\pi_0 - \partial_0 A_v)^2.
    \end{equation}
    Not only do $A_i,A_0$ vanish, but $\pi_i$ also decouples. The actions of boost generators are
    \begin{equation}
        \begin{aligned}
            &\left[B_k, A_v \right] = 0, \qquad \left[B_k, \pi_0 \right] = 0.
        \end{aligned}
    \end{equation}
    Since we can redefine a new scalar field $\chi =\pi_0 - \partial_0 A_v$ to absorb $A_v$, the Lagrangian includes purely a quadratic  term $\frac{1}{2}\chi^2$ without dynamics and thus the theory is somehow trivial.\par

    From the above examples, we see that we can always obtain new Carrollian invariant actions by starting from a Carrollian field theory, setting some field components vanishing, but ensuring the remaining field components to be in a sub-representation. These resulting theories cannot be read from Bargmann action directly. This paves a new way to find more Carrollian theories. It is more effective,  if the original representation is complicated, as there are more choices to get sub-representations.\par

\section{Discussions}\label{sec:Discussions}
    In this work, we tried to construct the Carrollian invariant field theories by restricting the parent Bargmann invariant theories to a null hyper-surface. Such null reduction guarantees the resulting theories to be Carrollian invariant. We mainly focused on the free massless scalar and $U(1)$ electromagnetic theories, and managed to reproduce the known electric sector and magnetic sector theories in the literature. The theories we constrcuted are  manifest off-shell Carrollian invariant.\par

    Another focus in this work is on the Carrollian conformal invariance. We found that for both the free scalar and $U(1)$ electromagnetic theories in $d=4$, their electric-sector and magnetic-sector theories are all Carrollian conformal invariant. We computed the 2-point correlators by using the path-integral formalism, and found the correlators to be consistent with the Ward identities. One remarkable point is that even in the simple theories studied in this work, the operators have constituted the generic representations of Carrollian conformal algebra\cite{Chen:2021xkw}. In the magnetic sector of free scalar, there appears the staggered structure, similar to the case in 2D BMS free scalar and free fermion\cite{Hao:2021urq,Yu:2022bcp,Hao:2022xhq}. More interestingly, in the magnetic sector of Carrollian electromagnetic theory, the gauge potentials form a chain representation, and the restricted field strength form a net representation of CCA. Another distinct feature in Carrollian electromagnetic theory is that both the gauge potential and the field strength are primary operators in $d=4$.\par

     The null-reduction method can be applied to other constructions. There are several directions worthy of pursuing:
    \begin{itemize}
        \item \textit{Non-conformal massive $\&$ interacting Carrollian theory.} Although the discussions in the present work cover only the free massless scalar and vector theories, there are no obstacles for the construction  to be generalized to other cases, including the massive scalar theories and the interacting theories, say Yang-Mills theory and (scalar) QED\cite{Chen:2023,Islam:2023rnc}. This is due to the fact that if we do not require the conformal symmetry, there are more options on the Bargmann actions, for example the massive theory or the theory with general interaction terms.\par

        \item \textit{Fermionic theory.} Recently, in \cite{Bagchi:2022eui} the Carrollian Clifford algebra were studied and the actions for the Carroll fermions were constructed. It would be interesting to reconsider the fermionic theory from the reduction of Bargmann fermion. Moreover, one may study the supersymmetric Carrollian field theory and QED.\par

        \item \textit{Higher-order derivative theory.} We expect that the null-reduction method can be applied to the construction of higher-derivative Carrollian field theory as well. In this case, the higher-order components in the expansion of the Bargmann field will become relevant in the construction.\par
        
        \item \textit{Carrollian gravity.} There already exist two approaches to Carrollian gravity. One is by gauging the Carrollian algebra \cite{Hartong:2015xda,Bergshoeff:2017btm,Figueroa-OFarrill:2022mcy}, and the other is by using the contraction of Lorentz theory \cite{Henneaux:2021yzg}. A comparison between the magnetic theory from Carrollian contraction and the construction from gauge procedure were made in \cite{Campoleoni:2022ebj}, while the electric sector is missing in the gauging description. In \cite{Hansen:2021fxi}, the authors claimed that the two sectors corresponds to the leading and next-to-leading order theory in the expansion of general relativity. The Bargmann reduction may provide another viewpoint of this issue, as it could in principle lead to both the electric and magnetic sectors at the same time.\par   
        
    \end{itemize}\par

\section*{Acknowledgments}
   We are grateful to Zhe-fei Yu, Pengxiang Hao, Hongjie Chen, Yijun He, Yunsong Wei for valuable discussions. The work is partially supported by NSFC Grant  No. 11735001, 12275004 .  \par
    
    \vspace{2cm}

\appendix
\renewcommand{\appendixname}{Appendix~\Alph{section}}

\section{Staggered modules in higher dimensional Carrollian CFT}\label{app:stagger}

The staggered modules (i.e. representations) appear in 2d Logarithmic CFTs, e.g. \cite{Rohsiepe:1996qj,Gaberdiel:2001tr,Kytola:2009ax,Creutzig:2013hma}, and 2d Carrollian CFTs\cite{Hao:2021urq,Yu:2022bcp,Hao:2022xhq}. In this appendix we briefly review the features of this type of modules and discuss its analog in higher dimensional Carrollian CFTs. We adopt a condensed notation without explicit indices on the generators of the conformal algebra or on the operators. The generators of the Carrollian conformal algebra contain dilatation $D$, generalized rotation $M$, translation $P$ and SCT $K$, with the following commutation relations
\begin{align}
&[D,P]=p_D P,\quad [D,K]=k_D K,\\
&[M,P]=p_M P,\quad [M,K]=k_M P,\quad
[P,K]=d D +m M
\end{align}
The actions on primary operators $\cO_i$ are $D\cO_i=\Delta_i \cO_i$ and $M\cO_i=\xi_i \cO_i$.

Mathematically, the staggered modules come from non-trivial module extensions. To understand the structure of a generic module of an algebra $\glie$, we can try breaking it into the simplest pieces - irreducible modules, $V_{N}\overset{?}{=} \bigoplus_{i=1}^{N} W_{i}$. But this direct sum decomposition cannot be achieved for non-semisimple Lie algebras, since it loses the track of the relations between different $W_{i}$-s. For this reason we need a method of sewing $W_{i}$-s back to $V_{N}$, and this leads to the problem of module extensions, see e.g. \cite{humphreys2008representations,weibel1995introduction}.

The standard way of decomposing $V_{N}$ is to choose a maximal submodule $V_{N-1} \subset V_{N}$ and to take the quotient $V_{N}/V_{N-1}=:W_{N}$, then by the maximality of $V_{N-1}$, $W_{N}$ is irreducible. Repeatedly we get a series of submodules, 
\begin{equation}
0=V_{0}\subset V_{1}\subset \dots V_{N-1}\subset V_{N},
\end{equation}
which is called the Jordan–Holder composition series of $V_{N}$. The irreducible modules $W_i=V_{i}/V_{i-1}$ are called the factors and $N$ is called the length of $V_N$. The composition series is not unique, but the length and the factors are invariants of $V_{N}$ itself. If relaxing the condition of maximality, the resulting composition series will be shorter than the Jordan–Holder one, and $W_i$ can be reducible modules.

On the other way, we can compose $W_{i}$-s into some bigger $V_{N}$. Let us start from $N=2$. More broadly, dropping the condition of irreducibility and considering two arbitrary modules $W_{1}$ and $W_{2}$, we find that the composed $V_2$ must satisfy the quotient condition $W_{2}=V_2/W_{1}$, or written in a short exact sequence,
\begin{equation}
\label{eq:shortes}
\begin{tikzcd}
0 \arrow[r] & W_{1}=:V_1 \arrow[r, "\iota"] & V_2 \arrow[r, "\pi"] & W_2 \arrow[r] & 0.
\end{tikzcd}
\end{equation}
In this notation the intertwinning map $\iota$ is injective and $\pi$ is surjective, and they characterize the ways of $W_{1}$ being embedded into $V_2$ and $W_{2}$ being projected from $V_2$. For the same pair $(W_1,W_2)$ there can be inequivalent $(\iota,\pi)$-s corresponding to not necessarily isomorphic $V_2$-s, and each triplet $(V_2,\iota,\pi)$ is called an extension of $W_{1}$ by $W_{2}$. For $N>2$, we can introduce new irreducible modules $W_{i}$ and repeat the preceding step recursively,
\begin{equation}
\begin{tikzcd}
0 \arrow[r] & V_{i-1} \arrow[r, "\iota_{i}"] & V_{i} \arrow[r, "\pi_{i}"] & W_{i} \arrow[r] & 0.
\end{tikzcd}
\end{equation}

In this decomposition-composition procedure, we find that  there can be other solutions of the constraint \eqref{eq:shortes} providing new modules of the algebra, besides the original module $V_{N}$. This phenomenon happens for the representations of non-semisimple Lie algebras and infinite dimensional representations of semi-simple or affine Lie algebras and Virasoro algebras. The former case appears in the Carrollian, Galilean and Schrodinger\footnote{%
The ``alien operators" introduced in \cite{Golkar:2014mwa} belong to a class of neutral operators in Schrodinger CFT, and could enter into the story of staggered modules.
} (conformal) field theories, and the latter one appears in relativistic CFTs. 

Now the remaining problem is to solve the module extensions of \eqref{eq:shortes}. It turns out that the equivalence classes of different extensions constitute a basis of the $\extension$ vector space $\extension_{\glie}(W_{2},W_{1})$, and the trivial extension $W_{1}\oplus W_{2}$ corresponds to the zero element, see e.g. chapter 3 \& 7 of \cite{weibel1995introduction}. 
The vector space $\extension$ is hard to compute, but in practice we only need to construct certain extensions according to the physical problem by constraining the undetermined coefficients in $(\iota,\pi)$.

Actually the module extension problem has already been encountered in the construction of finite dimensional modules of Carrollian rotation algebra \cite{Chen:2021xkw}. 
For example, the $4d$ electric vector $V=(0)\to(1)$ is an extension of $(1)$ by $(0)$. Here $(1)$ is the submodule $W_1$, and the morphisms $(\iota,\pi)$ are given by the actions of generators from $W_2=(0)$ to $\{(1),(0)\}$, which can be constrained by the commutation relations using the Wigner-Eckart theorem. In this simple example, the $\extension$ vector space can be computed as $\extension_{\glie}((0),(1))=\C$, hence the electric vector $(0)\to(1)$ is the only nontrivial extension. We leave the technical computation of $\extension_{\glie}((0),(1))=\C$ to the end of this appendix.\par

After introducing module extensions, we give a sketchy analysis of the staggered modules in the Logarithmic and Carrollian CFTs. For simplicity we focus on $N=2$ and assume that $W_1,\, W_2$ are singlet highest-weight modules, i.e. the corresponding primary operators $\cO_{1},\, \cO_{2}$ (indices omitted) are irreducible representations of the generalized rotation subalgebra\footnote{For reducible but indecomposable $\cO_{1},\, \cO_{2}$, this corresponds to the case $N>2$ and is more sophisticated than $N=2$ case.} $\{D,M\}$. 
By the surjectivity of $\pi$, we can choose the pre-images of $\cO_{2}$ in the extended module $V$, $\wave{\cO}_{2}\in V,\, \pi(\wave{\cO}_{2})=\cO_{2}$, and denote the conformal dimensions of $\cO_{1},\, \wave{\cO}_{2}$ as $\Delta_1,\, \Delta_2$ respectively\footnote{The choice of $\wave{\cO}_{2}$ admits gauge redundancies: for each $\cO\in W_1$ we have $\pi(\cO)=0$ and $\pi(\wave{\cO}_{2}+\cO)=\cO_2$.}.
To preserve the grading of the dilatation $D$, the difference $l=\Delta_2-\Delta_1\in\mathbb{Z}$ must be an integer. All the possible nontrivial extended modules $V$ can be cast into three types:

\begin{itemize}
    \item $l=0$. It can be shown in this case that for some generators $g\in \{D,M\}$ of the dilatation or the generalized rotation algebra, $\cO_{1}=g\wave{\cO}_{2}$. The resulting module corresponds to a rank-2 logarithmic multiplet in LogCFT, a boost multiplet in $2d$ Carrollian CFT, or a chain multiplet in higher dimensional Carrollian CFT. 
    \item $l<0$. In this case $\cO_{1}$ and $\wave{\cO}_{2}$ can be related by some lowering operators: $\cO_{1}=P^1\dots P^{|l|} \wave{\cO}_{2}$. Then $V$ itself is a reducible highest-weight module, and $\cO_{1}$ are singular in the sense that it is  simultaneously primary and descendent. 
    \item $l>0$. This type is called (strictly) staggered module and reveals new features comparing with the above  two types. There can exist singular vectors in $V$, and $\wave{\cO}_{2}$ can be neither primary nor descendent.
\end{itemize}

To obtain this result we need to consider how $W_2$ is ``staggered" with $W_1$. The action of conformal algebra on $W_2$ is determined by the operators $A=(D-\Delta_2)\wave{\cO}_{2},\, A'=(M-\xi_2)\wave{\cO}_{2},\, B=K\wave{\cO}_{2}, C=P\wave{\cO}_{2}$. The operators $C$ generate descendents and is irrelevant to the discussion. Noticing that the intertwinning map $\pi$ commutes with the action, we have $\pi(A)=(D-\Delta_2)\pi (\wave{\cO}_{2})=(D-\Delta_2)\cO_2=0$, hence by the exactness of \eqref{eq:shortes} $A\in \operatorname{im}(\pi)=\operatorname{ker}(\iota)=W_1$. Similarly $A',B\in W_1$. Then the extended module is illustrated by the following diagram:

\begin{equation}
\begin{tikzcd}[/tikz/column 1/.append style={anchor=base west}]
\Delta_2-1 &  & B  &  &                                                                   &  &   \\
           &  &    &  &                                                                   &  &   \\
\Delta_2   &  & A' &  & \mathcal{O}_2 \arrow[lluu, "K"'] \arrow[ll, "M"'] \arrow[rr, "D"] &  & A
\end{tikzcd}
\end{equation}

For $l=0$ there is no operator with conformal dimension less than $\Delta_2$, hence $B=0$ and $\wave{\cO}_{2}$ are primary operators in $V$. There are no other primary operators besides $\cO_{1}$, hence $\{A,A'\}$ are linear combinations of $\cO_{1}$. On the contrary, due to the irreducibility of $\cO_{1}$ under $\{D,M\}$, either $\cO_{1}\cap \{A,A'\}=0$ or $\cO_{1}\subset \{A,A'\}$. The former case implies $A=A'=0$, hence $V=W_1\oplus W_2$ is trivial. The latter case implies $\cO_{1}$ and $\{A,A'\}$ are linear combinations of each other, hence $\cO_{1}=g\wave{\cO}_{2}$ for some generator $g\in \{D,M\}$.
For $l<0$ the analysis is similar: if $V$ is non-trivial then $\cO_{1}$ are in the descendents of $\wave{\cO}_{2}$, hence are singular in $V$.

Now we consider the interesting case $l>0$ and assume the extension $V$ is non-trivial.
At the level $\Delta_2$ the operator number equals $\# (\cO_2)+ \# (P^{1}\dots P^{|l|}\cO_1)$, hence there must be operators not coming from the descendents of $\cO_1$. By the gauge redundancy $\wave{\cO}_{2}\to \wave{\cO}_{2}+\cO,\, \cO\in W_1$ we can choose the extra operators to be  $\wave{\cO}_{2}$. If $B\neq 0$ then $\wave{\cO}_{2}$ are neither primary nor descendent. There can be singular vectors from $\{A,A'\}$. 
Firstly, $KA=(D-\Delta_2+1)K\wave{\cO}_{2}=(D-\Delta_2+1)B=0$, hence $KA$ is singular in $V$ if $A\neq 0$.
Secondly, from $KA'=(M-\xi_2+k_M)B$, the matrix equation $(M-\xi_2+k_M)B=0$ can have non-vanishing solutions $B_0$, which further provides singular vectors $A'_0\subset \{A'\}$ with $KA'_0=(M-\xi+b)B_0=0$. For example, in $2d$ Carrollian CFT, supposing that the two singlet primaries are related by $\cO_1=L_1\wave{\cO}_2$, we have $M_1 A'=(M_0-\xi)M_1\wave{\cO}_2=0$ and $L_1 A'=(M_0-\xi)\cO_1+M_1 \wave{\cO}_2=0$, hence $A'=M_0\wave{\cO}_2$ is a singular vector if not vanishing.

Finally let us consider a special case $\cO_1=K\wave{\cO}_2$ and $K\cO_1=0$, i.e. $l=\Delta[K]$. This case is still broad enough to include the known staggered modules in the present work, in $2d$ Carrollian CFT and even in Schrodinger CFT. Writing $A,\, A'$ as linear combinations of $P\cO_1$, we have 
\begin{equation}
D\wave{\cO}_2=aP\cO_1,\quad M\wave{\cO}_2=bP\cO_1,
\end{equation}
subject to the matrix equation $a(p_M+\xi_1)P\cO_1=b(p_D+\Delta_1)P\cO_1$ from $[D,M]=0$.
To exploit the additional information $K\cO_1=0$, applying $K$ on both sides we get
\begin{equation}
(\Delta_1-k)\cO_1=a(d\Delta_1+m\xi_1)\cO_1,\quad (\xi_1-k)\cO_1=b(d\Delta_1+m\xi_1)\cO_1.
\end{equation}
These three matrix equations provide strong constraints on the undetermined parameter matrices $a,b$.

\subsection{Staggered (scalar, scalar) modules}

In this subsection we construct the staggered modules from two scalars $\phi,\pi$ with $l=1$. The Carrollian magnetic-sector scalar falls into this class.

The two highest-weight modules $W_1,\, W_2$ are generated by $\phi,\, \pi$ respectively, and we can make the most general ansatz respecting the conformal dimension and spin of $SO(3)$ without Levi-Civita tensor,
\begin{equation}
    \begin{aligned}
        &[D,\phi]= \Delta_\phi \phi, \qquad [J^i_{~j},\phi]=0, \qquad [B_i,\phi]=0, \qquad [K_\mu,\phi]=0,\\
        &[D,\pi]= \Delta_\pi \pi + c_0 \partial_0 \phi, \qquad [J^i_{~j},\pi]=0, \qquad [B_i,\pi]=c_1 \partial_i\phi, \\
        &[K_0,\pi]=c_2 \phi, \qquad [K_i,\pi]=0.
    \end{aligned}
\end{equation}
As a consistency check we unfreeze the condition $\Delta_\pi-\Delta_\phi=l=1$ and treat $\Delta_\pi$ as a free parameter. From the definition of the representation, the nontrivial constraints come from the commutators $[B_i,K_i],\, [D,B_i]$. The relation $[[B_i,K_i],\pi]=[B_i,[K_i,\pi]]-[K_i,[B_i,\pi]]$ gives $(c_2+2c_1\Delta_\phi)\phi=0$, and from $[D,B_i]$ we have $c_1(1+\Delta_\phi-\Delta_\pi)\partial_i \phi=0$. Solving them with respect to $\Delta_\phi$, we find the solutions can be classified as,
\begin{align}
&\text{trivial:} &&c_0=c_1=c_2=0,\quad \Delta_\pi \text{ is arbitrary}.\\
&\text{$B$-staggered:} &&c_0=0,\, c_2+2c_1\Delta_\phi=0,\,  \Delta_\pi=\Delta_\phi+1\\
&\text{$D$-staggered:} &&c_0\neq 0, \, c_1=c_2=0,\quad \Delta_\pi \text{ is arbitrary}.\\
&\text{$\{B,D\}$-staggered:} &&c_0\neq 0,\, c_2+2c_1\Delta_\phi=0,\,  \Delta_\pi=\Delta_\phi+1
\end{align}
For the first solution the modules generated by $\phi,\, \pi$ are decoupled and the extension is trivial. For the second solution, the independent parameter $c_1$ can be absorbed by the field renormalization $\phi\to a_1\phi,\, \pi \to a_2\pi$, hence the magnetic scalar is the only non-trivial extension. For the third and fourth solutions with $c_0\neq 0$, the free parameter $c_0$ indicates the dilatation $D$ contains Jordan blocks, which is similar to the staggered modules in Logarithmic CFTs. Besides, $[D-\Delta_\pi,\pi]=  c_0 \partial_0 \phi$ is a singular vector, consistent with the previous analysis that $A=(D-\Delta_2)\wave{\cO}_{2}$ is singular.

The structure of the extended module $V$ is illustrated as below. Notice that due to $[K_0,P_0]=0$, the operator $\partial_0 \phi$ is singular in $W_1$ and $V$, and $W_{1,2}=W_1/W_{1,1}$ is an additional quotient module.
\begin{equation}\label{eq:LargeSaggeredModule}
\begin{tikzcd}[/tikz/column 1/.append style={anchor=base west}]
\Delta_\phi     &  &                                               & \phi \arrow[ld, "P_0"', shift right] \arrow[d, "P_i", shift left]                           &  &                                                                                                                       \\
\Delta_\phi+1   &  & \partial_0 \phi \arrow[dd, "P^n_\mu", dashed] & \partial_i \phi \arrow[l, "B_i"] \arrow[u, "K_i", shift left] \arrow[dd, "P^n_\mu", dashed] &  & \pi \arrow[llu, "{K_0,c_2}"'] \arrow[ll, "{B_i,c_1}"] \arrow[dd, "P^n_\mu", dashed] \arrow[lll, "{D,c_0}"', bend left] \\
\vdots          &  &                                               &                                                                                             &  &                                                                                                                       \\
\Delta_\phi+n+1 &  & \partial^n \partial_0\phi \arrow[d, dashed]   & \partial^n \partial_i\phi \arrow[l, dashed] \arrow[d, dashed]                               &  & \partial^n\pi \arrow[ll, dashed] \arrow[d, dashed]                                                                    \\
\vdots          &  & \vdots                                        & \vdots                                                                                      &  & \vdots                                                                                                                \\
                &  & {W_{1,1}}                                     & {W_{1,2}}                                                                                   &  & W_2                                                                                                                  
\end{tikzcd}
\end{equation}

Finally if allowing the spatial Levi-Civita tensor, the ansatz should be modified by
\begin{equation}
    [J^i_{~j},\pi]=e_1 \epsilon^{i}_{j}\partial_0 \phi+d_1 \epsilon^{ik}_{j}\partial_k \phi, \qquad [B_i,\pi]=c_1 \partial_i\phi+ e_2 \epsilon^{ij}\partial_j\phi, \\
\end{equation}
where the $d$-term and $e$-terms only appear in $d=3,4$ respectively. Similar computation shows that for $d=4$ the $d$-term must vanish, while for $d=3$, only $e_2$ is forced to vanish, and $e_1$ is a free parameter. The extra staggered module in $d=3$ is
\begin{equation}
    \begin{aligned}
        &[D,\phi]= \Delta_\phi \phi, \qquad [J^i_{~j},\phi]=0, \qquad [B_i,\phi]=0, \qquad [K_\mu,\phi]=0,\\
        &[D,\pi]= (\Delta_\phi+1) \pi + c_0 \partial_0 \phi, \qquad [J^i_{~j},\pi]=e_1\epsilon_{j}^{i}\partial_0 \phi, \qquad [B_i,\pi]=c_1 \partial_i\phi, \\
        &[K_0,\pi]=-2c_1\Delta_\phi \phi, \qquad [K_i,\pi]=0.
    \end{aligned}
\end{equation}

\subsection{Computation of \texorpdfstring{$\extension_{\glie}((0),(j))$}~}

The traditional way of computing the vector space $\extension(W_2,W_1)$ is through the projective or injective resolution of modules. But unlike the case of semisimple Lie algebras, we find that the category of finite dimensional modules of $\isolie(3)$ does not have enough projective/injective modules, and infinite dimensional modules must enter into the game.

Instead, we can utilize the relation between $\extension$ and Lie algebra cohomology, $\extension_{\glie}(\C,M) \simeq H^{1}(\glie,M)$, see e.g. \cite{weibel1995introduction}. For $W_2=(0)=\C$ and $W_1=(1)=M$, the problem gets reduced to the computation of the first cohomology $H^{1}(\glie,M)$ of $\isolie(3)$. The cohomology of semi-direct product of Lie algebras can be calculated via the Hochschild-Serre spectral sequence \cite{hochschild1953cohomology1,hochschild1953cohomology}. We need only theorem 13 in \cite{hochschild1953cohomology}: supposing Lie algebra $\glie$ and its ideal subalgebra $\llie$ such that $\glie/\llie$ is semisimple, for the module $M$ of $\glie$, the cohomology is
\begin{equation}
H^{n}(\glie,M)\simeq \bigoplus_{i+j=n} H^{i}(\glie/\llie,\C)\otimes H^{j}(\llie,M)^{\glie},\label{cohomology}
\end{equation}
where $M^{\glie}\equiv H^0(\glie,M)$ means the invariant vectors in $M$: $M^{\glie}:=\{x\in M: g x=0, \forall g\in \glie\}$. 

In our case, $\glie=\isolie(3),\, \llie=\C^3,\, \glie/\llie=\solie(3)$ and $M$ is the singlet module $(1)$ of $\isolie(3)$. By the Wighthead lemma the first cohomology of semisimple Lie algebra vanishes $H^{1}(\solie(3),\C)=0$, hence there is only one term at the right-hand side of \eqref{cohomology}
\begin{equation}
H^{1}(\isolie(3),(1))=  H^{0}(\solie(3),\C)\otimes H^{1}(\C^3,(1))^{\glie}.
\end{equation}
The $H^1$ is just $\operatorname{Hom}$, hence $H^{1}(\C^3,(1))=\operatorname{Mat}_3(\C)$. Then $\operatorname{Mat}_3(\C)^{\glie}$ contains a constant diagonal matrix $c\cdot I$, which is invariant under the rotations. Finally with $H^{0}(\solie(3),\C)=\C$ we get $H^{1}(\isolie(3),(1))=\C$. Hence the electric vector is the only nontrivial extension of $(1)$ by $(0)$. With the same method we can show $\extension_{\glie}((0),(j))=0$ for $j\neq 1$, and this agrees with the result of the chain representations.

\section{Path-integral formalism for Carrollian theories}\label{app:PathInt}    

    In this section we present the path-integral formalism for free Carrollian field theories, including the scalar and electromagnetic field theories. In each theory, we discuss the electric sector and magnetic  sector separately. \par

\subsection{Electric sector of scalar}\label{appsubsec:ScalarElectricSector}
    The electric sector of Carrollian  scalar theory has the action
    \begin{equation}\label{eq:elec-action}
        \begin{aligned}
            S^\mathscr{C}_E = -\frac{1}{2}\int d^d x ~ \partial_t\phi\partial_t\phi
            =-\frac{1}{2}\int d^d x ~ \phi(-\partial_t^2)\phi
        \end{aligned}
    \end{equation}
up to a boundary term. Its corresponding generating functional is 
    \begin{equation}\label{eq:elec-functional}
        \begin{aligned}
            \mathcal{Z}^\mathscr{C}_E [j_\phi] &= N \int \mathcal{D}\phi\mathcal{D}\pi \exp \left(i (S^\mathscr{C}_E + \int d^d x j_\phi \phi)\right)\\
            &= N'  \exp \left(i \int d^d x ~\frac{1}{2} j_\phi(x) \Pi_E(x-y) j_\phi(y)\right)
        \end{aligned}
    \end{equation}
where $\Pi_E(x-y)=\int \frac{d^d p}{(2\pi)^d} e^{i(\vec p \cdot \vec x +\omega t)} \cdot \frac{1}{\omega^2}=-\frac{\abs{t}}{2} \delta^{(d-1)}(\vec x)$. Thus the 2-pt correlator is simply
    \begin{equation}\label{eq:elec-correlator}
        \begin{aligned}
            \left<\phi(x)\phi(y)\right> &=\left. (-i)^2\frac{1}{Z[0]} \frac{\delta^2}{\delta j_\phi(x)\delta j_\phi(y)}\mathcal{Z}^\mathscr{C}_E [J]\right|_{j_\phi=0}\\
            &=-i \Pi_E(x-y)
            =\frac{i\abs{t_x-t_y}}{2} \delta^{(d-1)}(\vec x-\vec y)
        \end{aligned}
    \end{equation}
    
\subsection{Magnetic sector of scalar}\label{appsubsec:ScalarMagneticSector}

    In order to perform the path-integral for the magnetic sector action, we need to rewrite the action in a standard balanced quadratic form:
    \begin{equation}\label{eq:mag-action}
        \begin{aligned}
         S^\mathscr{C}_M =-\frac{1}{2} \int d^d x ~ 2\pi\partial_t\phi + \partial_i\phi\partial_i\phi
         =-\frac{1}{2}\int d^d x \Phi^\dag \hat{A} \Phi
        \end{aligned}
    \end{equation}
where $ \Phi=(\phi,\pi)$ and the operator $\hat{A} = \begin{pmatrix} -\vec \partial^2 & -\partial_t \\ \partial_t & 0\end{pmatrix}$.\par

By adding a source term and then performing the Gaussian integral, we easily get the related generating functional $\mathcal{Z}[J]$ in the following form
    \begin{equation}
        \begin{aligned}\label{eq:mag-functional}
         \mathcal{Z}^\mathscr{C}_M [J] &= N \int \mathcal{D}\phi\mathcal{D}\pi \exp \left(i (S^\mathscr{C}_M + \int d^d x J \Phi)\right)\\
         &=N \int \mathcal{D}\phi\mathcal{D}\pi \exp \left(i \int d^d x \left(-\frac{1}{2}\Phi^\dag \hat{A} \Phi  +J\Phi\right)\right)\\
         &=N^\prime  \exp \left(i \int d^d x \frac{1}{2} J^\dag \hat{A}^{-1} J \right)\\
         &=N^\prime  \exp \left(i \int d^d x \int d^d y \frac{1}{2} J^\dag(x) \Pi_M (x-y) J(y) \right)\\
        \end{aligned}
    \end{equation}
where $J=(j_\phi,j_\pi)$, and $\Pi_M (x-y)$ is the Green function of $\hat{A}$, satisfying 
$\hat{A} \Pi(x-y)=\delta^{(d-1)}(\vec x)$. The Green function can be calculated by inverting $\hat{A}$ in the momentum space,
    \begin{equation}\label{eq:mag-green}
        \begin{aligned}
            \Pi_M(x-y)&=\int \frac{d^d p}{(2\pi)^d} e^{i(\vec p \cdot \vec x +\omega t)} \cdot
            \begin{pmatrix}
                {\vec p}^2 & -i\omega \\ i\omega &0
            \end{pmatrix}^{-1}\\
            &=\int \frac{d^d p}{(2\pi)^d} e^{i(\vec p \cdot \vec x +\omega t)} \cdot 
            \begin{pmatrix}
                 0& \frac{-i}{\omega} \\ \frac{i}{\omega} &\frac{-{\vec p}^2}{\omega^2}
            \end{pmatrix}.\\
        \end{aligned}
    \end{equation}
Then the 2-pt correlator can be read from the partition function,
    \begin{equation}\label{eq:mag-corr}
        \begin{aligned}
            \left<\Phi_i(x)\Phi_j(y)\right> &=\left. (-i)^2\frac{1}{Z[0]} \frac{\delta^2}{\delta J_i(x)\delta J_j(y)}\mathcal{Z}^\mathscr{C}_M [J]\right| _{J=0}\\
            &=-\frac{i}{2} \left(\Pi_{ij}(x-y)+\Pi_{ji}(y-x)\right)\\
            &=-\frac{i}{2}\int \frac{d^d p}{(2\pi)^d} e^{i(\vec p \cdot \vec x +\omega t)} \cdot \left(\Pi_{ij}(p)+\Pi_{ji}(-p)\right)\\
            &=\int \frac{d^d p}{(2\pi)^d} e^{i(\vec p \cdot \vec x +\omega t)} \cdot 
            \begin{pmatrix}
                 0& -\frac{1}{\omega} \\ \frac{1}{\omega} &i\frac{{\vec p}^2}{\omega^2}
            \end{pmatrix}\\
            &=\frac{i}{2}\begin{pmatrix}
                 0& -\mbox{Sign}(t_x-t_y)\delta^{(d-1)}(\vec x - \vec y) \\ \mbox{Sign}(t_x-t_y)\delta^{(d-1)}(\vec x-\vec y) &\abs{t_x-t_y}{\vec \partial}^2\delta^{(d-1)}(\vec x-\vec y)
            \end{pmatrix}.
        \end{aligned}
    \end{equation}
    This gives us the correlators in \eqref{eq:CorrelatorsOfCarrollianScalarMagneticSector}.
    \par

\subsection{Electric sector of \texorpdfstring{$U(1)$}~ theory}\label{appsubsec:U1ElectricSector}

    For the electric sector of $U(1)$, it is very similar to the magnetic sector of scalar. The action is
    \begin{equation}\label{app:u1-elec-action}
        \begin{aligned}
             S^\mathscr{C}_E 
                = -\frac{1}{2}\int d^d x ~ F_{0i}F_{0i} =  -\frac{1}{2}\int d^d x ~ \Phi^\dag \hat{B} \Phi
        \end{aligned}
    \end{equation}with $\Phi = (A_i, A_0)$ and 
    \begin{equation}
        \hat{B} = \begin{pmatrix}
        -\partial_0^2 \delta_{ij} &\partial_0 \partial_i\\
        ~\partial_0 \partial_j & -{\vec \partial}^2\\
        \end{pmatrix}.
    \end{equation}\par
    
    Then with the temporal gauge-fixing term $\mathcal{L}_{gf} = -\frac{1}{2\xi}(\partial_0 A_0)^2$, we may read the  correlators. In $d=4$, the correlators are
    \begin{equation}\label{app:u1-elec-corr-full}
        \begin{aligned}
            &\left<A_i(x)A_j(0)\right> = \left<A_j(x)A_i(0)\right>= \frac{i}{2}\delta_{ij}\abs{t}\delta^{(3)}(\vec x)+\frac{i\xi}{12} t^3 \mbox{Sign}(t) \partial_i\partial_j\delta^{(3)}(\vec x),\\
            &\left<A_i(x)A_0(0)\right> = \left<A_0(x)A_i(0)\right>= \frac{i\xi}{4}t^2 \mbox{Sign}(t)\partial_i\delta^{(3)}(\vec x),
            \quad \left<A_0(x)A_0(0)\right> = \frac{i\xi}{2}\abs{t}\delta^{(3)}(\vec x).\\
        \end{aligned}
    \end{equation}\par
    
    Moreover, in the Landau gauge $\xi=0$, the correlators containing $A_0$ vanish and the remaining ones are
    \begin{equation}
        \left<A_i(x)A_j(0)\right> = \frac{i}{2}\delta_{ij}\abs{t}\delta^{(3)}(\vec x). 
    \end{equation}\par

\subsection{Magnetic sector of \texorpdfstring{$U(1)$}~ theory}\label{appsubsec:U1MagneticSector}

    For the $U(1)$ magnetic sector, we can rewrite the action as 
    \begin{equation}\label{eq:u1-mag-action}
        \begin{aligned}
             S^\mathscr{C}_M 
                & = -\frac{1}{4}\int d^d x ~ F_{ij}F_{ij} + 4F_{0i}F_{vi} - 2F_{0v}^2\\
                & = -\frac{1}{4}\int d^d x ~ (\partial_i A_j - \partial_j A_i)^2 + 4(\partial_0 A_i - \partial_i A_0)(\pi_i - \partial_i A_v) -2 (\pi_0 - \partial_0 A_v)^2\\
                &=  -\frac{1}{2}\int d^d x ~ \Phi^\dag \hat{B} \Phi
        \end{aligned}
    \end{equation}
with $ \Phi=(A_v,A_i,A_0,\pi_i,\pi_0)$ and 
    \begin{equation}\label{eq:u1-mag-matrix}
        \hat{B} = \begin{pmatrix}
        \partial_0^2 & \partial_0\partial_j & -{\vec \partial}^2 & 0  & -\partial_0\\
        \partial_0\partial_i & -{\vec \partial}^2 \delta_{ij}+\partial_i\partial_j  & 0 & -\partial_0\delta_{ij} & 0\\
        -{\vec \partial}^2 & 0 & 0 & \partial_j & 0 \\
        0 & \partial_0\delta_{ij} & -\partial_i & 0 & 0\\
        \partial_0 & 0 & 0 & 0  & -1\\
        \end{pmatrix}.
    \end{equation}
    Since there are abundant gauge symmetries, this matrix is not invertible in the momentum space. In order to properly inverse it and perform the path-integral, we ought to include the gauge-fixing terms.
    
    The gauge fixing of magnetic $U(1)$ sector is tricky, especially if we want to keep the Carrollian conformal invariance. We actually need  two gauge-fixing terms for the first-order and second-order gauge transformations in (4.14).  Just as we prefer to choose Lorentz invariant gauge in the relativistic gauge theory, the ideal gauge-fixing term $\mathcal{L}_{gf} d^d x$ should be a Carrollian conformal invariant one in $d=4$. The choice of the first-order gauge fixing is simple by setting the temporal derivative of the time component of $A_\mu$ to zero,  $\partial_0 A_0 =0$, with the help of a $R_\xi$-type auxiliary field, leading to the gauge-fixing term $\mathcal{L}_1 = -\frac{1}{2 \xi_1} (\partial_0 A_0)^2$. However, it turns out to be impossible to find a usual quadratic gauge-fixing term for the second-order gauge transformation, which retains the invariances under both the boost and special conformal transformations  simultaneously.  Nevertheless, it is possible to find a Carrollian conformal invariant gauge-fixing term, if we allow for more general  terms. By direct calculation, we find that the gauge-fixing term \eqref{eq:covariant-gauge-fixing} turns up to be Carrollian conformal invariant up to total derivatives.\par
    
    Unlike 
    the usual gauge-fixing term of a quadratic form, which can be implemented into the path-integral by the standard Faddeev-Popov procedure, the gauge-fixing term \eqref{eq:covariant-gauge-fixing} here contains a part which is obviously not quadratic, and one may question whether such an exotic term will function properly as a gauge fixing term for the second-order gauge transformation. To show that \eqref{eq:covariant-gauge-fixing} can play the role of gauge fixing, we need to use a generalized ``Stueckelberg trick" as in Chapter 14.5 of \cite{Schwartz:2014sze}.  Shortly speaking, this trick works as follows. One may formally multiply and then divide the path-integral of the generating function by a suitable infinite function $f(\xi)$ which is an integration of auxiliary field, then in the multiplication one exchanges the orders of doing the integrations, and shift the auxiliary field to get the  gauge-fixing term $R_\xi$. To apply this trick to the case at hand, we need to select  $f(\xi)$ appropriately to make it consistent with our non-quadratic gauge-fixing term.
    We can consider the following function of $\xi_1,\xi_2$, which is an integration over two auxiliary fields $\alpha(x)$ and $\beta(x)$,
    \begin{equation}
        \begin{aligned}
        f(\xi_1,\xi_2) &= \int \mathcal{D}\alpha(x) \mathcal{D}\beta(x) \exp{-i\int d^d x \left(\frac{1}{2\xi_1}(\partial_0^2 \alpha)^2+\frac{1}{2\xi_2}\left(3(\partial_0\partial_i\alpha)^2+2\partial_0^2\alpha\partial_0\beta\right)\right)}\\
        &=\int \mathcal{D}\alpha(x) \mathcal{D}\beta(x) \exp{-i\int d^d x \left(\frac{1}{2\xi_1}(\partial_0^2 \alpha+\frac{\xi_1}{\xi_2}\partial_0\beta)^2+\frac{3}{2\xi_2}(\partial_0\partial_i\alpha)^2-\frac{\xi_1}{2\xi_2^2}(\partial_0\beta)^2\right)}.
        \end{aligned}\nonumber
    \end{equation}
    This integral is actually divergent, but is in a form of Gaussian integral if we treat $\partial_0\alpha(x)$, $\partial_i\alpha(x)$, and $\partial_0 \beta$ as independent integration variables. Then, we shift the auxiliary fields one by one 
    \begin{equation}
        \alpha(x) \to \alpha(x) - \frac{1}{\partial_0}A_0(x), \quad 
        \beta(x) \to \beta(x) - \frac{1}{\partial_0}\left((2\pi_0(x)-\partial_0 A_v(x) - \partial_i A_i(x)) 
        + \vec{\partial}^2 \alpha(x) \right),\nonumber
    \end{equation}
    where we have used the notation that $h(x)=\frac{1}{\partial_0} g(x)$ is the solution of $\partial_0 h(x) = g(x)$. 
    As the shift does not change the integral, we have 
    \begin{equation}\label{app:factor-mag-int}
        \begin{aligned}
         f(\xi_1,\xi_2) &=  \int \mathcal{D}\alpha(x) \mathcal{D}\beta(x) \exp \left\{-i\int d^d x \left(\frac{1}{2\xi_1}(\partial_0^2 \alpha - \partial_0 A_0)^2 \right. \right.\\
         & \left. \left.+\frac{1}{2\xi_2}\left(3(\partial_0\partial_i\alpha - \partial_i A_0)^2+2(\partial_0^2\alpha-\partial_0 A_0)(\partial_0\beta - (2\pi_0-\partial_0 A_v - \partial_i A_i) - \vec{\partial}^2 \alpha) )\right)\right) \right\}.\\
        \end{aligned}
    \end{equation}
    We can multiply and divide \eqref{app:factor-mag-int} when doing the path-integral, and use the ``Stueckelberg trick" to perform the gauge transformation shift, with $\alpha(x),\beta(x)$ being the gauge parameters, 
    \begin{equation}\label{app:u1-gauge-sym}
        \begin{aligned}
            &A_i(x) \to A_i(x) + \partial_i \alpha(x), \quad A_0(x) \to A_0(x) + \partial_0 \alpha(x),\\
            &A_v(x) \to A_v(x) + \beta(x), \quad \pi_i(x) \to \pi_i(x) + \partial_i \beta(x), \quad \pi_0(x) \to \pi_0(x) + \partial_0 \beta(x).\\
        \end{aligned}
    \end{equation}
    Since the measures $\mathcal{D}\alpha(x)\mathcal{D}\beta(x)\mathcal{D}\Phi(x)$, the action $S^\mathscr{C}_M[\Phi(x)]$, and gauge-invariant operators $\mathcal{O}_i$ are all invariant under the transformations, we have
    \begin{equation}
        \begin{aligned}
        \left<\mathcal{O}_1(x_1)\cdots\mathcal{O}_n(x_n)\right>&= \frac{1}{Z[0]}\left(\frac{1}{f(\xi_1,\xi_2)}\int\mathcal{D}\alpha \mathcal{D}\beta\right)\cross\int\mathcal{D}\Phi \mathcal{O}_1(x_1)\cdots\mathcal{O}_n(x_n)\\
        &\exp{i \left(S^\mathscr{C}_M -\frac{1}{2 \xi_1}\left(\partial_0 A_0\right)^2-\frac{1}{2 \xi_2}\left(3 \partial_i A_0 \partial_i A_0 + 2 \partial_0 A_0 \left(2\pi_0 - \partial_0 A_v - \partial_i A_i\right)\right)\right)}.
        \end{aligned}
    \end{equation}
    Therefore, we see that  the term \eqref{eq:covariant-gauge-fixing} appears in the action naturally. In other words, \eqref{eq:covariant-gauge-fixing} can be taken as the gauge-fixing term. With this term, it is not difficult to perform the standard inversion in the momentum space, and then use Fourier transform to calculate the correlators. The 2-point correlator of the field can be written as
    \begin{equation}
        \left<\Phi_i(x)\Phi_j(y)\right> =-\frac{i}{2} \left(\Pi_{ij}(x-y)+\Pi_{ji}(y-x)\right),
    \end{equation} 
    where $\Pi_{ij}(x-y)$ is the position space Green function, as in \eqref{eq:mag-green}.
It needs to be pointed out that even if we drop the $\xi_1$ term and only keep the $\xi_2$ term in $\mathcal{L}_{gf}$, the inversion is also possible and well-defined. 

In the end,  we find the 2-point correlators in the magnetc sector of $U(1)$ theory
    \begin{equation}\label{app:u1-mag-corr-full}
        \begin{aligned}
            &\left<A_v(x)A_v(0)\right> =-2i\left(1+\frac{\xi_2^2}{4\xi_1}\right) \abs{t} \delta^{(3)}(\vec x)-\frac{i\xi_2}{12}t^3 \mbox{Sign}(t)\vec{\partial}^2 \delta^{(3)}(\vec x),\\
            &\left<A_v(x)A_i(0)\right> = \left<A_i(x)A_v(0)\right> = \frac{i\xi_2}{4}t^2 \mbox{Sign}(t)\partial_i\delta^{(3)}(\vec x),\\
            &\left<A_v(x)A_0(0)\right> = \left<A_0(x)A_v(0)\right> = \frac{i\xi_2}{2}\abs{t}\delta^{(3)}(\vec x),\\
            &\left<A_v(x)\pi_i (0)\right> = -\left<\pi_i (x)A_v(0)\right>=\frac{3i}{2} \left(1+\frac{\xi_2^2}{3\xi_1}\right)\abs{t} \partial_i \delta^{(3)}(\vec x) + \frac{i\xi_2}{12}t^3\mbox{Sign}(t)\partial_i\vec{\partial}^2 \delta^{(3)}(\vec x),\\
            &\left<A_v(x)\pi_0 (0)\right> = -\left<\pi_0 (x)A_v(0)\right>=i\left(1+\frac{\xi_2^2}{2\xi_1}\right) \mbox{Sign}(t) \delta^{(3)}(\vec x)+\frac{i\xi_2}{4} t^2 \mbox{Sign}(t)\vec{\partial}^2 \delta^{(3)}(\vec x),\\
            &\left<A_i(x)\pi_j(0)\right> = -\left<\pi_j(x)A_i(0)\right> = -\frac{i}{2} \delta_{ij} \mbox{Sign}(t) \delta^{(3)}(\vec x)- \frac{i\xi_2}{4} t^2\mbox{Sign}(t)\partial_i\partial_j\delta^{(3)}(\vec x),\\
            &\left<A_i(x)\pi_0(0)\right> = -\left<\pi_0(x)A_i(0)\right> =\left<A_0(x)\pi_i(0)\right> = -\left<\pi_i(x)A_0(0)\right> = -\frac{i\xi_2}{2} \abs{t} \partial_i \delta^{(3)}(\vec x),\\
            &\left<A_0(x)\pi_0(0)\right> = -\left<\pi_0(x)A_0(0)\right> = -\frac{i\xi_2}{2} \mbox{Sign}(t) \delta^{(3)}(\vec x),\\
            &\left<\pi_i(x)\pi_j(0)\right> = \left<\pi_j(x)\pi_i(0)\right> = \frac{i}{2}\left(1+\frac{\xi_2^2}{\xi_1}\right)\abs{t}  \left( \partial_i\partial_j\delta^{(3)}(\vec x)+ \delta_{ij} \vec{\partial}^2 \delta^{(3)}(\vec x)\right) +\frac{i\xi_2}{12} t^3 \mbox{Sign}(t)\partial_i\partial_j\vec{\partial}^2 \delta^{(3)}(\vec x),\\
            &\left<\pi_i(x)\pi_0(0)\right> = \left<\pi_0(x)\pi_i(0)\right> = \frac{i}{2} \left(1+\frac{\xi_2^2}{\xi_1}\right)\mbox{Sign}(t) \partial_i \delta^{(3)}(\vec x) +\frac{i\xi_2}{4} t^2 \mbox{Sign}(t) \partial_i\vec{\partial}^2 \delta^{(3)}(\vec x),\\
            &\left<\pi_0(x)\pi_0 (0)\right> = i \left(1+\frac{\xi_2^2}{\xi_1}\right) \delta(t) \delta^{(3)}(\vec x) +\frac{i\xi_2}{2} t^2 \abs{t} \vec{\partial}^2 \delta^{(3)}(\vec x).
        \end{aligned}
    \end{equation}
    It can be checked directly  that for every choice of $\xi_1,\xi_2$, the Ward Identities are all satisfied, which reveals the Carrollian conformal invariance of the theory. Though these expressions look complicated, we can select the Landau-type gauge $\xi_2=0$ to simply them and obtain the nonvanishing correlators  listed in \eqref{eq:u1-mag-corr}. Here we list them again for completeness.
    \begin{equation}\label{app:u1-mag-corr}
        \begin{aligned}
            &\left<A_v(x)A_v(0)\right> =-2i \abs{t} \delta^{(3)}(\vec x), \quad \left<A_v(x)\pi_i (0)\right> = -\left<\pi_i (x)A_v(0)\right>=\frac{3i}{2} \abs{t} \partial_i \delta^{(3)}(\vec x),\\
            &\left<A_v(x)\pi_0 (0)\right> = -\left<\pi_0 (x)A_v(0)\right>=i \mbox{Sign}(t) \delta^{(3)}(\vec x),\\
            &\left<A_i(x)\pi_j(0)\right> = -\left<\pi_j(x)A_i(0)\right> = -\frac{i}{2} \delta_{ij} \mbox{Sign}(t) \delta^{(3)}(\vec x),\\
            &\left<\pi_i(x)\pi_j(0)\right> = \left<\pi_j(x)\pi_i(0)\right> = \frac{i}{2}\abs{t}  \left( \partial_i\partial_j\delta^{(3)}(\vec x)+ \delta_{ij} \vec{\partial}^2 \delta^{(3)}(\vec x)\right),\\
            &\left<\pi_i(x)\pi_0(0)\right> = \left<\pi_0(x)\pi_i(0)\right> = \frac{i}{2}\mbox{Sign}(t)  \partial_i \delta^{(3)}(\vec x), \quad \left<\pi_0(x)\pi_0 (0)\right> = i \delta(t) \delta^{(3)}(\vec x).
        \end{aligned}
    \end{equation}

\section{Ward identities and 2-point correlation functions}\label{app:2pt-ward}
    
    In this Appendix, we review the constraints on the 2-point correlation functions of the primary operators from the Ward identities of Carrollian conformal symmetries. There could be four classes of the correlators with different structures, which will be labeled by Case 1.1, Case 1.2, Case 2.1, and Case 2.2. It turns out that the correlators discussed in the main text belong to Case 2.1.  \par
    
    Similar to the case in CFT, the structure of 2-point correlation functions in CCFT is very much constrained by the Ward identities of the symmetries. For the Carrollian conformal symmetries, the corresponding Ward identities are listed in \eqref{app:ward-id}, 
    \begin{equation}\label{app:ward-id}
        \begin{aligned}
            P_\mu: & \quad (\partial^\mu_{1}+\partial^\mu_{2})\left<\mathcal{O}_1\mathcal{O}_2\right> =0,  \\
            D: & \quad x^\mu\partial^\mu \left<\mathcal{O}_1\mathcal{O}_2\right> + \Delta_1\left<\mathcal{O}_1\mathcal{O}_2\right> + \Delta_2\left<\mathcal{O}_1\mathcal{O}_2\right> =0,\\
            J_{ij}: & \quad (x^i\partial^j-x^j\partial^i)\left<\mathcal{O}_1\mathcal{O}_2\right> + \left<(J^{ij}\mathcal{O}_1)\mathcal{O}_2\right> + \left<\mathcal{O}_1 (J^{ij}\mathcal{O}_2)\right> = 0,\\
            B_i: & \quad x^i\partial_t \left<\mathcal{O}_1\mathcal{O}_2\right> + \left<[B_i,\mathcal{O}_1]\mathcal{O}_2\right> + \left<\mathcal{O}_1[B_i,\mathcal{O}_2]\right> = 0,\\
            K_0: & \quad \left( \left<[K_0,\mathcal{O}_1]\mathcal{O}_2\right> + \left<\mathcal{O}_1[K_0,\mathcal{O}_2]\right>\right) -x^i \left( \left<[B_i,\mathcal{O}_1]\mathcal{O}_2\right> - \left<\mathcal{O}_1[B_i,\mathcal{O}_2]\right>\right) = 0,\\
            K_i: & \quad \left( \left<[K_i,\mathcal{O}_1]\mathcal{O}_2\right> + \left<\mathcal{O}_1[K_i,\mathcal{O}_2]\right>\right) + x^i (\Delta_1-\Delta_2) \left<\mathcal{O}_1\mathcal{O}_2\right> \\
            &\qquad\quad  + x^j\left(\left<[J^i_{~j},\mathcal{O}_1]\mathcal{O}_2\right> - \left<\mathcal{O}_1[J^i_{~j},\mathcal{O}_2]\right> \right) + t \left( \left<[B_i,\mathcal{O}_1]\mathcal{O}_2\right> - \left<\mathcal{O}_1[B_i,\mathcal{O}_2]\right>\right) = 0.\\
        \end{aligned}
    \end{equation}
    It should be mentioned that we have used the techniques explained in the appendix of \cite{Chen:2021xkw} to simplify the expression for Carrollian special conformal transformation generators $K_0, K_i$.  These identities hold for all of the correlators appearing in this article. The one from the translational generator $P_\mu$ requires that 
    \begin{equation}
       \left<\mathcal{O}_1\mathcal{O}_2\right> = f(x^\mu), 
    \end{equation}
where $x^\mu=x^\mu_1-x^\mu_2$. 
    
    As shown in \cite{Chen:2021xkw}, by solving the Ward identities, the 2-point correlators of the operators in a CCFT is generically composed of two independent types, one being of the power-law form, the other being proportional to the Dirac $\delta$-function.  In \cite{Chen:2021xkw}, the authors have discussed the one of the power-law form in detail. In this appendix, we mainly focus on the 2-point correlators for the primary operators in chain representations, and pay more attention to the correlators which appear as the generalized functions\footnote{A nice introduction to the generalized functions can be found in \cite{gel1964properties}.} in general $d$ dimensions, including the Dirac $\delta$-functions.  The techniques used here is similar to the ones in \cite{Chen:2021xkw}, and we strongly recommend the reader to find more details there.\par

    It should also be stressed that here we only consider the correlators of the primary operators. Some operators in the staggered modules, like $\pi$ in the magnetic scalar theory, are special in the sense that they are neither primary ($KO\neq 0$) nor descendent,  and their correlators can not be constrained by the discussions here. Even though these operators do obey some Ward identities from their transformation laws, which help us to determine their correlators, there is short of general rules on the correlators of these operators.\par
    
    For the primary operators $\cO_1,\cO_2$, their 2-point correlation function $f=\left<\cO_1 \cO_2\right>$ is a homogeneous function by using the Ward identity of $D$,
    \begin{equation}
        D: \quad (t\partial_t+x^i\partial_i) f(t,\vec x) + (\Delta_1+\Delta_2) f(t,\vec x) =0.\label{DWard}
    \end{equation}
    The solution to this equation  is a combination of two independent solutions, the power-law functions and the generalized functions like the (derivatives of) Dirac $\delta$-distribution. For example, the one-dimensional version of this differential equation is
    \begin{equation}
         x\partial f(x) + \lambda f(x) = 0, \end{equation}
         with the solution being
                \begin{equation}\label{eq:1DHomogeneousGeneralizedFunction}
                f(x) = c_1 x^{-\lambda} + c_2 \partial^{(\lambda-1)}\delta(x),           \end{equation}
    where $c_i$ are constants, and $c_2 \neq 0$ for $\lambda = 1,2,...,$. In the Carrollian case, $t$ direction and $x_i$ directions could be considered separately, and thus the solution to \eqref{DWard} is simply 
    \begin{equation}
        f(t,\vec x) = g(t) g(\vec x),
    \end{equation}
    where $g(t)$ and $g(\vec x)$ are the homogeneous generalized functions  of the form \eqref{eq:1DHomogeneousGeneralizedFunction}. \par
    
    Another important constraint is from the Ward identity of $B_i$ on the lowest-level correlators $f=\left<\cO_1 \cO_2\right>$:
    \begin{equation}
        B_i: \quad x^i\partial_t f(t,\vec x) =0.
    \end{equation}
    By the ``lowest-level", we mean $[B_i, \cO_1]=[B_i, \cO_2]=0$. Considering the fact $x\delta(x) = 0$, we find four independent solutions, 
    \begin{equation}
        \begin{aligned}
            \partial_t f=0: 
                &\quad \left\{\begin{aligned} 
                    &f(t,\vec x)\propto P(\vec x), && && \textbf{(Case 1.1)}\\ 
                    &f(t,\vec x)\propto \prod_i \partial_i^{n_i}(\vec\partial^2)^n\delta^{(d-1)}(\vec x), && \Delta_1+\Delta_2 = d-1+\sum_i n_i + 2n, && \textbf{(Case 1.2)}
                \end{aligned}\right.\\[10pt]
            x^i f=0:        
                &\quad \left\{\begin{aligned} 
                    &f(t,\vec x)\propto P(t)\delta^{(d-1)}(\vec x), && && \qquad\quad \textbf{(Case 2.1)}\\ 
                    &f(t,\vec x)\propto \partial_t^{n_t}\delta(t)\delta^{(d-1)}(\vec x), && \Delta_1+\Delta_2 = d+ n_t, && \qquad\quad \textbf{(Case 2.2)} 
                \end{aligned}\right.
        \end{aligned}\nonumber
    \end{equation}
    where both $P(t)$ and $P(\vec x)$ are the power-law functions, and Case 1.2 appears for $\Delta_1+\Delta_2 = d-1, d, d+1,...$ and Case 2.2 appears for $\Delta_1+\Delta_2 = d, d+1, d+2,...$. In fact, the correlators of the primary operators in this paper belong to Case 2.1.
    
    The Case 1.1 with $f(t,\vec x)\propto P(\vec x)$ being the power-law function has been discussed in \cite{Chen:2021xkw}. In the rest of this section, we first repeat the constraints in Case 1.1 and then discuss the other situations. \par
    
\subsection{Case 1.1 and Case 1.2}\label{app:2ptCase1.1and1.2}
As shown in \cite{Chen:2021xkw},  the chain representations can have the following forms
    \begin{equation}
        \begin{aligned}
            (j)&\\
            (j)\rightarrow&(j), \qquad j\neq 0\\
            (0)\rightarrow(1)&\rightarrow(0), \\
            \cdots\rightarrow(j)\rightarrow(j+1)&\rightarrow(j+2)\rightarrow\cdots,\\
            \cdots\rightarrow(j)\rightarrow(j-1)&\rightarrow(j-2)\rightarrow\cdots.\\
        \end{aligned}
    \end{equation}
    For Case 1.1, the correlators could be the power-law functions of $x^\mu$, and the non-vanishing 2-point correlators only appear in the case that $\cO_1, \cO_2$ have (partially) inverse structure, and the selection rule is $\Delta_1=\Delta_2$. The correlator takes the form
    \begin{equation}
        \left<\cO_1\cO_2\right>=\frac{C~(t/|\vec x|)^r~I^{m_1,m_2}_{j_1,j_2}}{|\vec x|^{(\Delta_1+\Delta_2)}}\delta_{\Delta_1,\Delta_2},
    \end{equation}
    where $I$ is a rank-$0$ homogeneous function of $x_i$ representing the tensor structure of $O_i$. \par
    
    For Case 1.2, $\Delta_1+\Delta_2\ge d-1 \in \mathbb{Z}$, there exists another  solution for the lowest-level 2-point correlators, 
    \begin{equation}
    f(t,\vec x)\propto \prod_i\partial_i^{n_i}(\vec\partial^2)^n\delta^{(d-1)}(\vec x), \hspace{3ex} \sum_i n_i = \Delta_1+\Delta_2 - (d-1) -2n,\hspace{2ex}n_i\in \mathbf{N}^+.
    \end{equation} 
     For the higher-level correlators, the solutions are of the form $f^\prime(t,\vec x)\propto t^r \prod_i\partial_i^{n^\prime_i}(\vec\partial^2)^n\delta^{(d-1)}(\vec x)$ with $\sum_i n^\prime_i -2n -r = \Delta_1+\Delta_2 - (d-1), n^\prime_i\in \mathbf{N}^+$. The full restriction on the 2-point correlators in Case 1.2 is similar to Case 1.1, except the case that one of the operators is a scalar,  which will be discussed separately later. The reason that the selection rule is (almost) the same is that the power laws are proportional to (derivatives of) Dirac $\delta$-functions under canonical regularization  \cite{gel1964properties}:
    \begin{equation}
        \begin{aligned}
            \frac{2}{\Omega_{(d-1)}} \left.\frac{r^\lambda}{\Gamma\left(\frac{\lambda+d-1}{2}\right)}\right|_{\lambda=-(d-1)-2k}=\frac{(-1)^k (d-2)!}{2^k k! (d-1+2k-2)!}(\vec\partial^2)^k\delta^{(d-1)}(\vec x)
        \end{aligned}
    \end{equation}
    for $k=0,1,2,...$, with $r^2=\sum_i x_i^2$. As a result, most of the constraints from the Ward identities are the same as the ones in Case 1.1. Thus if $\Delta_1+\Delta_2\ge d-1 \in \mathbb{Z}$ and $\Delta_1=\Delta_2$, the correlators are non-vanishing for $\cO_1$ and $\cO_2$ in partially inverse representations, and the structures of the correlators are of the form
    \begin{equation}
        \left<\cO_{1,l_1}^{\{s_1\}} \cO_{2,l_2}^{\{s_2\}}\right>=C~t^r~(D_{s_1}D_{s_2}(\vec\partial^2)^n\delta^{(d-1)}(\vec x)-\text{traces}), \quad \text{with } D_{s_i}=\partial_{s_{i,1}}\cdots\partial_{s_{i,l_i}}
    \end{equation}
    The explicit selection rule is rather tedious, and we do not repeat them here. The interested readers may refer \cite{Chen:2021xkw} for detailed discussions. \par
    
    The exceptional situation in Case 1.2 is when one of the primary operators is in scalar representation $(0)$. In this case, there is one additional set of the selection rules, due to the special property of Dirac $\delta$-function.   In the following, we explain how this additional selection rule emerges and show the structure of the correlators in this situation.   Firstly, for the simplest case that both $\cO_1$ and $\cO_2$ are scalars with $\Delta_1+\Delta_2 = d-1$, the correlator is $f=\left<\cO_1 \cO_2\right> \propto \delta^{(d-1)}(\vec x)$ in Case 1.2. It is known that
    \begin{equation}
        \begin{aligned}
            \textbf{Case 1.1:} \quad &x_i f\propto\frac{x_i}{r^{(d-1)}} \neq 0, \\
            \textbf{Case 1.2:}\quad &x_i f\propto x_i \delta^{(d-1)}(\vec x) = 0, 
        \end{aligned}
    \end{equation}
    which makes the constraints from the Ward identities of $K_i$ on $f$ for Case 1.1 and 1.2 different, 
    \begin{equation}
        \begin{aligned}
            \textbf{Case 1.1:} \quad &\text{solution: } f=\frac{C_1}{r^{(d-1)}},  &&\text{constraint: }  \Delta_1=\Delta_2 = \frac{d-1}{2}, \\
            \textbf{Case 1.2:} \quad &\text{solution: } f=C_2\delta^{(d-1)}(\vec x) , &&\text{constraint: } \Delta_1+\Delta_2 = d-1. 
        \end{aligned}
    \end{equation}
    Thus for Case 1.2, we have the selection rule
    \begin{equation}
        \left<\cO_1 \cO_2\right>=C ~ \delta^{(d-1)}(\vec x), \hspace{3ex} \cO_1, \cO_2\in (0), \qquad \Delta_1+\Delta_2 = d-1.
    \end{equation}\par
    
    Next, we consider the case that $\cO_1$ is in more complicated chain representation. In the case that $\cO_1\in(j)$ is a symmetric traceless tensor (STT) with spin $j$, $\cO_2\in (0)$ is a scalar. Using the fact $x_i\partial_i\delta^{(d-1)}(\vec x) = -\delta^{(d-1)}(\vec x)$, we find that the restrictions from the Ward identities of $K_i$ are 
    \begin{equation}
        \left<\cO_1^{\{s_1,...,s_j\}} \cO_2\right>=C ~ (\partial_{s1} \cdots\partial_{s_{j}}\delta^{(d-1)}(\vec x)-\text{traces}), \qquad\Delta_1=1,\quad\Delta_2 = d-2+j.
    \end{equation}
    The ``traces" term is the trace of $\partial_{s1} \cdots\partial_{s_{j}}\delta^{(d-1)}(\vec x)$, and subtracting this term makes the correlators respect the traceless condition of $\cO_1$. Moreover for $\cO_1\in(j)_2\to(j)_1$ and $\cO_2\in (0)$, we have\footnote{Here we use subscripts to distinguish different sectors of $(j)_2\to(j)_1$ with the same spin $j$. Similar notation for  $(0)_3\to(1)_2\to(0)_1$ will appear below. } 
    \begin{equation}
        \begin{aligned}
            \left<\cO_{1,(j)_2}^{\{s_1,...,s_{j}\}} \cO_2\right>=C ~ (\partial_{s_1}\cdots\partial_{s_{j}}\delta^{(d-1)}(\vec x)-\text{traces}), \qquad\left<\cO_{1,\text{others}} \cO_2\right>=0,\\ \qquad \Delta_1=1,\quad\Delta_2 = d-2+j,
        \end{aligned}
    \end{equation}
    For $\cO_1$ being a decreasing chain, $O_1\in(j+n)\to(j+n-1)\cdots\to(j+1)\to(j)$ and $O_2\in (0)$, we have 
    \begin{equation}
        \left<\cO_{1,l_1=j+r}^{\{s_1,...,s_{l_1}\}} \cO_2\right>=\frac{C~t^r}{r!} ~ (\partial_{s_1}\cdots\partial_{s_{l_1}}\delta^{(d-1)}(\vec x)-\text{traces}), \qquad \Delta_1=1,\Delta_2 = d-2+j,
    \end{equation}
    where $\cO_{1,l_1}$ is the spin-$l_1$ part of $\cO_1$. For $\cO_1$ in an increasing chain representation, $\cO_1\in(j)\to(j+1)\cdots\to(j+n-1)\to(j+n)$, and $\cO_2\in (0)$, the correlators vanish except for the highest-rank sector in $\cO_1$. Namely, we have 
    \begin{equation}
        \left<\cO_{1,(j)}^{\{s_1,...,s_j\}} \cO_2\right>=C ~ (\partial_{s_1}\cdots\partial_{s_{j}}\delta^{(d-1)}(\vec x)-\text{traces}), \quad\left<\cO_{1,\text{others}} \cO_2\right>=0 , \quad \Delta_1+\Delta_2 = d-1+j.
    \end{equation}
    And finally, for $\cO_1\in(0)_3\to(1)_2\to(0)_1$ and $\cO_2\in (0)$, we have 
    \begin{equation}
        \left<\cO_{1,(0)_3} \cO_2\right>=C ~ \delta^{(d-1)}(\vec x), \qquad\left<\cO_{1,\text{others}} \cO_2\right>=0 , \qquad \Delta_1+\Delta_2 = d-1.
    \end{equation}
    We have presented all the exceptional cases involving a scalar primary operator. Here we only discuss the case that the other operator belong to a chain representation, and we do not discuss the case that the other operator is in a net-like representation. \par
    
\subsection{Case 2.1 and 2.2}
    
    Case 2.1 and Case 2.2 come from the fact that $x^i\delta^{(d-1)}(\vec x)=0$ solves the equation of the  Ward identities of $B_i$.  The selection rules for these two cases are very different from the ones in Case 1.1 and Case 1.2. \par
    
    First we consider Case 2.1 with the operators being the symmetric traceless tensors (STTs) and in the singlet representations $(j)$. Since the spacial dependence in the correlators is always $\delta^{(d-1)}(\vec x)$, the only possible non-vanishing  lowest-level correlator is from the case that $\cO_1$ and $\cO_2$ have the same spin, $l_1=l_2$. It can be checked that the Ward identities of $K_i$ are manifestly satisfied using the fact that $x^i\delta^{(d-1)}(\vec x)=0$, and there is no selection rule on $\Delta_1$ and $\Delta_2$. Therefore we have 
    \begin{equation}\label{eq:Case2.1SingletCorrelators}
        \begin{aligned}
            &\left<\cO_1 \cO_2\right>=C ~ t^{(d-1-\Delta_1-\Delta_2)}\delta^{(d-1)}(\vec x), &&l_1=l_2=0,\\
            &\left<\cO_1^{i_1} \cO_2^{j_1}\right>=C ~\delta^{i_1}_{j_1} t^{(d-1-\Delta_1-\Delta_2)}\delta^{(d-1)}(\vec x), &&l_1=l_2=1,\\
            &\left<\cO_1^{i_1 i_2} \cO_2^{j_1 j_2}\right>=C \left(\delta^{i_1}_{j_1}\delta^{i_2}_{j_2}+\delta^{i_1}_{j_2}\delta^{i_2}_{j_1}-\frac{2}{d-1}\delta^{i_1 i_2}\delta_{j_1 j_2}\right) t^{(d-1-\Delta_1-\Delta_2)}\delta^{(d-1)}(\vec x), &&l_1=l_2=2,\\
            &\qquad\qquad\qquad\qquad\qquad\qquad\qquad\qquad\qquad\qquad\vdots \\
            &\left<\cO_1^{i_1\cdots i_s} \cO_2^{j_1\cdots j_s}\right>=C \left(\delta^{i_1}_{(j_1}\cdots\delta^{i_s}_{j_s)}-\text{trace}\right) t^{(d-1-\Delta_1-\Delta_2)}\delta^{(d-1)}(\vec x), &&l_1=l_2=s.\\
        \end{aligned}
    \end{equation}
    The ``trace" term is to cancel the trace of $O_1$ indices and the trace of $O_2$ indices, as both $\cO_1$ and $\cO_2$ are STTs. The coefficient $C$'s are undetermined constants. \par
    
    For the chain representations, there are very limited restrictions for the correlators being non-vanishing. The calculations show that if two chain representations have the same sub-sector, the correlators of the operators in these subs-sectors and in the higher levels are non-vanishing. In other words, if 
    \begin{equation}
        \begin{aligned}
            \cO_1 &\in \cdots \to(j_{n+1})\to(j_{n})\to(j_{n-1})\to\cdots,\\
            \cO_2 &\in \cdots \to(j_{m+1})\to(j_{m})\to(j_{m-1})\to\cdots,\qquad \text{with } j_{n}=j_{m}
        \end{aligned}
    \end{equation}
    then 
    \begin{equation}
        \left<\cO_{1,l_1=j_{\ge n}} \cO_{2,l_2=j_{\ge m}}\right>\neq 0. 
    \end{equation}
    For the chains, there are the selection rules on $\Delta_1$ and $\Delta_2$, but the specific selection rule must be discussed case by case. For examples, we have
    \begin{equation}\label{eq:Case2.1CorrelatorofChain1}
        \begin{aligned}
            &\left<\cO_{1,l_1}^{\{s_1,...,s_{l_1}\}} \cO_{2,l_2}^{\{r_1,...,r_{l_2}\}}\right>\\
                &=C \frac{(d-1-\Delta_1-\Delta_2)! ~ t^{(d-1-\Delta_1-\Delta_2+l_1+l_2)}}{(d-1-\Delta_1-\Delta_2+l_1+l_2)!} (\partial_{s1}  \cdots \partial_{s_{l_1}}\partial_{r1} \cdots\partial_{r_{l_2}}\delta^{(d-1)}(\vec x) -\text{traces})\\
            &\qquad\qquad\qquad\qquad\qquad\qquad\text{for } \cO_1, \cO_2\in\cdots\to(2)\to(1)\to(0), \quad\text{with } \Delta_1=\Delta_2=1.
        \end{aligned}
    \end{equation}
    \begin{equation}
        \begin{aligned}
            &\left<\cO_{1,l_1}^{\{s_1,...,s_{l_1}\}} \cO_{2,l_2}^{\{r_1,...,r_{l_2}\}}\right>\\
                &=C \frac{(d-1-\Delta_1-\Delta_2)! ~ t^{(d-1-\Delta_1-\Delta_2+l_1+l_2-2)}}{(d-1-\Delta_1-\Delta_2+l_1+l_2)!} \left(\delta_{(s_1}^{(r_1} \partial_{s2}  \cdots \partial_{s_{l_1})}\partial^{r2} \cdots\partial^{r_{l_2})}\delta^{(d-1)}(\vec x) -\text{traces}\right)\\
            &\qquad\qquad\qquad\qquad\qquad\qquad\text{for } \cO_1, \cO_2\in\cdots\to(3)\to(2)\to(1), \quad\text{with } \Delta_1=\Delta_2=0.
        \end{aligned}
    \end{equation}
    Especially, we have 
    \begin{equation}\label{eq:Case2.1Correlatorof010Reps}
        \begin{aligned}
            &\begin{aligned}
                &\left<\cO_{1,(0)_3} \cO_{2,(0)_3}\right> = C~ \frac{t^{(d-1-2\Delta+2)}}{(5-2\Delta)(\Delta-3)} \partial^2\delta^{(d-1)}(\vec x)\\
            \end{aligned}\\[10pt]
            &\begin{aligned}
                &\left<\cO_{1,(0)_3} \cO_{2,(1)_2}^{r}\right> = C~ \frac{t^{(d-1-2\Delta+1)}}{(\Delta-3)} \partial_r\delta^{(d-1)}(\vec x)\\
                &\left<\cO_{1,(1)_2}^{s} \cO_{2,(0)_3}\right> = C~ \frac{t^{(d-1-2\Delta+1)}}{(\Delta-3)} \partial_s\delta^{(d-1)}(\vec x)\\
            \end{aligned}\\[10pt]
            &\begin{aligned}
                &\left<\cO_{1,(0)_3} \cO_{2,(0)_1}\right> = C~ t^{(d-1-2\Delta)} \delta^{(d-1)}(\vec x)\\
                &\left<\cO_{1,(1)_2}^{s} \cO_{2,(1)_2}^{r}\right> = C~ \frac{1-\Delta}{\Delta-3}t^{(d-1-2\Delta)}\delta_{sr}\delta^{(d-1)}(\vec x)\\
                &\left<\cO_{1,(0)_1} \cO_{2,(0)_3}\right> = C~ t^{(d-1-2\Delta)} \delta^{(d-1)}(\vec x) \qquad\qquad\qquad\qquad \left<\text{others}\right>=0\\
            \end{aligned}\\[10pt]
            &\qquad\qquad\qquad\qquad\qquad\text{for } \cO_1, \cO_2\in(0)_3\to(1)_2\to(0)_1, \text{ with } \Delta_1=\Delta_2=\Delta\\
        \end{aligned}
    \end{equation}\par

    The selection rule for Case 2.2 is the same as the ones for Case 2.1. Different from the relation between Case 1.1 and 1.2, there is no exceptional situation. The analog of the exceptional case in Case 1.2 is when $\Delta_1+\Delta_2= d$ with the correlator $f\propto \delta(t)\delta^{(d-1)}(\vec x)$, but the constraint from the  Ward identities of $K_i$ gives similar selection rules for Case 2.1 and 2.2.

    The correlators appeared in the main text are all of Case 2.1. 
    The primary operator in the electric scalar theory is the field $\phi$, and the correlator is $\left<\phi(x)\phi(0)\right>=\frac{i}{2}\abs{t} \delta^{(d-1)}(\vec x).$ It can be checked that the correlator satisfies the Ward identities, no matter if the temporal part is in power of $t$ or $\abs{t}$, and this correlator matches the form of \eqref{eq:Case2.1SingletCorrelators}. 
    Similar to the electric scalar theory, the magnetic scalar theory have the primary operator $\phi$ with $\left<\phi(x)\phi(0)\right>=0$, which obviously matches the form of \eqref{eq:Case2.1SingletCorrelators}.
    The primary operators in the electric sector of electromagnetic theory are $A_\mu = (A_0, A_i)$. They are in $(1)\to(0)$ representation and the corresponding correlators are \eqref{app:u1-elec-corr-full}. These correlators have the same form with \eqref{eq:Case2.1CorrelatorofChain1}. 
    Finally, the fundamental operators $A_\alpha = (A_v, A_i, A_0)$ in the magnetic sector of electromagnetic theory are primary operators , which are in $(0)\to(1)\to(0)$ representation.  Their correlators are in \eqref{app:u1-mag-corr-full} which match the ones in \eqref{eq:Case2.1Correlatorof010Reps} with $\Delta_1=\Delta_2=1$. \par

\bibliographystyle{JHEP}
\bibliography{refs.bib}
\end{document}